\def\mynocite#1{\relax}  

\documentstyle[prd,eqsecnum,aps,verbatim,macros,floats]{revtex}
\input psfig
\pssilent
\renewcommand{\thetable}{\arabic{table}}
\begin{document}

%
\title {$t\bar t$ production cross section in $p \bar p$ collisions at $\sqrt{s}
= 1.8$ TeV }
%
\author{                                                                      
V.M.~Abazov,$^{23}$                                                           
B.~Abbott,$^{58}$                                                             
A.~Abdesselam,$^{11}$                                                         
M.~Abolins,$^{51}$                                                            
V.~Abramov,$^{26}$                                                            
B.S.~Acharya,$^{17}$                                                          
D.L.~Adams,$^{56}$                                                            
M.~Adams,$^{38}$                                                              
S.N.~Ahmed,$^{21}$                                                            
G.D.~Alexeev,$^{23}$
% ...
\\                                                                            
\vskip 0.30cm                                                                 
\centerline{(D\O\ Collaboration)}                                             
\vskip 0.30cm                                                                 
}                                                                             
\address{                                                                     
\centerline{$^{1}$Universidad de Buenos Aires, Buenos Aires, Argentina}       
\centerline{$^{2}$LAFEX, Centro Brasileiro de Pesquisas F{\'\i}sicas,         
                  Rio de Janeiro, Brazil}                                     
\centerline{$^{3}$Universidade do Estado do Rio de Janeiro,                   
                  Rio de Janeiro, Brazil}                                     
\centerline{$^{4}$Institute of High Energy Physics, Beijing,                  
                  People's Republic of China}                                 
\centerline{$^{5}$Universidad de los Andes, Bogot\'{a}, Colombia}             
\centerline{$^{6}$Charles University, Center for Particle Physics,            
                  Prague, Czech Republic}                                     
\centerline{$^{7}$Institute of Physics, Academy of Sciences, Center           
                  for Particle Physics, Prague, Czech Republic}               
\centerline{$^{8}$Universidad San Francisco de Quito, Quito, Ecuador}         
\centerline{$^{9}$Institut des Sciences Nucl\'eaires, IN2P3-CNRS,             
                  Universite de Grenoble 1, Grenoble, France}                 
\centerline{$^{10}$CPPM, IN2P3-CNRS, Universit\'e de la M\'editerran\'ee,     
                  Marseille, France}                                          
\centerline{$^{11}$Laboratoire de l'Acc\'el\'erateur Lin\'eaire,              
                  IN2P3-CNRS, Orsay, France}                                  
\centerline{$^{12}$LPNHE, Universit\'es Paris VI and VII, IN2P3-CNRS,         
                  Paris, France}                                              
\centerline{$^{13}$DAPNIA/Service de Physique des Particules, CEA, Saclay,    
                  France}                                                     
\centerline{$^{14}$Universit{\"a}t Mainz, Institut f{\"u}r Physik,            
                  Mainz, Germany}                                             
\centerline{$^{15}$Panjab University, Chandigarh, India}                      
\centerline{$^{16}$Delhi University, Delhi, India}                            
\centerline{$^{17}$Tata Institute of Fundamental Research, Mumbai, India}     
\centerline{$^{18}$Seoul National University, Seoul, Korea}                   
\centerline{$^{19}$CINVESTAV, Mexico City, Mexico}                            
\centerline{$^{20}$FOM-Institute NIKHEF and University of                     
                  Amsterdam/NIKHEF, Amsterdam, The Netherlands}               
\centerline{$^{21}$University of Nijmegen/NIKHEF, Nijmegen, The               
                  Netherlands}                                                
\centerline{$^{22}$Institute of Nuclear Physics, Krak\'ow, Poland}            
\centerline{$^{23}$Joint Institute for Nuclear Research, Dubna, Russia}       
\centerline{$^{24}$Institute for Theoretical and Experimental Physics,        
                   Moscow, Russia}                                            
\centerline{$^{25}$Moscow State University, Moscow, Russia}                   
\centerline{$^{26}$Institute for High Energy Physics, Protvino, Russia}       
\centerline{$^{27}$Lancaster University, Lancaster, United Kingdom}           
\centerline{$^{28}$Imperial College, London, United Kingdom}                  
\centerline{$^{29}$University of Arizona, Tucson, Arizona 85721}              
\centerline{$^{30}$Lawrence Berkeley National Laboratory and University of    
                  California, Berkeley, California 94720}                     
\centerline{$^{31}$University of California, Davis, California 95616}         
\centerline{$^{32}$California State University, Fresno, California 93740}     
\centerline{$^{33}$University of California, Irvine, California 92697}        
\centerline{$^{34}$University of California, Riverside, California 92521}     
\centerline{$^{35}$Florida State University, Tallahassee, Florida 32306}      
\centerline{$^{36}$University of Hawaii, Honolulu, Hawaii 96822}              
\centerline{$^{37}$Fermi National Accelerator Laboratory, Batavia,            
                   Illinois 60510}                                            
\centerline{$^{38}$University of Illinois at Chicago, Chicago,                
                   Illinois 60607}                                            
\centerline{$^{39}$Northern Illinois University, DeKalb, Illinois 60115}      
\centerline{$^{40}$Northwestern University, Evanston, Illinois 60208}         
\centerline{$^{41}$Indiana University, Bloomington, Indiana 47405}            
\centerline{$^{42}$University of Notre Dame, Notre Dame, Indiana 46556}       
\centerline{$^{43}$Iowa State University, Ames, Iowa 50011}                   
\centerline{$^{44}$University of Kansas, Lawrence, Kansas 66045}              
\centerline{$^{45}$Kansas State University, Manhattan, Kansas 66506}          
\centerline{$^{46}$Louisiana Tech University, Ruston, Louisiana 71272}        
\centerline{$^{47}$University of Maryland, College Park, Maryland 20742}      
\centerline{$^{48}$Boston University, Boston, Massachusetts 02215}            
\centerline{$^{49}$Northeastern University, Boston, Massachusetts 02115}      
\centerline{$^{50}$University of Michigan, Ann Arbor, Michigan 48109}         
\centerline{$^{51}$Michigan State University, East Lansing, Michigan 48824}   
\centerline{$^{52}$University of Nebraska, Lincoln, Nebraska 68588}           
\centerline{$^{53}$Columbia University, New York, New York 10027}             
\centerline{$^{54}$University of Rochester, Rochester, New York 14627}        
\centerline{$^{55}$State University of New York, Stony Brook,                 
                   New York 11794}                                            
\centerline{$^{56}$Brookhaven National Laboratory, Upton, New York 11973}     
\centerline{$^{57}$Langston University, Langston, Oklahoma 73050}             
\centerline{$^{58}$University of Oklahoma, Norman, Oklahoma 73019}            
\centerline{$^{59}$Brown University, Providence, Rhode Island 02912}          
\centerline{$^{60}$University of Texas, Arlington, Texas 76019}               
\centerline{$^{61}$Texas A\&M University, College Station, Texas 77843}       
\centerline{$^{62}$Rice University, Houston, Texas 77005}                     
\centerline{$^{63}$University of Virginia, Charlottesville, Virginia 22901}   
\centerline{$^{64}$University of Washington, Seattle, Washington 98195}       
}                                                                             

\maketitle
%
\begin{abstract}
Results are presented on a measurement of the $t \bar t$ pair production
cross section in $p \bar p$ collisions at $\sqrt{s} = 1.8$ TeV from nine
independent decay channels. The data were collected by the D\O\ 
experiment during the 1992--1996 run of the Fermilab Tevatron Collider. 
A total of 80 candidate events is observed with an expected background 
of $38.8 \pm 3.3$ events. For a top quark mass of 172.1 GeV/$c^2$,
the measured cross section is 
$5.69 \pm 1.21 ({\rm stat}) \pm 1.04 ({\rm sys})$ pb.
\end{abstract}
%


\twocolumn
\tableofcontents
\newpage
\section {Introduction}
\label{intro}
\label{sec:intro}
\def\ttb{t\bar t}
\def\ppb{p\bar p}
\def\aplan{$\cal{A}$}                   
\def\MEt{{\mbox{$E\kern-0.57em\raise0.19ex\hbox{/}_{T}$}} }
\def\MEx{{\mbox{$E\kern-0.57em\raise0.19ex\hbox{/}_{x}$}}\ }
\def\MEy{{\mbox{$E\kern-0.57em\raise0.19ex\hbox{/}_{y}$}}\ }
\def\ra{\rightarrow}
\def\et{\mbox{$E_T$}}
\def\met{\mbox{$\rlap{\kern0.25em/}E_T$}}
\def\metc{\mbox{$\rlap{\kern0.25em/}E_T^{\rm cal}$}}
\def\gev{\mbox{$\mbox{ GeV}$~}}
\def\gevn{\mbox{$\mbox{ GeV}$}}
\def\gevc{\mbox{$\mbox{ GeV}/c$~}}
\def\gevcn{\mbox{$\mbox{ GeV}/c$}}
\def\gevcc{\mbox{$\mbox{GeV}/c^2$~}}
\def\gevccn{\mbox{$\mbox{GeV}/c^2$}}
\def\pt{\mbox{$p_T$}}
\def\zmumu{\mbox{$Z \rightarrow \mu\mu $~}}
\newcommand{\emeas}{$E_{\rm meas}^{\rm jet}$}
\newcommand{\eptcl}{$E_{\rm ptcl}^{\rm jet}$}
\newcommand\metcal{\mbox{${\hbox{$E$\kern-0.6em\lower-.1ex\hbox{/}}}_T^{\text{cal}}$}}%

The observation of the top quark by the CDF and D\O\ collaborations 
in the spring of 1995 \cite{cdfdisc,d0disc} was the culmination of a 
long and intensive 
search that began following the discovery of the $\tau$ lepton in 1976 
\cite{taudisc} and the bottom ($b$) quark in 1977 \cite{bdisc}. 
The discovery of these two 
particles gave a firm foundation to the existence of a third family, 
originally proposed by Kobayashi and Maskawa in 1973 to account for the
occurrence of $CP$ violation within the standard model \cite{KM}.
The $b$ quark was shown to possess a charge of 
$Q_b = -{1 \over 3}e$~\cite{qb1,qb2,qb3} and a weak isospin of 
$I_3 = -{1 \over 2}$ 
\cite{i31,i32,i33}. Within the standard model (SM), this demanded the 
existence of a partner to the $b$ quark with a charge of 
$+{2 \over 3}e$ and a 
weak isospin of $+{1 \over 2}$. This partner is called the ``top'' quark. 

\begin{figure}
\vbox{
\vskip -0.2 in
\centerline{\hskip 0.2 in \psfig{figure=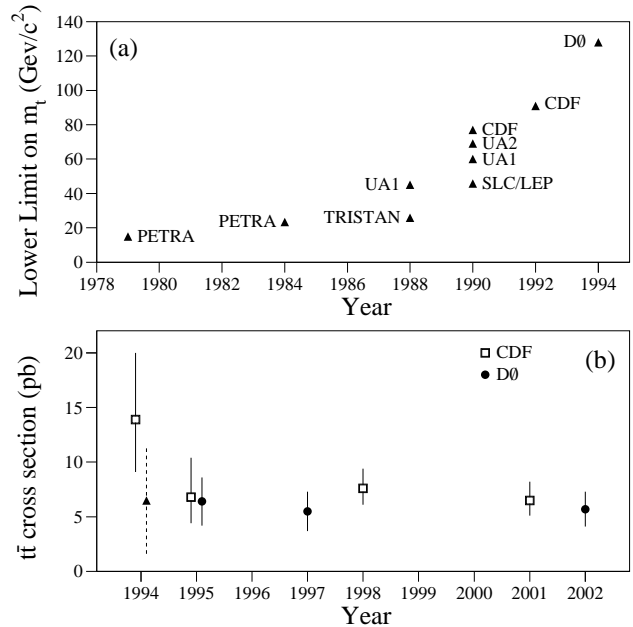,width=3.6in}}
\caption{(a) Lower limit on the top quark mass from 1978 to 1994
 \protect\cite{petra1,petra2,tristan,markII,opal,ua11988,ua11990,ua21990,cdf1990,cdf1992,d01994}. (b) Published $\ttb$ quark cross section results from 1994 
 to 2001~\protect\cite{cdf1994,cdfdisc,d0disc,d01997,cdf1998,cdf2001}.
The solid triangle marker with the dashed line uncertainty corresponds to the 
unpublished D\O\ $\ttb$ cross section in mid-1994~\protect\cite{glasgowd0}.
\label{fig:top_his}}}
\end{figure}

Initial searches for the top quark were carried out at $e^+e^-$
colliders. These searches looked for a narrow resonance (if a bound $\ttb$ 
state was produced), an increase in the rate of $e^+e^- \ra {\rm hadrons}$
(if a bound $\ttb$ state was not produced), or events with more
spherical angular distributions which differentiate top quark events from
the more planar angular distributions expected from the lighter quarks.
As shown in Fig.~\ref{fig:top_his}(a), experiments at the 
PETRA~\cite{petra1,petra2}, TRISTAN~\cite{tristan}, and 
SLC/LEP~\cite{markII,opal} colliders raised the lower limit on 
the top quark mass ($m_t$)
from 15 \gevcc in 1979 to 45.8 \gevcc in 1990. 
In the late 1980's, in the absence of a signal, the focus of the top quark 
search shifted from $e^+e^-$ colliders to $\ppb$ colliders and higher
center-of-mass energies. Unlike $e^+e^-$ colliders, 
$\ppb$ colliders cannot provide direct limits on the mass of the top quark,
but rather upper limits on the $\ttb$ production cross section. 
By assuming a relationship between mass and cross section (as 
provided by SM theory), these cross section upper limits can be turned into
lower limits on the mass. The UA1 collaboration provided the first such 
limit in 1988, setting a lower bound on the top quark mass of 
45 \gevccn~\cite{ua11988}. This limit was followed in 1990 by an updated 
limit from UA1 (60 \gevccn)~\cite{ua11990} and new limits from UA2 and CDF 
(69~\cite{ua21990} and 77~\cite{cdf1990}~\gevcc respectively). In 1992,
CDF raised the lower limit on the top quark mass to 91~\gevccn~\cite{cdf1992}, 
and in 1994, D\O\ set a lower bound of 128~\gevccn~\cite{d01994}. 

The first evidence for $\ttb$ production was claimed by the CDF collaboration
in April of 1994~\cite{cdf1994}. With an integrated luminosity of 19.3 $\ipb$, 
CDF observed twelve candidate events with an expected background of about six 
events and estimated a 0.26\% probability for the background to fluctuate to 
at least twelve events. The excess was assumed to be due to $\ttb$ production 
and the cross section was determined to be 
$\sigma_{\ttb} = 13.9^{+6.1}_{-4.8}$~pb for $m_t = 174$ \gevccn.
The D\O\ analysis in mid-1994~\cite{glasgowd0} based on
13.5 $\ipb$ yielded 7 events with an expected background
of $3.2\pm 1.1$ events.  The D\O\ and CDF sensitivities (expected number
of events for a given cross section) and expected significance
(signal to background ratio) were the same.   The small excess
seen in D\O, if interpreted as being due to $\ttb$ production,
gave a cross-section of $6.5\pm 4.9$ pb for $m_t = 180$ \gevccn. 
At the time of the top quark 
discovery the following year, the CDF and D\O\ collaborations reported $\ttb$ 
production cross sections of 
$\sigma_{\ttb} = 6.8^{+3.6}_{-2.4}$~pb for $m_t = 176$ \gevcc \cite{cdfdisc} 
and $\sigma_{\ttb} = 6.4\pm 2.2$~pb for $m_t = 199$ \gevcc \cite{d0disc}, 
respectively. 
These results were updated by D\O\ (1997) and CDF (1998) to 
$\sigma_{\ttb} = 5.5\pm 1.8$~pb~\cite{d01997} for $m_t = 173.3$ \gevcc and
$\sigma_{\ttb} = 7.6^{+1.8}_{-1.5}$~pb~\cite{cdf1998} for $m_t = 175$ \gevccn, 
respectively. In 2001, 
the CDF collaboration reported 
$\sigma_{\ttb} = 6.5^{+1.7}_{-1.4}$~pb for $m_t = 175$ \gevcc \cite{cdf2001} 
as their final $\ttb$ 
production cross section based on the 1992--1996 run of the Tevatron. 
The corresponding result from the D\O\ collaboration, reported in this 
article, is $\sigma_{\ttb} = 5.7\pm 1.6$~pb for $m_t = 172.1$ \gevccn.

At the Tevatron center-of-mass energy of 1.8 TeV, 
top quarks can be produced singly or in pairs. 
The two cross sections are of similar 
magnitude~\cite{singtop} but single top quark events are much more difficult 
to distinguish from background and have not yet been 
observed~\cite{d0singt1,d0singt2}. This paper is thus concerned only with 
$t \bar t$ pair production. 

\begin{figure}[t]
\vbox{
\vskip 0.1 in
\centerline{\psfig{figure=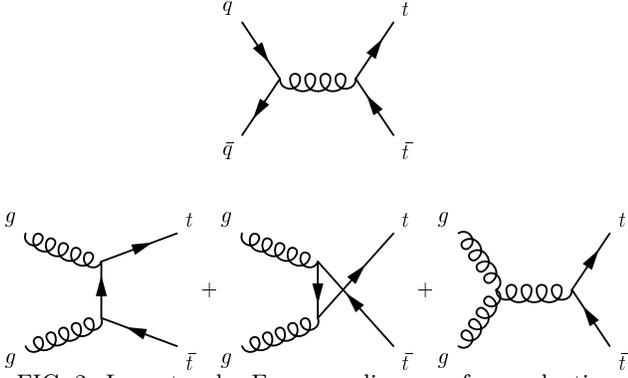,width=3.3in}}
\caption{Lowest order Feynman diagrams for production of $t\overline{t}$
pairs at the Tevatron. At Tevatron energies, the diagram involving 
quark-antiquark fusion dominates over those involving gluon-gluon fusion.
\label{fig:ttbfeyn}}}
\end{figure}

The $p\bar p \ra t\bar t$ production cross section can be factorized
in terms of the parton-parton cross section and the parton distribution
functions for the proton and anti-proton, and is written~\cite{css}
\begin{eqnarray}
\sigma(p\bar p \ra t\bar t) =  \nobreak \nonumber \hskip 2in \\ 
 \sum_{i,j} \int dx_i dx_j f_i^p(x_i, \mu^2) f_j^{\bar p}(x_j, \mu^2) 
{\hat{\sigma}}_{ij}(\hat{s},\mu^2,m_t),
\label{eq:sigmattb}
\end{eqnarray}
where the summation indices $i$ and $j$ run over the light quarks and 
gluons, $x_i$ and $x_j$ are the momentum fractions of the partons involved
in the $p\bar p$ collision, $f_i^p(x_i, \mu^2)$ and $f_j^{\bar p}(x_j, \mu^2)$
are the parton distribution functions, and 
${\hat{\sigma}}_{ij}(\hat{s},\mu^2,m_t)$
is the parton-parton cross section at $\hat{s} = x_i x_j s$. 
The renormalization and factorization scales, typically chosen to be the
same value $\mu$, are arbitrary parameters with dimensions of energy. 
The former is introduced by the renormalization procedure and the latter
by the splitting of the cross section into perturbative ($\hat{\sigma}$) and
nonperturbative ($f^p,f^{\bar p}$) parts. An exact calculation of the 
cross section would be independent of the choice of $\mu$, but current
calculations are performed to finite order in perturbative QCD and are thus
dependent on $\mu$, which is usually taken to be of the order of $m_t$.
Theorists typically estimate the uncertainty introduced by truncating the
perturbation expansion by varying $\mu$ over some arbitrary range, usually 
$m_t/2 < \mu < 2m_t $ (the range used for all theoretical cross sections
referred to in this paper).

In leading-order QCD (LO), ${\cal O}(\alpha_s^2)$, $t \bar t$ production 
proceeds through $q\bar q \ra t\bar t$ and $gg \ra t\bar t$ 
processes (see Fig.~\ref{fig:ttbfeyn}). At $\sqrt{s}=1.8$ TeV, the 
$q\bar q \ra t\bar t$ process dominates, contributing 90\% of the cross
section with the $gg \ra t\bar t$ process contributing only 10\%.
The first calculations of the LO cross section $\hat{\sigma}$ were performed
in the late 1970's 
\cite{georgi78,jones78,gluck78,babcock78,hagiwara79,combridge79}. 
Calculations of the $t\bar t$ production cross section at next-to-leading 
order (NLO), ${\cal O}(\alpha_s^3)$, began
to appear in the late 1980's 
\cite{dawson88,dawson89,beenakker89,beenakker91,mangano92,altarelli88,ellis91}.
The 1990's saw the introduction of calculations which attempt to estimate the 
contribution of the higher order terms through a technique known as 
{\sl resummation}, in which the sums of the dominant logarithms from soft 
gluon emission to all orders in perturbation theory are calculated, thus 
reducing the dependence of the cross section on the value of $\mu$. The first 
such calculations~\cite{laenen,berger3} summed only leading-log (LL) 
contributions. Increased precision was soon achieved through 
calculations~\cite{bonciani,kidonakis99} which incorporated summations 
through next-to-leading-log (NLL) contributions. The most recent 
calculations~\cite{kidonakis0105,kidonakis} sum contributions through 
next-to-next-to-leading-log (NNLL). Although the NLL and NNLL calculations 
have reduced the scale dependence, kinematic-induced ambiguities lead to 
estimated uncertaintied of about 7\% (these latter uncertainties are not
included in the theoretical cross section predictions given in this paper).

In the SM, the top quark is expected to decay predominantly into a $W$ boson 
and a $b$ quark. Decay mechanisms whereby the top quark decays into a
charged Higgs boson are not considered here, but are investigated
in Refs.~\cite{d0chiggs,d0chiggs2,cdfchiggs}.
The channels in which the top quark is sought are thus determined by the decay
modes of the two $W$ bosons in the $t \bar t$ event. The $W$ boson can decay
leptonically into an electron, muon, or a $\tau$ lepton (and associated
neutrino), and hadronically into $u\overline{d}$, $u\overline{s}$, 
$u\overline{b}$, $c\overline{d}$, $c\overline{s}$, or $c\overline{b}$ pairs. 

\begin{figure}
\vbox{
\vskip 0.05in
\centerline{\hskip -0.25 in \psfig{figure=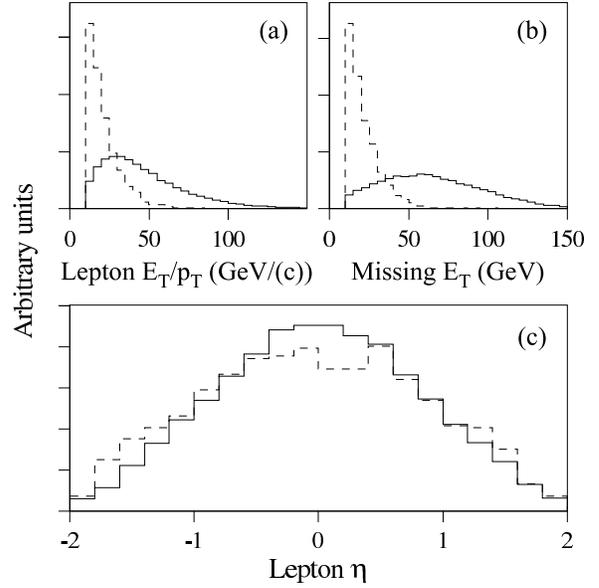,width=3.0in}}
\vskip 0.07in
\caption{ Expected distributions for $e\mu$ dilepton events of (a) electron
$E_T$ or muon $p_T$, (b) $\met$, and 
(c) lepton $\eta \equiv {\rm tanh}^{-1}({\rm cos}\theta)$ 
(two entries per event). 
The solid histograms are $t\bar t \ra e\mu + X$
signal events (generated with \progname{herwig}~\protect\cite{herwig} with 
$m_t$ = 175 \gevcc
for $p\bar p$ collisions at $\sqrt{s}=1.8$ TeV). The dashed histograms are
$Z+{\rm jets}\ra \tau\tau + {\rm jets} \ra e\mu + {\rm jets}$ events 
(also generated with \progname{herwig}). All histograms are normalized to
unity and all events are required to have 
$p_T^{\ell} > 10 \gevcn, \ \met > 10$ GeV,
and at least two jets with $E_T > 15$ GeV and $|\eta| < 2.0$.
\label{fig:int_dil_lep}}}
\end{figure}

\begin{figure}
\vbox{
\vskip 0.1cm
\centerline{\hskip -0.25in \psfig{figure=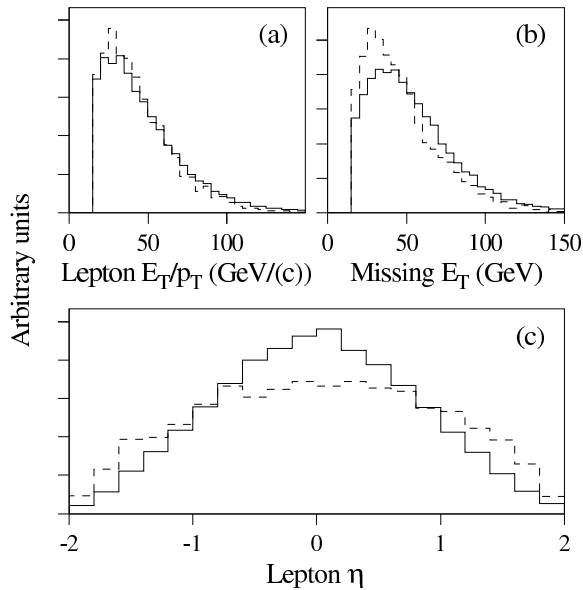,width=3.0in}}
\vskip 0.07in
\caption{ Expected distributions for lepton+jets events of (a) electron
$E_T$ and muon $p_T$ (two entries per event),
(b) $\met$, and (c) lepton $\eta$. The solid histograms are $t\bar t$ 
signal events (generated with \progname{herwig} with $m_t$ = 175 \gevcc
for $p\bar p$ collisions at $\sqrt{s}=1.8$ TeV). The dashed histograms are
$W+\geq 4$ jet events (generated with 
\progname{vecbos}~\protect\cite{vecbos}). 
All histograms are normalized to unity and all events are required to have 
$p_T^{\ell} > 15 \gevcn, \ \met > 15$ GeV, and at least four jets with 
$E_T > 15$ GeV and $|\eta| < 2.0$.
\label{fig:int_lj_lep}}}
\end{figure}

\begin{figure}
\vbox{
\centerline{\hskip -0.25in \psfig{figure=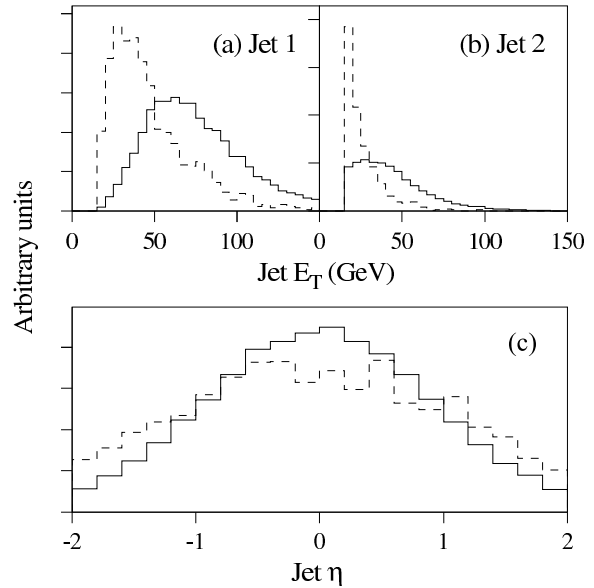,width=3.0in}}
\vskip 0.07in
\caption{ Expected distributions for $e\mu$ dilepton events of (a and b) 
the transverse energies of the two leading jets and (c) the jet $\eta$ 
(two entries per event). The solid histograms are $t\bar t \ra e\mu + X$
signal events (generated with \progname{herwig} with $m_t$ = 175 \gevcc
for $p\bar p$ collisions at $\sqrt{s}=1.8$ TeV). The dashed histograms are
$Z+{\rm jets}\ra \tau\tau + {\rm jets} \ra e\mu + {\rm jets}$ events 
(also generated with \progname{herwig}). All histograms are normalized to
unity and all events are required to have 
$p_T^{\ell} > 10 \gevcn, \ \met > 10$ GeV, and at least two jets with 
$E_T > 15$ GeV and $|\eta| < 2.0$.
\label{fig:int_dil_jet}}}
\end{figure}

\begin{figure}[t]
\vbox{
\vskip 0.1cm
\centerline{\hskip -0.25in \psfig{figure=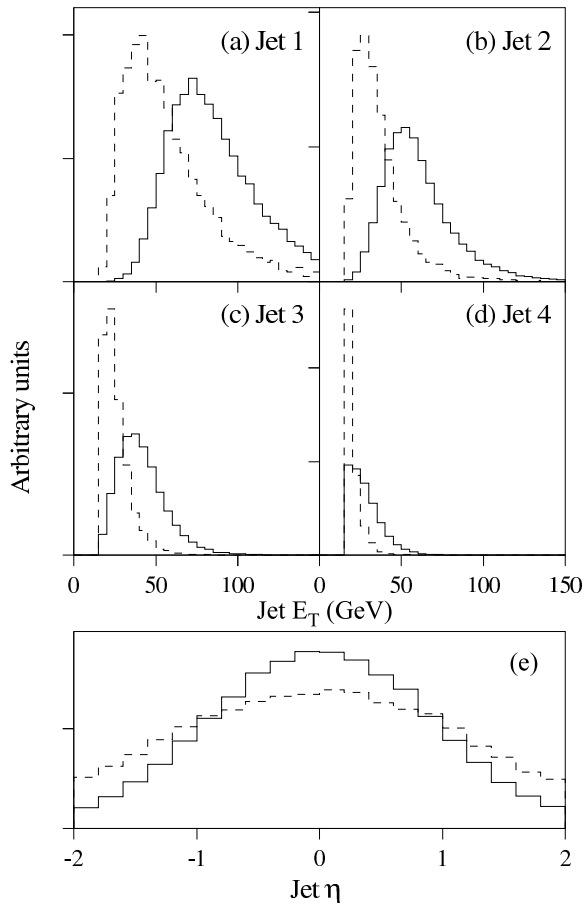,width=3.0in}}
\vskip 0.07in
\caption{ Expected distributions for lepton+jets events of (a--d) the
transverse energies of the four leading jets and (e) the jet $\eta$ 
(four entries per event). The solid histograms are $t\bar t$ 
signal events (generated with \progname{herwig} with $m_t$ = 175 \gevcc
for $p\bar p$ collisions at $\sqrt{s}=1.8$ TeV). The dashed histograms are
$W+\geq 4$ jet events (generated with \progname{vecbos}). All histograms are 
normalized to unity and all events are required to have 
$p_T^{\ell} > 15 \gevcn, \ \met > 15$ GeV, and at least four jets with 
$E_T > 15$ GeV and $|\eta| < 2.0$.
\label{fig:int_lj_jet}}}
\end{figure}

The channels can be classified as follows: the dilepton channel where
both $W$ bosons decay leptonically into an electron or a muon 
($ee$, $\mu\mu$, $e\mu$), the lepton + jets channel where one of the $W$ 
bosons decays leptonically and the other hadronically ($e$+jets, $\mu$+jets), 
and the all-jets channel where both $W$ bosons decay
hadronically. This paper will focus primarily on the dilepton and lepton + jets
channels. The all-jets channel is discussed in detail in Ref.~\cite{alljets} 
and is only summarized here. The $\ttb$ channels containing a tau lepton are 
not 
explicitly considered, although events containing $\tau \ra e\nu\bar\nu$ and 
$\tau \ra \mu\nu\bar\nu$ decays do contribute to the efficiency of all 
channels 
containing an electron or a muon. Similarly, the inability to distinguish 
between a hadronic tau decay and a hadronic jet, contributes to the efficiency
of the lepton+jets channels. As is indicated in 
Figs.~\ref{fig:int_dil_lep}--\ref{fig:int_lj_jet}, the leptonic channels are 
characterized by high transverse-momentum ($p_T$) 
leptons and jets as well as missing transverse momentum 
($\met$) due to high $p_T$ neutrinos (see Sec.~\ref{misset}). 
The plots show the distributions 
of several kinematic quantities expected from $t\bar t$ decay compared with 
those expected from the leading background for the $e\mu$ 
(Figs.~\ref{fig:int_dil_lep} and \ref{fig:int_dil_jet}) and lepton+jets
(Figs.~\ref{fig:int_lj_lep} and \ref{fig:int_lj_jet}) channels. 
Initial search strategies are based on previous studies and 
analyses~\cite{topstrat,cdf1994,prd1}.

The paper is structured as follows: Sec.~\ref{d0det} gives a brief
overview of the D\O\ detector and indicates those aspects which were
employed in the dilepton and lepton+jets analyses. Section~\ref{triggers} 
describes the triggers used in the first stage of the event selection.
Event reconstruction and particle identification are the subjects of 
Sec.~\ref{reco}. Section~\ref{mc} discusses the simulation of the 
$t \bar t$ signal and background. The dilepton channels are described in 
Sec.~\ref{dilep} and the lepton+jets channels are described in 
Sec.~\ref{ljets_intro}. The all-jets channel is described briefly in 
Sec.~\ref{alljets}. Section~\ref{syserr} discusses the 
systematic uncertainties. The $t\bar t$ cross section 
results are summarized and tabulated in Sec.~\ref{crsec} and the 
conclusions to be drawn from the combined analyses are presented in 
Sec.~\ref{concl}. Appendix~\ref{escale} describes the corrections applied 
to the jet energy scale; Appendices~\ref{mrveto} and \ref{mrrecov} describe 
the main-ring veto and recovery; Appendix~\ref{emuNN} presents an independent 
Neural Network based analysis of the $e\mu$ channel; and 
Appendix~\ref{correrr} describes in detail the handling of the uncertainties 
and the correlations between them.

%
\section {The D\O\ detector}
\label{d0det}

\begin{figure}
\vbox{
\vskip 0.2cm
\centerline{\psfig{figure=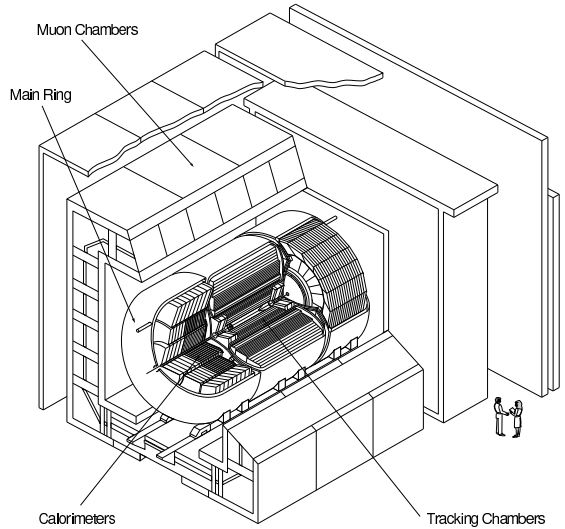,width=3.0in}}
\vskip -2.2cm
\caption{Cutaway view of the D\O\ detector, showing the tracking chambers,
calorimetry, and muon system.
\label{fig:det}}}
\end{figure}

D\O\ is a multipurpose detector designed to study  $p \bar p$ collisions at
high energies. The detector was commissioned at the Fermilab Tevatron
Collider during  the summer of 1992. The work presented here is based on 
approximately 125 pb$^{-1}$ of data recorded between August 1992 
and February 1996. A full description of the detector may be found in 
Ref.~\cite{d0nim}. This section describes briefly those properties of the 
detector that are relevant for the $t\bar t$ production cross section 
measurements.

Spatial coordinates are specified in a system with the origin at the center
of the detector and the positive $z$-axis pointing in the direction of the
proton beam. The $x$-axis points radially out of the Tevatron ring and 
the $y$-axis points upward. Due to the approximate cylindrical symmetry of
the detector, it is also convenient to use the variables $r$ (the
perpendicular distance from the beamline), $\phi$ (the azimuthal angle
with respect to the $x$-axis), and $\theta$ (the polar angle with
respect to the $z$-axis). The polar direction is usually described by the
pseudorapidity, defined as 
$\eta \equiv {\rm tanh}^{-1}({\rm cos} \theta$).

In the previous section it was noted that the final state from $t\bar t$
decay may contain electrons, muons, jets, and neutrinos. The D\O\ detector
was designed to identify and measure the energy/momentum of all of these
objects. As shown in Fig.~\ref{fig:det}, the detector has three major 
subsystems: the central tracking chambers, a uranium liquid-argon 
calorimeter, and a muon spectrometer. The detector design was optimized for 
high-resolution, nearly hermetic calorimetry that provides the sole measurement
of the energies of electrons and jets. Due to the compact design of the
calorimeter, the inner tracking volume is relatively small, and there is no
central magnetic field. 

The central tracking detectors measure the trajectories of
charged particles and aid in the identification of electrons. The former 
function is performed using three wire-chamber systems, and the latter
by a transition-radiation detector (TRD). The three wire-chamber systems 
consist
of two concentric cylindrical chambers centered on the interaction point and 
a set of two forward drift chambers that are situated at the ends of the 
cylinder. These chambers provide charged-particle 
tracking over the region $|\eta| < 3.2$, measuring the trajectories of 
charged particles with a resolution of 2.5 mrad in $\phi$ and 28 mrad in 
$\theta$. The position of the interaction vertex 
along the beam direction ($z$) can be determined with a resolution of 8 mm. 
These chambers also measure the track ionization for distinguishing 
singly charged particles and $e^+e^-$ pairs from photon conversions.
Concentric with, and radially between, the two central chambers is the TRD.
By measuring the amount of radiation 
emitted by single isolated particles as they pass through many thin sheets 
of polypropylene, this detector aids in the separation of electrons 
from charged pions and $\pi^{\pm}/\gamma$ overlaps (since the amount of 
emitted transition radiation is proportional to the value of $E/m$ for 
the particle). This device provides a factor of ten rejection of pions
while retaining 90\% of isolated electrons.

Surrounding the central tracking system is the calorimeter, which is composed
of plates of uranium and stainless steel/copper absorber surrounded by 
liquid argon as the sensitive ionization medium.
The calorimeter is divided into three parts, the central calorimeter (CC),
$|\eta| \leq 1.2$, and
two end calorimeters (EC), which together cover the pseudorapidity range
$|\eta| < 4.2$. Each consists of an inner electromagnetic (EM) section, 
a fine hadronic (FH) section, and a coarse hadronic (CH) section, housed in a 
steel cryostat. Each EM section is 21 radiation lengths deep and 
is divided into four longitudinal segments (layers). The hadronic sections 
are 7--9 nuclear interaction lengths deep and are divided into four (CC) or 
five (EC) layers. The outer layer of each hadronic calorimeter is known as the 
``outer hadronic layer''. The calorimeter is transversely segmented
into pseudo-projective towers with 
$\Delta\eta\times\Delta\phi$  = $0.1 \times 0.1$. The third layer
of the EM calorimeter, in which the maximum of EM showers is expected, 
is segmented twice as finely into cells with 
$\Delta\eta\times\Delta\phi$  = $0.05 \times 0.05$. With this fine 
segmentation, the azimuthal position resolution for electrons with energy
above 50 GeV is about 2.5 mm. The energy resolution is
$\sigma(E)/E = 15\%/ \sqrt{E\hbox{(GeV)}} \oplus 0.4\%$
for electrons. For charged pions the resolution is about
$50\%/ \sqrt{E\hbox{(GeV)}}$ and for jets it is about 
$80\%/ \sqrt{E\hbox{(GeV)}}$~\cite{d0nim}. 
For minimum bias data, the resolution for each component of
{\hbox{$\rlap{\kern0.25em/}E_T$}},
{\hbox{$\rlap{\kern0.25em/}E_{x}$}} and {\hbox{$\rlap{\kern0.25em/}E_{y}$}},
has been measured to be $1.08\hbox{ GeV} + 0.019 (\Sigma E_{T})$, 
where $\Sigma E_{T}$ is the scalar sum of the transverse energies in all
calorimeter cells. In order to improve the energy resolution for jets that 
straddle two cryostats, an inter-cryostat detector (ICD) made of 
scintillator tiles is situated in the space between the EC and CC cryostats.
In addition, separate single-cell structures called ``massless gaps'' (MG) 
are installed in the intercryostat region in both the CC and EC calorimeters.

The D\O\ muon detection systems cover $|\eta|\leq 3.3$.
Since muons from top quark decays predominantly populate the central
region, this work uses only the wide-angle muon spectrometer (WAMUS) which
consists of four planes of proportional drift tubes (PDT) in front of
magnetized iron toroids with a magnetic field of 1.9~T and two groups of three
planes each of proportional drift tubes behind the toroids.
The magnetic field lines and the wires in the drift tubes are oriented
transversely to the beam direction.
The WAMUS covers the region $|\eta| < 1.7$ over the entire azimuth, with the
exception of the central region below the calorimeter ($|\eta| < 1$,
$225^{\circ}< \phi  < 315^{\circ}$), where the inner layer is missing to make
room for the calorimeter support-structure. The WAMUS system is
divided into the {\em central iron} (CF), $|\eta| \leq 1.0$, and 
{\em end iron} (EF), $1.0 < |\eta| \leq 1.7$, regions. As will be discussed in
Sec.~\ref{muid}, the EF region was used for only part of the Run 1 data set.
The total thickness of the material in the calorimeter and iron toroids 
varies between 13 
and 19 interaction lengths, making background from hadronic punchthrough 
negligible. 
The tracking volume is small, thereby reducing backgrounds to prompt muons 
from in-flight decays of $\pi$ and $K$ mesons.
The muon momentum $p$ is measured from its deflection 
angle in the magnetic field of the toroid.
The momentum resolution is limited by multiple Coulomb scattering in the 
material traversed, the position resolution in the muon chambers, and
uncertainty in the magnetic field integral. The typical resolution in $1/p$
is approximately Gaussian and given by 
\begin{equation}
\delta(1/p) = 0.18(p-2)/p^2 \oplus 0.003
\label{eq:mumomres}
\end{equation}
(with $p$ in GeV/$c$).

As shown in Fig.~\ref{fig:det}, a separate synchrotron, the Main Ring, 
sits above the
Tevatron and passes through the forward muon system and the outer hadronic
section of the calorimeters. During data taking, it was used to accelerate
protons for antiproton production. Losses from the Main Ring 
can deposit energy in the calorimeters and muon system, increasing the 
instrumental background. As discussed below (Secs.~\ref{triggers}, 
\ref{dilep}, and \ref{ljets_intro}), these 
``Main-Ring events'' are removed during the initial selection of all channels.
Nevertheless, as discussed in Appendix~\ref{mrrecov}, and Secs.~\ref{ee} and 
\ref{ljets_topo}, several analyses have been able
to recover some, or all, of these events.

%
\section{Triggers}
\label{triggers}
During normal operation, the Tevatron maintains two counter-rotating beams,
one consisting of six bunches of protons and the other consisting of six
bunches of antiprotons. Proton and antiproton bunches collide at the D\O\ 
interaction region every 3.5 $\mu$s (286 kHz).
The D\O\ trigger system is used to select the interesting events and 
reduce this
to a rate of approximately 3--4 Hz, suitable for recording on tape.

The D\O\ trigger system is composed of three hardware stages (level 0, 
level 1, and level 1.5) and one software stage (level 2)~\cite{d0nim,prd1}.  
The first stage (level 0) consists of hodoscopes of scintillation 
counters mounted close to the beam on the inner surfaces of the
end-calorimeter cryostats and registers
hits consistent with a $p\bar p$ interaction. This stage is typically
used as an input to level 1, but level 0 is not
required to fire before an event can proceed to the next stage.
In addition, level 0 is used to measure the luminosity.
The next stage (level 1) forms fast analog sums of the transverse
energies in calorimeter towers. These towers have a size of 
$\Delta\eta\times\Delta\phi = 0.2\times 0.2$ and are segmented 
longitudinally into electromagnetic and hadronic sections. Based on
these sums and patterns of hits in the muon spectrometer, the level 1
trigger decision takes place within the space of a single beam crossing, 
unless a level 1.5 decision is required (see below).
Events accepted at level 1 are digitized and passed on
to the level 2 trigger which consists of a farm of 48 general-purpose
processors. Software filters running on these processors make the final
trigger decision.

At both level 1 and level 2, the triggers are defined in terms of
specific objects: electron/photon, muon, jet, $\MEt$. 
Tables~\ref{tab:triggers_top_elex}--\ref{tab:triggers_top_met} 
show the triggers used for $t\bar t$ event selection. 
Table~\ref{tab:tagratettrigs}
shows the triggers used for the muon tag-rate studies discussed in 
Sec.~\ref{ljets_mutag}. As noted above, level 0 is treated as an 
input term to level 1. Level 1 triggers that
do not demand a level 0 pass are denoted ``NoL0.''

At level 1, the triggers for electrons (and photons) require the transverse
energy in the EM section of the calorimeter to be above programmed
thresholds: $E_T \equiv E\sin\theta>T$, where $E$ is the energy 
deposited in the tower,
$\theta$ its angle with the beam as viewed from the center of the detector
($z = 0$), and
$T$ a programmable threshold. The level 2 electron triggers exploit the full
segmentation of the EM calorimeter to identify electron showers. 
Using the trigger towers above threshold at level 1 as seeds,
the algorithm forms clusters that include all cells in the four EM
layers and the first FH layer in a region of 
$\Delta\eta\times\Delta\phi = 0.3\times 0.3$, centered on the highest $E_T$
tower. It checks the shower shape against criteria on the fraction of the 
energy found in the different EM layers. The $E_T$ of the electron is computed 
based on its energy and the $z$ position of the interaction vertex as 
determined from the timing of hits in the level 0 hodoscopes. The level 2 
algorithm can also apply an
isolation requirement or demand an associated track in the central detector.

During the later portion of the run, the level 1.5 trigger processor became
available for selecting electrons and photons. For this purpose, the $E_T$ 
of each EM trigger tower 
passing the level 1 threshold is summed with the neighboring tower that has
the most energy and a cut is made on this sum. The hadronic portions of the
two towers are also summed and the ratio of the EM transverse energy to the 
total transverse energy of the two towers is required to be greater than
0.85. The use of a level 1.5 electron trigger is indicated in 
Tables~\ref{tab:triggers_top_elex}--\ref{tab:tagratettrigs} 
as an ``EX'' tower in the level 1 column.

\widetext
\begin{table*}
\caption{Electron triggers used in collection of the 
$t\bar t$ signal sample.
Column 1 gives the trigger name, column 2 gives the run period for which it 
was applied, column 3 gives the exposure in $\ipb$ (see text for definition), 
columns 4 and 5 give the level 1 and level 2 definitions, and
column 6 lists the channels that used each trigger. 
See Appendix~\ref{mrrecov} for definitions of the MR veto terms: 
\progname{GB}, \progname{MRBS}, \progname{ML}, and \progname{GC}. 
Channel names are defined in Secs.~\ref{dilep} and \ref{ljets_intro}.}
\squeezetable  
\tighten
\begin{tabular}{l|c|c|c|c|l}
Name & Run    & Expsr.   & Level 1 & Level 2 & Used by \\
     &        & ($\ipb$) & & & \\
%
\cline{1-6}
\progname{ele-high} 
             &  1a  & 11.0 & 1 EM tower, $\et > 10\gevn$
                           & 1 isolated $e$, $\et > 20\gevn$
                           & $e+\jets$/topo \\
             &      &      & GB   
                    &      &            \\
%
\cline{1-6}
\progname{ele-jet}  
             &  1a  & 14.4 & 1 EM tower, $\et > 10\gevn$, $|\eta| < 2.6$
                           & 1 $e$, $\et > 15\gevn$, $|\eta| < 2.5$
                           & $ee$, $e\mu$, $e\nu$ \\
             &      &      & 2 jet towers, $\et > 5\gevn$
                           & 2 jets ($\Delta R=0.3$), $\et > 10\gevn$, $|\eta| < 2.5$
                           & $e+\jets$    \\
             &      &      & MRBS
                           & $\metcal > 10\gevn$ 
                           & $e+\jets/\mu$ \\
%
\cline{1-6}
\progname{ele-jet-high}
             &  1b  & 98.0 & 1 EM tower, $\et > 12\gevn$, $|\eta| < 2.6$ 
                           & 1 $e$, $\et > 15\gevn$, $|\eta| < 2.5$
                           & $ee$, $e\mu$, $e\nu$ \\
             &      &      & 2 jet towers, $\et > 5\gevn$, $|\eta| < 2.0$   
                           & 2 jets ($\Delta R=0.3$), $\et > 10\gevn$, $|\eta| < 2.5$
                           & $e+\jets$/topo \\
             &      &      & ML
                           & $\metcal > 14\gevn$ 
                           & $e+\jets/\mu$ \\
%
\cline{1-6}
\progname{ele-jet-high}
             &  1c  & 1.9  & 1 EM tower, $\et > 12\gevn$, $|\eta| < 2.6$ 
                           & 1 $e$, $\et > 15\gevn$, $|\eta| < 2.5$
                           & $ee$, $e\mu$, $e\nu$ \\
             &      &      & 2 jet towers, $\et > 5\gevn$, $|\eta| < 2.0$   
                           & 2 jets ($\Delta R=0.3$), $\et > 10\gevn$, $|\eta| < 2.5$
                           & $e+\jets/\mu$ \\
             &      &      & ML
                           & $\metcal > 14\gevn$ 
                           &                \\
%
\cline{1-6}
\progname{ele-jet-higha}
             &  1c  & 11.0 & 1 EM tower, $\et > 12\gevn$, $|\eta| < 2.6$
                           & 1 $e$, $\et > 17\gevn$, $|\eta| < 2.5$
                           & $ee$, $e\mu$, $e\nu$ \\
             &      &      & 2 jet towers, $\et > 5\gevn$, $|\eta| < 2.0$   
                           & 2 jets ($\Delta R=0.3$), $\et > 10\gevn$, $|\eta| < 2.5$
                           & $e+\jets/\mu$ \\
             &      &      & 1 EX tower, $\et > 15\gevn$
                           & $\metcal > 14\gevn$ 
                           &                \\
             &      &      & ML                        
                           &                    
                           &                \\
%
\cline{1-6}
\progname{em1-eistrkcc-ms}
             &  1b  & 93.4 & 1 EM tower, $\et > 10\gevn$          
                           & 1 isolated $e$ w/track, $\et > 20\gevn$
                           & $e\nu$ \\
             &      &      & 1 EX tower, $\et > 15$\gevn
                           & $\metcal > 15\gevn$
                           & $e+\jets$/topo \\
             &      &      & GC, NoL0                         
                           &                   
                           &           \\
%
\cline{1-6}
\progname{mu-ele}
             &  1a  & 13.7 & 1 EM tower, $\et >  7\gevn$          
                           & 1 $e$, $\et > 7\gevn$
                           & $e\mu$ \\
             &      &      & 1 $\mu$, $|\eta| < 2.4$
                           & 1 $\mu$, $\pt > 5\gevcn$, $|\eta|<2.4$
                           &               \\
             &      &      & MRBS
                           & 
                           &                     \\
\cline{2-6}
             &  1b  & 93.9 & 1 EM tower, $\et >  7\gevn$          
                           & 1 $e$, $\et > 7\gevn$, $|\eta|<2.5$
                           & $e\mu$          \\
             &      &      & 1 MX $\mu$, $|\eta| < 2.4$
                           & 1 $\mu$, $\pt > 8\gevcn$, $|\eta|<2.4$
                           &                 \\
             &      &      & GC
                           & 
                           &                  \\
%
\cline{1-6}
\progname{mu-ele-high}
             &  1c  & 10.6 & 1 EM tower, $\et >  10\gevn$, $|\eta|<2.5$
                           & 1 $e$, $\et > 10\gevn$, $|\eta|<2.5$
                           & $e\mu$            \\
             &      &      & 1 MX $\mu$, $|\eta| < 2.4$
                           & 1 $\mu$, $\pt > 8\gevcn$, $|\eta|<1.7$
                           &                    \\
             &      &      & GC
                           & 
                           &                  \\
%
\end{tabular}
\label{tab:triggers_top_elex}
\end{table*}
\narrowtext

\widetext
\begin{table*}
\caption{Muon+jet triggers used in collection of the $t\bar t$ signal sample.
Column 1 gives the trigger name, column 2 gives the run period for which it
was applied, column 3 gives the exposure in $\ipb$ (see text for definition),
columns 4 and 5 give the level 1 and level 2 definitions, and
column 6 lists the channels that used each trigger. See Appendix~\ref{mrrecov} 
for definitions of the MR veto terms: 
\progname{GB} and \progname{GC}. Channel names are defined in 
Secs.~\ref{dilep} and \ref{ljets_intro}.}
\squeezetable  
\tighten
\begin{tabular}{l|c|c|c|c|l}
Name & Run    & Expsr.   & Level 1 & Level 2 & Used by \\
     &        & ($\ipb$) & & & \\
%
\cline{1-6}
\progname{mu-jet-high}
             &  1a  & 10.2 & 1 $\mu$, $|\eta| < 2.4$
                           & 1 $\mu$, $\pt > 8\gevcn$, $|\eta| < 1.7$
                           & $e\mu$, $\mu\mu$ \\
             &      &      & 1 jet tower, $\et > 5\gevn$    
                           & 1 jet ($\Delta R=0.7$), $\et > 15\gevn$ 
                           & $\mu+\jets$/topo \\
             &      &      & GB
                           &
                           & $\mu+\jets/\mu$ \\
\cline{2-6}
             &  1b  & 66.4 & 1 $\mu$, $\pt > 7\gevcn$, $|\eta| < 1.7$
                           & 1 $\mu$, $\pt > 10\gevcn$, $|\eta| < 1.7$, scint
                           & $e\mu$, $\mu\mu$  \\
             &      &      & 1 jet tower, $\et > 5\gevn$, $|\eta| < 2.0$
                           & 1 jet ($\Delta R=0.7$), $\et > 15\gevn$, $|\eta| < 2.5$
                           & $\mu+\jets$/topo \\
             &      &      &  GC
                           &
                           &  $\mu+\jets/\mu$ \\
%
\cline{1-6}
\progname{mu-jet-cal}
             &  1b  & 88.0 & 1 $\mu$, $\pt > 7\gevcn$, $|\eta| < 1.7$             
                           & 1 $\mu$, $\pt > 10\gevcn$, $|\eta| < 1.7$ 
                           & $\mu\mu$ \\
             &      &      & 1 jet tower, $\et > 5\gevn$, $|\eta| < 2.0$
                           & cal confirm, scint
                           & $\mu+\jets$/topo \\
             &      &      & GC
                           & 1 jet ($\Delta R=0.7$), $\et > 15\gevn$, $|\eta| < 2.5$
                           & $\mu+\jets/\mu$ \\
%
\cline{1-6}
\progname{mu-jet-cent}
             &  1b  & 48.5 & 1 $\mu$, $|\eta| < 1.0$                
                           & 1 $\mu$, $\pt > 10\gevcn$, $|\eta| < 1.0$, scint
                           & $e\mu$, $\mu\mu$ \\
             &      &      & 1 jet tower, $\et > 5\gevn$, $|\eta| < 2.0$   
                           & 1 jet ($\Delta R=0.7$), $\et > 15\gevn$, $|\eta| < 2.5$
                           & $\mu+\jets$/topo \\
             &      &      & GC
                           &
                           & $\mu+\jets/\mu$ \\
\cline{2-6}
             &  1c  &  8.9 & 1 $\mu$, $|\eta| < 1.0$                
                           & 1 $\mu$, $\pt > 12\gevcn$, $|\eta| < 1.0$, scint
                           & $e\mu$, $\mu\mu$ \\
             &      &      & 1 jet tower, $\et > 5\gevn$, $|\eta| < 2.0$    
                           & 1 jet ($\Delta R=0.7$), $\et > 15\gevn$, $|\eta| < 2.5$
                           &                 \\
             &      &      & 2 jet towers, $\et > 3\gevn$
                           & \\
             &      &      & GC
                           & \\
\cline{1-6}
\progname{mu-jet-cencal}
             &  1b  & 51.2 & 1 $\mu$, $|\eta| < 1.0$                           
                           & 1 $\mu$, $\pt > 10\gevcn$, $|\eta| < 1.0$ 
                           & $\mu\mu$ \\
             &      &      & 1 jet tower, $\et > 5\gevn$, $|\eta| < 2.0$   
                           & cal confirm, scint
                           & $\mu+\jets$/topo \\
             &      &      & GC
                           & 1 jet ($\Delta R=0.7$), $\et > 15\gevn$, $|\eta| < 2.5$
                           & $\mu+\jets/\mu$ \\
\cline{2-6}
             &  1c  & 11.4 & 1 $\mu$, $|\eta| < 1.0$                      
                           & 1 $\mu$, $\pt > 12\gevcn$, $|\eta| < 1.0$ 
                           & $e\mu$, $\mu\mu$  \\
             &      &      & 1 jet tower, $\et > 5\gevn$, $|\eta| < 2.0$    
                           & cal confirm, scint
                           &                 \\
             &      &      & 2 jet towers, $\et > 3\gevn$
                           & 1 jet ($\Delta R=0.7$), $\et > 15\gevn$, $|\eta| < 2.5$ \\
             &      &      & GC
                           & \\
%
\end{tabular}
\label{tab:triggers_top_mujet}
\end{table*}
\narrowtext

\widetext
\begin{table*}
\caption{Jet triggers used in collection the $t\bar t$ signal sample.
Column 1 gives the trigger name, column 2 gives the run period for which it
was applied, column 3 gives the exposure in $\ipb$ (see text for definition),
columns 4 and 5 give the level 1 and level 2 definitions, and
column 6 lists the channels that used each trigger. 
See Appendix~\ref{mrrecov} for definitions of the MR veto terms: 
\progname{ML}, \progname{MB}, and \progname{MRBS}. 
The lepton+jets channels are defined in Sec.~\ref{ljets_intro}.}
\squeezetable  
\tighten
\begin{tabular}{l|c|c|c|c|l}
Name & Run    & Expsr.   & Level 1 & Level 2 & Used by \\
     &        & ($\ipb$) & & & \\
%
\cline{1-6}
\progname{jet-3-mu}
             &  1b  & 11.9 & 3 jet towers, $\et > 5\gevn$                 
                           & 3 jets ($\Delta R=0.7$), $\et > 15\gevn$, $|\eta| < 2.5$
                           & $\mu+\jets$/topo  \\
             &      &      & $\metcal > 20\gevn$
                           & $\metcal > 17\gevn$
                           & $\mu+\jets/\mu$ \\
             &      &      & ML
                           & 
                           &                 \\
%
\cline{1-6}
\progname{jet-3-miss-low}
             &  1b  & 57.8 & 3 large tiles, $\et > 15$, $|\eta| < 2.4$       
                           & 3 jets ($\Delta R=0.5$), $\et > 15\gevn$, $|\eta| < 2.5$
                           & $\mu+\jets$/topo  \\
             &      &      & 3 jet towers, $\et > 7\gevn$, $|\eta| < 2.6$
                           & $\metcal > 17\gevn$
                           & $\mu+\jets/\mu$  \\
             &      &      & MB
                           & 
                           &                 \\
%
\cline{1-6}
\progname{jet-3-l2mu}
             &  1b  & 25.8 & 3 large tiles, $\et > 15$, $|\eta| < 2.4$           
                           & 1 $\mu$, $\pt > 6\gevcn$, $|\eta| < 1.7$ 
                           & $\mu+\jets$/topo  \\
             &      &      & 3 jet towers, $\et > 7\gevn$, $|\eta| < 2.6$   
                           & cal confirm, scint
                           & $\mu+\jets/\mu$ \\
             &      &      & MB
                           & 3 jets ($\Delta R=0.5$), $\et > 15\gevn$, $|\eta| < 2.5$ \\
             &      &      & 
                           & $\metcal > 17\gevn$ \\
%
\cline{1-6}
\progname{jet-multi}
             &  1a  & 14.6 & 4 jet towers, $\et > 5\gevn$
                           & 5 jets ($\Delta R=0.3$), $\et > 10\gevn$, $|\eta| < 2.0$
                           & all-jets \\
             &      &      & MRBS
                           & 
                           &                       \\
\cline{2-6}
             &  1b  & 96.6 & 3 large tiles, $\et > 15\gevn$, $|\eta| < 2.4$
                           & 5 jets ($\Delta R=0.3$), $\et > 10\gevn$, $|\eta| < 2.5$
                           & all-jets \\
             &      &      & 3 jet towers, $\et > 7\gevn$, $|\eta| < 2.6$    
                           & $\Sigma \et > 100\gevn$ for jets with $|\eta|<2.5$
                           &                   \\
             &      &      & and 1 jet tower, $\et > 3\gevn$
                           & \\
             &      &      & ML
                           & \\
\cline{2-6}
             &  1c  & 11.3 & 3 large tiles, $\et > 15\gevn$, $|\eta| < 2.4$
                           & 5 jets ($\Delta R=0.3$), $\et > 10\gevn$, $|\eta| < 2.5$
                           & all-jets \\
             &      &      & 3 jet towers, $\et > 7\gevn$, $|\eta| < 2.6$    
                           & $\Sigma \et > 120\gevn$ for jets with $|\eta|<2.5$
                           &                   \\
             &      &      & and 1 jet tower, $\et > 3\gevn$
                           & \\
             &      &      & ML
                           & \\
\end{tabular}
\label{tab:triggers_top_jetx}
\end{table*}
\narrowtext

\widetext
\begin{table*}
\caption{$\MEt$ triggers used in collection of the $t\bar t$ signal sample.
Column 1 gives the trigger name, column 2 gives the run period for which it
was applied, column 3 gives the exposure in $\ipb$ (see text for definition),
columns 4 and 5 give the level 1 and level 2 definitions, and
column 6 notes that these triggers were used only by the $e\nu$ channel.
See Appendix~\ref{mrrecov} for definitions of the MR veto terms: 
\progname{MRBS} and \progname{GB}. 
The $e\nu$ channel is defined in Sec.~\ref{dilep}.}
\squeezetable  
\tighten
\begin{tabular}{l|c|c|c|c|l}
Name & Run    & Expsr.   & Level 1 & Level 2 & Used by \\
     &        & ($\ipb$) & & & \\
%
\cline{1-6}
\progname{missing-et}
             &  1a  & 13.7 & $\metcal > 30\gevn$ 
                           & $\metcal > 35\gevn$
                           & $e\nu$ \\
             &      &      &  1 jet tower, $\et > 5\gevn$, $|\eta| < 2.6$ 
                           & 
                           &                      \\
             &      &      & MRBS
                           &                       \\
\cline{2-6}
             &  1b  & 83.6 & $\metcal > 40\gevn$
                           & $\metcal > 40\gevn$
                           & $e\nu$ \\
             &      &      & 1 jet tower, $\et > 5\gevn$, $|\eta| < 2.6$
                           & 
                           &                       \\
             &      &      & GB
                           &                       \\
%
\cline{1-6}
\progname{missing-et-high}
             &  1c  & 0.7  & $\metcal > 50\gevn$
                           & $\metcal > 50\gevn$
                           & $e\nu$ \\
             &      &      & 1 jet tower, $\et > 5\gevn$, $|\eta| < 2.6$
                           & 
                           &                        \\
             &      &      & GB
                           &                        \\
\end{tabular}
\label{tab:triggers_top_met}
\end{table*}
\narrowtext

\widetext
\begin{table*}
\caption{Triggers used to study the $\ell$+jets/$\mu$ backgrounds and tag 
rate function (see Sec.~\ref{ljets_mutag}).
Column 1 gives the trigger name, column 2 gives the run period for which it
was applied, column 3 gives the exposure in $\ipb$ (see text for definition),
columns 4 and 5 give the level 1 and level 2 definitions, and
column 6 notes that these triggers were used only for $\ell$+jets background
studies. 
See Appendix~\ref{mrrecov} for definitions of the MR veto terms: 
\progname{GB}, \progname{MRBS}, \progname{ML}, and \progname{GC}. 
The lepton+jets channels are defined in Sec.~\ref{ljets_intro}.}
\squeezetable  
\tighten
\begin{tabular}{l|c|c|c|c|l}
Name & Run    & Expsr.   & Level 1 & Level 2 & Used by \\
     &        & ($\ipb$) & & & \\
      %
\cline{1-6}
\progname{jet-min}
             &  1b  & 0.007& 1 jet tower, $\et > 3\gevn$
                           & 1 jet ($\Delta R=0.3$), $\et > 20\gevn$
                           & $\ell+\jets/\mu$      \\
             &      &      & GB
                           & prescale = 20
                           & bkg                \\
%
\cline{1-6}
\progname{jet-3-mon}
             &  1b  & 0.92 & 2 jet towers, $\et > 5\gevn$
                           & 3 jets ($\Delta R=0.3$), $\et > 10\gevn$
                           & $\ell+\jets/\mu$     \\
             &      &      & and 1 jet tower, $\et > 3\gevn$
                           & prescale = 5
                           & bkg               \\
             &      &      & GB
                           & 
                           &                     \\
%
\cline{1-6}
\progname{jet-4-mon}
             &  1b  & 4.6  & 2 jet towers, $\et > 5\gevn$
                           & 4 jets ($\Delta R=0.3$), $\et > 10\gevn$
                           & $\ell+\jets/\mu$      \\
             &      &      & and 1 jet tower, $\et > 3\gevn$
                           & 
                           & bkg                   \\
             &      &      & GB
                           &                
                           &                 \\
%
\cline{1-6}
\progname{jet-multi}
             &  1a  & 14.6 & 4 jet towers, $\et > 5\gevn$
                           & 5 jets ($\Delta R=0.3$), $\et > 10\gevn$, $|\eta| < 2.0$
                           & $\ell+\jets/\mu$     \\
             &      &      & MRBS
                           & 
                           & bkg                  \\
\cline{2-6}
             &  1b  & 96.6 & 3 large tiles, $\et > 15\gevn$, $|\eta| < 2.4$
                           & 5 jets ($\Delta R=0.3$), $\et > 10\gevn$, $|\eta| < 2.5$
                           & $\ell+\jets/\mu$      \\
             &      &      & 3 jet towers, $\et > 7\gevn$, $|\eta| < 2.6$    
                           & $\Sigma \et > 100\gevn$ for jets with $|\eta|<2.5$
                           & bkg              \\
             &      &      & and 1 jet tower, $\et > 3\gevn$
                           & \\
             &      &      & ML
                           & \\
%
\cline{1-6}
\progname{ele-1-mon}
             &  1b  & 3.1  & 1 EM tower, $\et > 7\gevn$, $|\eta| < 2.5$ 
                           & 1 $e$, $\et > 16\gevcn$
                           & $\ell+\jets/\mu$     \\
             &      &      & 1 jet tower, $\et > 3\gevn$
                           & 
                           & bkg                 \\
             &      &      & GC
                           &                      \\
%
\cline{1-6}
\progname{gis-dijet}
             &  1b  & 93.5 & 1 EM tower, $\et > 10\gevn$, $|\eta| < 2.5$ 
                           & 1 isolated $e/\gamma$, $\et>15\gevcn$, $|\eta|<2.0$
                           & $\ell+\jets/\mu$      \\
             &      &      & 1 jet tower, $\et > 3\gevn$
                           & 3 jets ($\Delta R=0.7$), $\et>15\gevn$, $|\eta|<2.0$
                           & bkg              \\
             &      &      & GC
                           & $\Sigma \et > 70\gevn$ for jets with $|\eta|<2.0$
                           &                   \\
%
\cline{1-6}
\progname{em1-eistrkcc-esc}
             &  1b  & 91.9 & 1 EM tower, $\et > 10\gevn$, $|\eta| < 2.5$ 
                           & 1 $e$ (no shape cuts), $\et > 16\gevn$
                           & $\ell+\jets/\mu$      \\
             &      &      & 1 jet tower, $\et > 3\gevn$
                           & and 1 isolated $e$ w/track, $\et > 20\gevn$ 
                           & bkg                  \\
             &      &      & GC
                           & 
                           &                       \\
\end{tabular}
\label{tab:tagratettrigs}
\end{table*}
\narrowtext

Muon triggers make use of hit patterns in the muon chambers at level 1
and provide the number of muon candidates in different regions of the
muon spectrometer. The algorithm searches for hit patterns consistent 
with a muon originating from the nominal vertex ($z = 0$). A level 1.5
processor is also available and can be used to place a $p_T$ requirement
on the candidates (at the expense of a slightly increased dead time). 
The use of a level 1.5 muon trigger is indicated in 
Tables~\ref{tab:triggers_top_elex}--\ref{tab:tagratettrigs} as an
``MX'' muon in the level 1 column.

At level 2, muon tracks are reconstructed using the muon PDT hits and 
the $z$ position of the interaction vertex from level 0.   
Valid muon track selection is based on the muon $p_T$ and quality 
requirements (similar to those of Sec.~\ref{muqual}).
The level 2 muon trigger can also require the presence of a minimum 
ionizing particle trace in the calorimeter cells along the track.
This requirement is indicated in 
Tables~\ref{tab:triggers_top_elex}--\ref{tab:tagratettrigs} 
by ``cal confirm.''
In addition, in between Run 1a and Run 1b, layers of scintillator were 
added to the exterior of the central muon system to veto cosmic rays.   
The muon triggers indicated by ``scint'' required the scintillator 
timing to be in a window of 30 ns before to 70 ns after the 
beam crossing.

Jet triggers use projective towers of energy deposition in the calorimeter
similar to the EM trigger towers but including energy from the hadronic 
portion of the calorimeter. Level 1 jet triggers require the sum of the
transverse energy in the EM and FH sections of a trigger tower (jet tower) 
to be above
programmed thresholds: $E\sin\theta>T$, where $E$ is the energy deposit 
in the tower, $\theta$ its angle with the beam as seen from the center of 
the detector ($z = 0$), and $T$ a programmable threshold.
Alternatively, level 1 can sum the transverse energies within ``large tiles''
of size $0.8 \times 1.6$ in $\eta\times\phi$ and cut on these sums.
The level 2 jet algorithm begins with an $E_T$-ordered list of towers
that are above threshold at level 1. A level 2 jet is formed by placing a
cone of radius $\Delta R =\sqrt{\Delta\eta^2 + \Delta\phi^2}$ around 
the seed tower
from level 1. If another seed tower lies within the jet cone, then it is
passed over and not allowed to seed a new jet. Using the vertex position
measured by the level 0 hodoscopes, the summed $E_T$ in all of 
the towers included in the jet defines the jet $E_T$. If any two jet cones
overlap, then the towers in the overlap region are added into the jet 
candidate that was formed first. 

$\metcal$, the missing transverse energy as measured in the calorimeter
(see Sec.~\ref{misset} for definition), 
can be computed at both level 1 and level 2. At level 1,
the $z$ position is assumed to be $z=0$. At level 2, the vertex position
from level 0 is used. In the offline reconstruction, the determination of
$\metcal$ uses the $z$ position as determined by the tracking system.
Therefore, the resolution of $\metcal$ at the trigger level is significantly
poorer than that in the offline reconstruction.

As noted in Sec.~\ref{d0det}, the Main Ring passes directly
through a portion of the outer hadronic calorimeter and muon system.
Particles lost from the Main Ring can affect the measurements in these 
subsystems. Several schemes were employed at the trigger level to reduce 
or eliminate these effects; these are described in Appendix~\ref{mrveto}.

In addition to the complications introduced by the Main Ring, there are
also effects due to multiple interactions. At the mean luminosity 
(7.5 $\times 10^{30} /{\rm cm}^2/{\rm s}$), there are
on average 1.3 interactions per bunch crossing. Since the cross section for
the production of high-$p_T$ interactions is small compared to that for
minimum bias, it is very unlikely that more than one high-$p_T$ interaction
will be present in any given bunch crossing. These additional minimum-bias
interactions are usually not included in the Monte Carlo models, but do
contribute to mismeasurement of the primary interaction vertex, and therefore
to mismeasurement of lepton and jet transverse energies/momenta. The systematic
uncertainty due to multiple interactions is discussed in Sec.~\ref{ss:mierr}.

The Run 1 data were 
acquired in three separate run periods: Run 1a from 1992--1993, Run 1b
from 1994--1995, and Run 1c from 1995--1996. The period appropriate to
each trigger is given in the second column of 
Tables~\ref{tab:triggers_top_elex}--\ref{tab:tagratettrigs}.

The integrated luminosity ${\cal L}$ was determined from the 
counting rate in the level 0 hodoscopes ($R_{L0}$) as 
\begin{equation} 
{\cal L} = \frac{-{\rm ln}(1 - \tau R_{L0})}{\tau \sigma_{L0}}
\end{equation} 
where $\tau = 3.5 \ \mu s$ is the time interval between beam crossings
and $\sigma_{L0}$ is the effective $p \bar p$ cross section subtended 
by the level 0 counters. As described in detail in Ref.~\cite{wwidth}, 
$\sigma_{L0} = 43.1 \pm 1.9 {\rm \ mb}$ is obtained from the level 0 
trigger efficiency and geometrical acceptance, and from a ``world average'' 
$p\bar p$ total inelastic cross section of $57.39 \pm 1.56$ mb based on 
results from the CDF~\cite{cdflum}, E710~\cite{e7101990}, and 
E811~\cite{e811} collaborations at Fermilab. 
The level 0 trigger efficiency is determined using 
samples of data collected from triggers on random beam crossings and 
the geometrical acceptance from Monte Carlo studies.
It should be noted that the CDF
luminosity determinations are based solely on its own measurement of the 
$p\bar p$ inelastic cross section. As a result, luminosities
reported by CDF are 6.2\% lower than those currently reported by D\O, and 
consequently, all CDF cross sections are {\it ab initio} 6.2\% larger
than all D\O\ cross sections. Earlier D\O\ cross sections (and all previous
D\O\ $t\bar t$ cross sections) were based on a $p\bar p$ inelastic cross 
section determined only from the CDF and E710 
measurements and are 3.2\% lower than current D\O\ cross sections.

The integrated luminosity (exposure) seen by each 
of the triggers is given in the third column, labelled ``Expsr.,'' of
Tables~\ref{tab:triggers_top_elex}--\ref{tab:tagratettrigs}. These values
include luminosity losses due to Main-Ring vetos and prescale factors 
(if appropriate), but do not include the loss to the offline 
\progname{good-beam} requirement or losses from runs  
rejected at later stages of the analysis (see Appendix~\ref{mrveto} for
a discussion of the Main-Ring veto schemes).

%
\section {Event reconstruction}
\label{reco}
\subsection {Electron identification}
\label{e-id}

Electrons and positrons are identified by the distinctive pattern of energy 
that 
electromagnetic showers deposit in the calorimeter and by the presence of a 
track from the interaction vertex to the cluster of hit calorimeter 
cells.
The algorithm for clustering calorimeter energy and quantities used to
distinguish electrons from backgrounds are described in Ref.~\cite{prd1}.
The present analysis includes two additional features:
the separation between electrons and backgrounds has been improved by the 
introduction of a multivariate discriminant,
and, for the dilepton channels, use is made of information from the TRD.

The electromagnetic energy scale was calibrated using $Z\rightarrow ee$, 
$J/\psi \rightarrow ee$, and $\pi^0 \rightarrow \gamma\gamma$ decays to a 
precision of 0.08\% at $E=M_Z/2$ and to 0.6\% at $E=20$ GeV 
\cite{elecres1,elecres2}.

The complete set of identification variables, efficiencies, and 
misidentification rates
is discussed below. Unless otherwise indicated, electrons specified to be in 
the CC region of the detector span the range $ 0 \leq |\eta| \leq 1.2 $ and 
electrons specified to be in the EC region of the detector span the range 
$ 1.2 < |\eta| \leq 2.0 $ (with the region between the cryostats, 
$1.2<|\eta|<1.5$, having only a minimal acceptance).
Since the central tracking system does not measure
the charge of particles, it is not possible to distinguish between electrons 
and positrons. Therefore, for the remainder of this paper, ``electron'' shall 
be used to indicate both electrons and positrons.

\subsubsection{Electromagnetic energy fraction}

Electromagnetic energy clusters are formed by combining calorimeter towers
using
a nearest-neighbor algorithm with EM tower seeds. The electromagnetic
energy fraction $f_{\rm{EM}}$ of a cluster is the ratio of its energy found 
in EM calorimeter cells to its total energy. All electron 
candidates are required to have $f_{\rm{EM}} \geq 0.9$.

\subsubsection{Isolation fraction ($\cal I$)}

Electron showers are compact and mostly contained in the
core of EM cells within a radius $R=0.2$ in ($\eta$,$\phi$)
around the shower center. The isolation fraction $\cal I$ is defined
as the ratio of energy in non-core EM and FH cells ($E_{\rm tot}$) 
within a cone of 0.4 around the center to energy in the EM cluster 
core ($E_{\rm EM}$)
\begin{equation}
{\cal I} = { {E_{\rm tot}(0.4)-E_{\rm EM}(0.2)}\over{E_{\rm EM}(0.2)} }.
\end{equation}
This quantity tends to be substantially smaller for
electrons from the decay of $W$ and $Z$ bosons than for the background,
most of which originates from hadronic jets where the electron candidate
is usually accompanied by nearby energetic particles.

\subsubsection{Covariance matrix ($\chi^2_e$)}

A covariance matrix is used to compute a $\chi^2$ variable ($\chi^2_e$) 
representing the consistency of the cluster shape 
with that of an electron shower. The covariance
matrix uses 41 variables: the fractions of energy deposited in
the first, second, and fourth layers of the EM calorimeter; 
the fractions of energy in each cell of the third EM layer lying in 
a six by six array around the tower containing the highest energy cell;
the logarithm of the cluster energy; and the $z$ position of the interaction
vertex. The elements of the covariance matrix depend on $\eta$ and 
were determined using the \progname{d\o geant}~\cite{d0geant} model of the 
detector (see Sec.~\ref{mc}).

\subsubsection{Cluster-track match significance ($\sigma_{\rm trk}$)}

Calorimeter clusters are required to lie along the trajectories of 
charged particle tracks reconstructed
in one of the inner tracking chambers. The cluster-track match
significance $\sigma_{\rm trk}$ is a measure of the distance between the
cluster centroid and the intersection of the extrapolated track to the
third layer of the EM calorimeter.

\subsubsection{Track ionization ($dE/dx$)}

Photons that convert to $e^{+}e^{-}$ pairs before the calorimeter 
produce pairs of tracks that match an EM cluster well and are too close
together to be resolved. Such double tracks can be identified by
the amount of ionization produced along the track ($dE/dx$); photon 
conversions typically
deposit twice the charge expected from one minimum ionizing particle.

\begin{figure}[t]
\vbox{
\vskip -0.8cm
\centerline{\hskip 0.2 in \psfig{figure=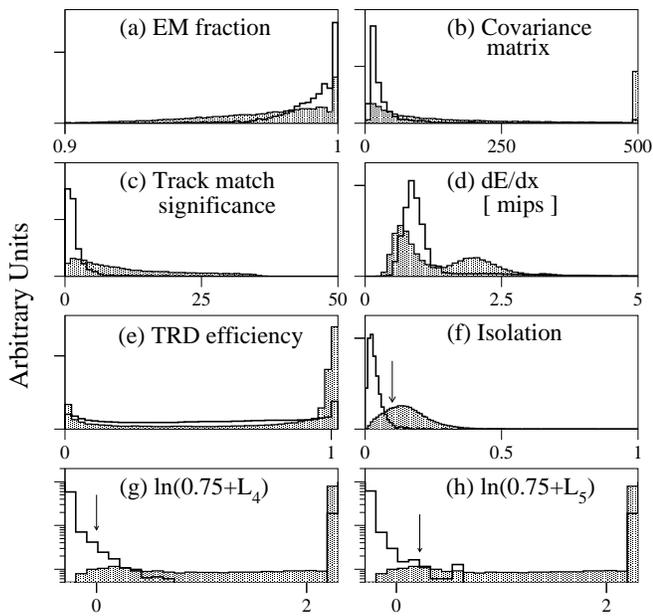,width=3.75in}}
\caption{(a) -- (e) Electron identification variables used in the 
$L_4$ and $L_5$ likelihood ratios,
(f) Isolation, and (g) and (h) 4-variable and 5-variable likelihood ratios.
The open histograms are from electron candidates from 
$Z \rightarrow e^+e^-$ events 
and the shaded histograms are from electron candidates from EM clusters in 
inclusive jet data (mainly background). Arrows indicate the position of the
cuts on isolation, $L_4$, and $L_5$. All quantities are for the CC region 
of the detector only.
\label{fig:eid}}}
\end{figure}

\subsubsection{TRD efficiency ($\epsilon_t$)}

The response of the TRD is characterized by
the variable $\epsilon_t$:
\begin{equation}
\epsilon_t \left( \Delta E \right) =  
         { \int_{\Delta E}^\infty {\partial N \over \partial E } 
         \left( E \right) dE \over \int_0^\infty {\partial N \over \partial E }
         \left( E \right) dE },
\label{eq:trdeff}
\end{equation}
where $\Delta E$ is the difference between the total energy recorded in the 
TRD ($E$) and that recorded in the
layer with the largest signal (this is done to reduce sensitivity to
$\delta$-rays) and $\partial N \over \partial E $ is the electron 
energy 
spectrum from a sample of $W \rightarrow e\nu$ events \cite{ewv,bkthesis}.
Hadrons generally deposit energy mainly in a single layer (giving a small value
for $\Delta E$) and electrons deposit energy more evenly (giving a larger
value for $\Delta E$). Therefore, hadrons tend to have values of 
$\epsilon_t$ near unity whereas the distribution from electrons is roughly 
uniform over the allowed range from 0 to 1.

\subsubsection{Likelihood ratio ($L_4, L_5$)}

In order to attain the maximum background rejection while keeping
a high efficiency for real electrons, the variables $f_{\rm{EM}}$, 
$\chi_e^2$, $\sigma_{\rm trk}$, and $dE/dx$ are combined into an 
approximate four-variable likelihood ratio $L_4$ for the hypotheses that a 
candidate electron is 
signal or background. Similarly, the variables $f_{\rm{EM}}$, $\chi_e^2$,
$\sigma_{\rm trk}$, $dE/dx$, and $\epsilon_t$ are combined into an approximate 
five-variable likelihood ratio $L_5$. These likelihood ratios are defined
using the Neyman-Pearson test for 
signal ($e$) and background ($b$) hypotheses, where an EM cluster is 
considered to be an electron if it satisfies 
\begin{equation}
L_n \equiv \frac{p_n(x|b)}{p_n(x|e)} < k,
\end{equation}
where $x$ is the vector of observables, $p_n(x|H)$ is the probability density
for $x$ if the hypothesis $H$ is true, and $k$ is the cutoff value.
The probability densities are computed by forming the joint likelihood of
the four or five variables:
\begin{eqnarray}
 p_4(x|H) &=& p(f_{\rm{EM}}|H) \cdot p(\chi_e^2|H) \cdot 
            p(\sigma_{\rm trk}|H) \nonumber \\
          & & \cdot p(dE/dx|H), \\
 p_5(x|H) &=& p(f_{\rm{EM}}|H) \cdot p(\chi_e^2|H) \cdot 
            p(\sigma_{\rm trk}|H) \nonumber \\
          & & \cdot p(dE/dx|H) \cdot p(\epsilon_t|H),
\end{eqnarray}
where $p(y|H)$ is the probability density for a single variable $y$ if
the hypothesis $H$ is true.
These signal and background hypotheses are constructed respectively
from inclusive $Z\ra e^+e^-$ data and inclusive jet production.

The distributions associated with all the above variables for electrons in
the CC region of the detector are shown in Fig.~\ref{fig:eid}.

\subsubsection{Selection}

Based on these quantities, four classes of electron candidates
are defined:
\begin{enumerate}
\item {\sl extra-loose} electrons are defined as objects satisfying
      $f_{EM} \geq 0.9$, $\cal{I}$$ < 0.3$, and $\chi_e^2 < 300$.
\item {\sl minimal} electrons are defined as objects satisfying
      $f_{EM} \geq 0.9$ and $\cal{I}$$ < 0.1$.
\item {\sl loose} electrons are defined as the subset of the extra-loose
      sample that satisfies the additional requirements $\cal{I}$$ < 0.1$ 
      and $L_5 < 0.5 $ for CC and EC clusters.
\item {\sl tight} electrons are defined as the subset of the extra-loose
      sample that satisfies the additional requirements $\cal{I}$$ < 0.1$ 
      and $L_4 < 0.25 (0.3)$ for CC (EC) clusters.
\end{enumerate}

The loose definition is used for the final selection in the dilepton 
channels ($ee$, $e\mu$, $e\nu$). The tight definition is used for the final 
selection in the $e$+jets channels.

\subsubsection{Efficiency}
\label{eid-eff}

The efficiencies for electron identification are 
obtained by using the $Z\rightarrow ee$ mass peak. The procedure
is based on a sample of events from the \progname{em1-eistrkcc-esc} trigger 
(see Table~\ref{tab:tagratettrigs}) that has two reconstructed 
electromagnetic clusters, each with $E_T \geq 20$ GeV. From this sample,
one of the electron candidates, denoted as the ``tag,'' is required to
be a good electron ($\chi^2_e \leq 100$, ${\cal I} \leq 0.15$). If the other
electromagnetic cluster, denoted as the ``probe,'' satisfies 
${\cal I} \leq 0.1$, then the invariant mass of the pair, 
$m({\rm tag}, {\rm probe})$, is recorded. This is done separately for probes 
in the CC and EC regions of the calorimeter. The number of entries in the 
$Z$ boson mass window, $80~\gevcc < m({\rm tag},{\rm probe})<~100~\gevccn$, 
with background subtracted, and in the instrumented region of the 
central tracking system, defines the number of true
electron probes \cite{ptthesis}. The track finding efficiency 
$\varepsilon_{\rm trk}$ is defined as the ratio of the number of 
true electron probes with a track to the total number of true electron
probes. This efficiency varies with the number of interactions per event
(see Secs.~\ref{triggers} and \ref{ss:mierr}).
Typical values are $82.7\pm1.1$~\% for electrons in the CC and 
$85.2 \pm 1.0$~\% in the EC. The electron identification efficiencies, given 
in Table~\ref{tab:eid}, are defined by the ratio of the 
number of true electron probes with a reconstructed track that pass the 
given identification requirements to the total number of true electron 
probes with a reconstructed track. 
These efficiencies do not include geometric factors due to 
uninstrumented fiducial regions of detector. The geometrical acceptance
for electrons in the D\O\ detector is ($87.6 \pm 0.5$)\% in the CC and 
($79.2 \pm 1.4$)\% in the EC.

\begin{table}[thp]
\caption{Definition of loose and tight electron identification 
criteria and the corresponding efficiencies (Eff) and misidentification 
rates ($R_{\rm mis})$.
\label{tab:eid}}
\vskip 0.5cm
\begin{tabular}{l|c|c|c|c}
           & \multicolumn{2}{c|}{ Loose }  & \multicolumn{2}{c}{ Tight }  \\

Region   & CC              & EC             & CC            & EC            \\ 
\hline 
Def      & $L_5<.5$        & $L_5<.5$       & $L_4<.25$     & $L_4<.3$      \\
Eff(\%)  & $88.0 \pm 1.6 $ & $63.8 \pm 2.3$ & $81.1 \pm 1.0$ 
                                                            & $51.4 \pm 1.8$\\
$R_{\rm mis}$(\%)& $4.6 \pm 0.1$  & $8.0 \pm 0.1$   & $2.2 \pm 0.1$ & $4.0 \pm 0.3$ \\
\end{tabular}
\end{table}

\subsubsection{Misidentification rate ($R_{\rm mis}$)}

The electron misidentification rates ($R_{\rm mis}$) given in 
Table~\ref{tab:eid} are measured
from a sample of QCD multijet events that contained one electromagnetic
cluster passing the extra-loose electron identification requirements 
defined above. From this sample of extra-loose
electron candidates, the fraction passing the loose/tight electron 
identification is obtained 
separately for the CC and EC regions of the
calorimeter and defined to be the rate for an extra-loose electron candidate
to pass the loose/tight criteria. Note that the 
multijet backgrounds due to electron misidentification are handled 
differently in the $e$+jets analyses and are discussed in 
Secs.~\ref{ljets_topo} and \ref{ljets_mutag}.

\subsection {Muon identification}
\label{muid}

Muon tracks are reconstructed using the muon system PDTs. Additional 
information about the interaction vertex, matching tracks in the central 
tracker, and minimum ionizing traces left in the calorimeter is also available.

As noted in Sec.~\ref{d0det}, the decay products from
the $t\bar t$ pair are emitted at central rapidities and the 
muon identification is therefore restricted to the central 
(WAMUS) portion of the D\O\ muon system, $|\eta| \leq 1.7$. 
Due to inefficiencies caused by
radiation damage, the forward muon region (EF) with $1.0\leq|\eta|\leq 1.7$ 
was not used in these
analyses for Run 1a ($\approx 10 \ipb$) or the early part of Run 1b 
($\approx 49 \ipb$). The chambers were subsequently cleaned 
and returned to full efficiency for the remainder of Run 1b and Run 1c.
In the discussion below, the pre-cleaning period of Run 1b is denoted as
``{\sl preclean}'' and the post-cleaning period as ``{\sl postclean}''.

Several categories of muons are used in the analyses.
The primary categories correspond to the selection of {\sl isolated} muons 
arising 
dominantly from $W\rightarrow\mu\nu$ decay and {\sl non-isolated}  
({\sl tag}) muons from $b\rightarrow\mu+X$ decays.
Isolation implies a separation of the muon track from nearby jet activity.
Isolated muons fall into two categories, {\sl tight} and {\sl loose}.
The selection requirements for the three types vary slightly over time
and are summarized in Tables~\ref{tab:muid1a} -- \ref{tab:muid1bpost}
for Run 1a, Run 1b(preclean), and Run 1b+c(postclean) respectively. 
Tight and loose muons share most requirements except that tight muons
have the additional requirements of an impact parameter cut and a minimum 
magnetic field
path length (see below). The $p_T$ and $\Delta R(\mu,{\rm jet})$ 
requirements for isolated
muons are characteristic of what is expected from $W\rightarrow\mu\nu$ decay.
The selection requirements for tag muons are very similar to those for 
loose-isolated muons except for the lower momentum threshold of 
$p_T \geq 4\gevc$ and the non-isolation 
requirement of $\Delta R(\mu,{\rm jet}) < 0.5$.
These $p_T$ and $\Delta R$ requirements select muons characteristic of 
those expected from heavy-flavor decays.

The momentum of the muon is computed from the deflection of 
the muon trajectory in the magnetized toroid. The momentum calculation uses a 
least squares method that considers seven parameters: four describing 
the position and angle of the track before the calorimeter 
(in both the bend and non-bend views), two describing the effects
due to multiple scattering, and the inverse of the muon momentum $1/p$.
This seven-parameter fit is applied to sixteen data points: vertex position
measurements along the $x$ and $y$ directions, the angles and positions of 
track segments before and after the calorimeter and outside of the iron, 
and two angles (one in the bend view, one in the non-bend view) representing 
the deflection due to multiple Coulomb scattering of the muon in the 
calorimeter. Energy loss corrections are applied using the restricted 
energy loss formula parametrized in \progname{geant} \cite{geant}.

The muon momentum resolution depends on the amount of material traversed, 
the magnetic field integral, and the precision of the measurement of the 
muon bend angle. As noted in Sec.~\ref{d0det}, the resolution function
shown in Eq.~\ref{eq:mumomres}, was 
based on studies of $Z \rightarrow \mu\mu$ data.
The first term in the resolution function reflects
multiple Coulomb scattering in the iron toroids and is the dominant
effect for low momentum muons. The second term reflects the resolution 
of the muon position measurements. The muon momentum scale was calibrated 
using $J/\psi \rightarrow \mu\mu$ and $Z \rightarrow \mu\mu$ candidates and
has an uncertainty of 2.5 \%.

The complete set of identification variables and misidentification rates
is discussed below.

\subsubsection{Muon quality ($Q$)}
\label{muqual}

For each found track in the muon system, the reconstruction makes a set of 
cuts on the number of modules hit, the impact parameters, 
and the hit residuals.
The number of cuts which a track fails is defined as the muon quality, $Q$
(for a perfect track $Q=0$). A similar parameter is also produced by the
level 2 trigger. If a track fails more than one (CF) or any (EF) of the cuts 
on the above 
quantities, then it is of insufficient quality and is rejected.
Tracks that have hits only in the inner layer of the muon system 
(inside the toroid) are also rejected. This eliminates almost all 
hadronic punchthrough from the calorimeter into the muon system.   
If a muon track is not bent by the toroid, muon momentum 
cannot be measured (as is the case for tracks which only have hits in the
inner layers).

\subsubsection{Calmip/MTC requirement}

As a muon passes through the calorimeter it deposits energy through
ionization along its path. This minimum ionizing trace should match to the
track found in the muon and central tracking systems and can serve as 
a very powerful tool for the rejection of backgrounds.
During the course of the run this was used in two ways. For Run 1a, 
it is accomplished by checking the energy in the calorimeter towers along
the expected path of the muon: For events in which a track match is 
found in the central tracking system within $\Delta \eta \leq 0.45$ and 
$\Delta \phi \leq 0.45$ of the 
muon track, an energy deposit of at least 0.5 GeV is required in the 
calorimeter towers along the track plus its two nearest neighbor towers; 
for muons without a central detector track match, at least 1.5 GeV is 
required (to allow 
for tracking inefficiencies in the region near $|\eta| \approx$ 1 where  
the coverage of the central tracking system is incomplete).
This requirement is denoted by ``calmip'' in Tables~\ref{tab:muid1a} -- 
\ref{tab:muid1bpost}. For data from Runs 1b and 1c, a more sophisticated
procedure is employed. This procedure, denoted ``MTC,'' is based on muon 
tracking in the calorimeter. The track from the muon system is used to 
define a path through the calorimeter to the position of the interaction
vertex. A $5\times 5$ road of calorimeter cells is defined along this
path. Any cell with an energy two standard deviations above the
noise level is counted as hit. The longest contiguous set of hit cells
constitutes the calorimeter track. Muon candidates are required to have
tracks with hits in at least 70\% of the possible layers in the
hadronic calorimeter. If a track does not have hits in all 
the layers, then it is also required that at least one of the nine
central cells in the outermost layer of the $5\times 5$ road be 
hit~\cite{ptthesis}.
These requirements reject both combinatoric background and cosmic rays.
The MTC criteria cannot be used on the Run 1a data because the required
 information
is not supplied by the 1a reconstruction. For the $\mu$+jets channels
(which uses the tight muon identification criteria) the Run 1a raw data was 
reprocessed, incorporating the needed information. 
Thus, in Table~\ref{tab:muid1a},
MTC refers to the tight identification and the tag identification for 
the $\mu$ + jets channels and calmip refers to the loose identification 
and the tag identification for the $e$+jets/$\mu$ channel.

\subsubsection{Impact parameter (IP)}

An impact parameter requirement for the muon trajectory relative to the
interaction vertex provides further rejection against 
cosmic rays and misreconstructed muons.  Here $IP_{BV}$ and $IP_{NB}$ 
are the two-dimensional distances-of-closest approach between the muon 
and its associated 
vertex in the bend and non-bend projections respectively. 
These are combined in quadrature, 
$IP \equiv \sqrt{IP_{BV}^2 + IP_{NB}^2}$, and $IP$ is required to be less
than 20 cm. 

\subsubsection{Minimum magnetic path length ($\int Bdl$)}

Muons that pass through the thinner part of the iron toroid near
$|\eta|\approx0.9$ have poorer momentum resolution and may also be 
contaminated by a small background from punchthrough.
Excluding these thin regions, the punchthrough fraction is $<2$ \% 
and is essentially negligible for muons with $p_T > 5$ \gevcn.
The $\int \vec{B} \times \vec{dl}$ requirement
ensures that muons traverse enough
field ($\geq 1.83$ Tm) to provide an acceptable $p_T$ measurement.


\begin{table}
\caption{Definitions of and identification efficiencies for loose, tight, 
and tag CF ($|\eta| \leq 1.0$) muons 
for Run 1a. For calmip/MTC: $e\mu$, $\mu\mu$ (loose) and $e$ + jets/$\mu$ 
(tag) use calmip; $\mu$ + jets/topo (tight) and 
$\mu$ + jets/$\mu$ (tight and tag) reprocessed the 1a data and therefore 
use MTC. The two efficiencies given for tag muons reflect inclusion of 
calmip or MTC requirements respectively.
\label{tab:muid1a}}
\vskip 0.5cm
\begin{tabular}{l|c|c|c}
                   & \multicolumn{3}{c}{$\mu$ id Run 1a (CF)} \\
       definition: & Loose  & Tight & Tag     \\
\cline{2-4}
 $p_{T}^{\mu} \geq $         & 15          & 20            & 4        \\ 
 $Q$ $\leq $                 & 1           & 1             & 1        \\ 
 calmip/MTC                  & yes         & yes           & yes      \\
 $IP \leq$                   & --          & 20 cm         & --       \\
 $\int Bdl \geq $            & --          & 1.83 Tm      & --       \\ 
 $\Delta R({\mu,{\rm jet}})$  & $\geq 0.5$  & $\geq 0.5$    & $< 0.5$  \\ 
\hline
 Eff (\%)                    & $64 \pm 6$ & $46\pm7$      & $80\pm6/77\pm6$ \\
\end{tabular}
\end{table}

\begin{table}
\caption{Definitions of and identification efficiencies for loose, tight, 
and tag CF ($|\eta| \leq 1.0$) muons for Run 1b(preclean).
\label{tab:muid1bpre}}
\vskip 0.5cm
\begin{tabular}{l|c|c|c}
      & \multicolumn{3}{c}{$\mu$ id Run 1b preclean (CF) } \\
 definition: & Loose  & Tight & Tag     \\
\cline{2-4}
    $p_{T}^{\mu} \geq $         & 15          & 20            & 4        \\ 
    $Q$ $\leq $                 & 1           & 1             & 1        \\ 
    MTC                         & yes         & yes           & yes      \\
    $IP \leq$                   & --          & 20 cm         & --       \\
    $\int Bdl \geq $            & --          & 1.83 Tm      & --       \\ 
    $\Delta R({\mu,{\rm jet}})$  & $\geq 0.5$  & $\geq 0.5$    & $< 0.5$  \\ 
\hline
    Eff (\%)                    & $65\pm5$    & $46\pm7$      & $76\pm6$  \\
\end{tabular}
\end{table}

\begin{table}
\caption{Definitions of and identification efficiencies for loose, tight, 
and tag muons for CF ($|\eta| \leq 1.0$) and EF ($ 1.0 < |\eta| \leq 1.7$) 
regions for run 1b+c (postclean).
\label{tab:muid1bpost}}
\vskip 0.5cm
\begin{tabular}{l|cc|cc|cc}
 & \multicolumn{6}{c}{$\mu$ id Run 1b+c postclean } \\
 & \multicolumn{2}{c|}{Loose} & \multicolumn{2}{c|}{Tight} & \multicolumn{2}{c}{Tag} \\
 definition:   &  CF     &    EF    &    CF    &    EF   &    CF    &    EF   \\
\cline{2-7}
$p_{T}^{\mu} \geq $  & \multicolumn{2}{c|}{15} & \multicolumn{2}{c|}{20} & 
\multicolumn{2}{c}{4}   \\ 
  $Q$ $\leq $ &   1  &  0   &   1   &  0    &  1  & 0  \\ 
  MTC  & \multicolumn{2}{c|}{yes} & \multicolumn{2}{c|}{yes} &\multicolumn{2}{c}{yes} \\
  $IP \leq$                   & --& --& \multicolumn{2}{c|}{20 cm}        & --& --\\
  $\int Bdl \geq $            & --& --& \multicolumn{2}{c|}{1.83 Tm}     & --& --\\ 
  $\Delta R({\mu,{\rm jet}})$  & \multicolumn{2}{c|}{$\geq 0.5$} & \multicolumn{2}{c|} 
{$\geq 0.5$}    & \multicolumn{2}{c}{$< 0.5$}  \\ 
\hline
  Eff (\%)      & $73\pm3$ & $68\pm5$  & $49\pm7$ & $52\pm16$ & $84\pm4$  & $62\pm15$  \\
\end{tabular}
\end{table}

\subsubsection{Isolation}

A muon is considered isolated if it is well separated from any 
reconstructed jet. Isolation, or $\Delta R(\mu,{\rm jet})$, is the distance 
in $(\eta,\phi)$ space between a muon and the nearest 0.5 cone jet with
$E_T\geq$ 8 GeV.

\subsubsection{Efficiency}
\label{muid-eff}

The total muon-finding efficiency is the product of the muon geometrical 
acceptance and the muon identification efficiency. The muon geometrical 
acceptance is ($73.7 \pm 0.4$)\% for the CF and ($64.1 \pm 1.1$)\% for the EF.
The total muon-finding efficiency is well-modeled by a modified version of
\progname{d{\o}geant}. These modifications include input from measured muon
resolutions and efficiencies of the PDTs. The muon identification
efficiency is obtained from this modified version of \progname{d{\o}geant},
but is further corrected to account for time dependent detector 
inefficiencies and incorrect modeling of the muon track finding efficiency.
As can be seen in Tables~\ref{tab:muid1a} -- 
\ref{tab:muid1bpost}, the muon identification efficiency varies with muon 
category and run period.

\subsection {Jets}
\label{jets}
Jets are reconstructed using a cone algorithm~\cite{prd1,ajm,rva} with 
cone sizes, 
$\Delta R (\equiv \sqrt{\Delta \eta^2 + \Delta \phi^2})$, of 0.3 and
0.5. The cone size of $\Delta R = 0.3$ is used only in the level-2 trigger 
and for certain aspects
of the all-jets analysis (see Sec.~\ref{alljets}); all other analyses 
use a cone size of $\Delta R = 0.5$ to
maximize the efficiency for reconstructing $t \bar t$ events. 
The algorithm is executed 
as follows. Starting from an $E_T$-ordered list of calorimeter towers,
the towers within $\Delta R \approx 0.3$ and with $E_T > 1$ GeV are 
grouped into preclusters. The energy within a given cone (0.3 and 0.5 for
the analyses presented here) centered on the precluster is summed, and a new
``$E_T$-weighted'' center is obtained. Starting with this new center, the 
process is repeated until the center stabilizes. A jet is required to have
$E_T > 8$ GeV. If two jets share energy, they are combined or split,
depending on the fraction of $E_T$ shared relative to the $E_T$ of the lower 
$E_T$ jet. If the shared fraction exceeds 50\%, the jets are combined.

\begin{figure}
\vbox{
\centerline{\hskip -0.1 in \psfig{figure=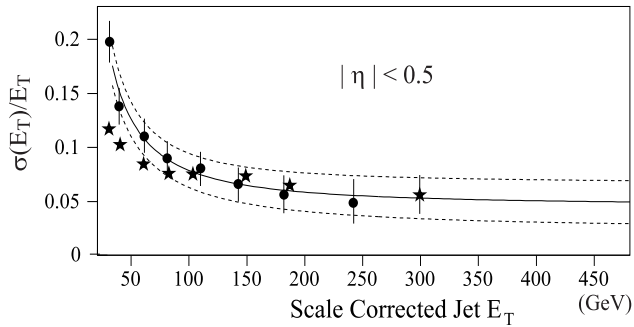,height=1.7in,width=3.3in}}
\vskip 0.2cm
\caption{ Jet $E_T$ fractional resolution for $|\eta| < 0.5$. The circles
and solid line correspond to the nominal resolution; the dotted lines are the
systematic uncertainty on the resolution measurement. The stars 
correspond to resolutions obtained from  
\progname{HERWIG}+\progname{d{\o}geant} Monte Carlo and are used at high 
$E_T$ where dijet data are not available.
\label{fig:jetres}}}
\end{figure}

The jet energy resolution is obtained from studies of $E_T$ balance in 
dijet and photon+jet data
in different $\eta$ regions~\cite{prd1}. As shown in
Fig.~\ref{fig:jetres}, the fractional resolution ($\sigma(E_T)/E_T$) in 
the central region varies from 20\% at a jet $E_T$ of 30 GeV to 8\% at a 
jet $E_T$ of 100 GeV. Resolutions in the other detector regions are similar.
The absolute jet energy scale is discussed in the following section.

\subsection {Missing $E_T$ (\hbox{$\rlap{\kern0.25em/}E_T$})}
\label{misset}

Neutrinos escape the detector without interacting. Similarly, muons
pass through the calorimeter depositing very little energy. The presence of
a high-energy neutrino can be inferred from an imbalance in 
transverse energy/momentum as measured in the calorimeter and muon systems.

Missing transverse energy in the
calorimeter, ${\hbox{$\rlap{\kern0.25em/}E_T^{\rm cal}$}}$, is defined as
\begin{equation}
{\hbox{$\rlap{\kern0.25em/}E_T^{\rm cal}$}} = 
\sqrt{{\hbox{$\rlap{\kern0.25em/}E_{x}^{\rm cal}$}}^2
+{\hbox{$\rlap{\kern0.25em/}E_{y}^{\rm cal}$}}^2},
\end{equation}
where
\begin{equation}
{\hbox{$\rlap{\kern0.25em/}E_{x}^{\rm cal}$}} = 
-\sum_i E_i {\rm sin}(\theta _i) {\rm cos}(\phi _i)  -
\sum_j \Delta E_{x}^{j},
\end{equation}
and
\begin{equation}
{\hbox{$\rlap{\kern0.25em/}E_{y}^{\rm cal}$}} = 
-\sum_i E_i {\rm sin}(\theta _i) {\rm sin}(\phi _i) -
\sum_j \Delta E_{y}^{j}.
\end{equation}
The first sum is over all cells in the calorimeter and ICD, and the second 
sum is over the corrections in $E_T$ applied to all electrons and jets in the 
event (see Sec.~\ref{escale}). The missing transverse energy 
(\hbox{$\rlap{\kern0.25em/}E_T$}) resolution of the calorimeter is
parameterized as~\cite{prd1}
\begin{equation}
\sigma(\metcal) = 1.08 {\rm \ GeV} + 0.019 \sum E_T,
\label{eq:metres}
\end{equation}
where $\sum E_T$ is the scalar $E_T$, which is defined to be the 
scalar sum of all calorimeter cell $E_T$ values.

For events that contain muons, the transverse momentum of the muon is
subtracted from $\metcal$ to compute the total missing  $E_T$:
\begin{equation}
{\hbox{$\rlap{\kern0.25em/}E_{x}$}} = 
{\hbox{$\rlap{\kern0.25em/}E_{x}^{\rm cal}$}} - 
\sum_i p^{\mu _i}_{x},
\end{equation}
\begin{equation}
{\hbox{$\rlap{\kern0.25em/}E_{y}$}} = 
{\hbox{$\rlap{\kern0.25em/}E_{y}^{\rm cal}$}} - 
\sum_i p^{\mu _i}_{y}.
\end{equation}

\section{Event simulation}
\label{mc}

The $t\bar t$ signal efficiencies and several rare background rates 
are computed via Monte Carlo methods. The primary event generator for
the signal is \progname{HERWIG}~\cite{herwig} with 
CTEQ3M~\cite{cteq3m} parton distribution functions (pdf). 
Tests were also performed with three values of $\Lambda_{\rm QCD}$, 
and using the MRSA$^{\prime}$ pdfs~\cite{mrsap}, 
but no significant variation in $t\bar t$ acceptance was seen.
\progname{HERWIG} chooses the momenta out of the initial hard scattering
according to matrix element calculations and models initial and final 
state gluon emission using leading-log QCD evolution~\cite{dglap}. 
Each top quark is then made 
to decay into a $W$ boson and a $b$ quark, and the final state partons
are hadronized into jets. Underlying spectator 
interactions are also included in the model. As a cross-check, 
acceptances were also computed using the \progname{ISAJET}~\cite{isa} 
event generator 
(also using the CTEQ3M pdfs), and the difference 
between the two is incorporated into the systematic uncertainties 
on a per channel basis (see Secs.~\ref{ss:ttmcerr} -- \ref{ss:ttmcbtagerr} 
for details). 
\progname{ISAJET} also chooses the momenta out of the hard scattering based
on matrix element calculations, but models the initial and final
state gluon emission using Feynman-Field fragmentation~\cite{feynfield}. 

\progname{HERWIG} was chosen as the
primary generator because it provides good agreement with data in D\O's 
color coherence~\cite{colorcoh} and jet-shape~\cite{d0jetshape} 
analyses. As discussed in 
Sec.~\ref{crsec}, within available statistics, 
the leptonic top candidates found in
the current analysis are in good agreement with expectations from 
\progname{HERWIG}. However, it should be noted that version 5.7 of 
\progname{HERWIG} (the version used for the present analyses) is 
based on leading-log matrix elements, and is therefore not in complete 
agreement with higher-order predictions~\cite{orr1,orr2}.

\progname{HERWIG} and \progname{ISAJET} samples were generated with top
quark masses between 90 and 230 \gevccn. To increase event-processing 
efficiency, two samples were made for each mass and generator: (1) one 
in which both of the $W$ bosons were required to decay leptonically
($e,\mu,\tau$), from which only those that resulted in a final state of
$ee$, $\mu\mu$, or $e\mu$ were kept, and (2) one in which one of the $W$
bosons was forced to decay leptonically ($e,\mu,\tau$), from which 
those with no final state electrons or muons were rejected, as were one-half 
of the dilepton events (in order to preserve the proper branching
ratios).

For the dilepton channels, backgrounds from $Z\ra \tau\tau$, $Z\ra \mu\mu$, 
$WW$, $WZ$, and Drell-Yan production 
are determined with \progname{PYTHIA}~\cite{pythia} and with 
\progname{ISAJET},
and the difference used as a measure of systematic uncertainty.

Background levels from $W$+jets production are determined from data. However, 
as discussed in Sec.~\ref{ljets_topo}, shape information from the 
\progname{VECBOS}~\cite{vecbos} Monte Carlo program is used to 
determine the survival probability for the latter stages of the 
$\ell$+jet/topo analyses. For the $\mu$+jets/$\mu$ analysis 
(see Sec.~\ref{ljets_mutag}), \progname{VECBOS} is used to determine the 
$Z \ra \mu\mu$ 
background. In both cases, the CTEQ3M~\cite{cteq3m} pdfs
are used. \progname{VECBOS} incorporates the exact tree-level matrix
elements for $W$ and $Z$ boson production, with up to four additional
partons, and supplies the final state partons. In order to include the
effects of additional radiation and underlying events, and to model the
hadronization of the final state particles, the \progname{VECBOS} output
is passed through \progname{HERWIG}'s QCD evolution and fragmentation 
stages. Since \progname{HERWIG} requires information about the color labels
of its input partons, both programs were modified to assign
color and flavor to the generated partons. The flavors are assigned 
probabilistically by keeping track of the relative weights of each
diagram contributing to the process. The color labels are assigned randomly.
To estimate the systematic uncertainty, samples were also generated 
using \progname{ISAJET}, instead of \progname{HERWIG}, to fragment the 
\progname{VECBOS} partons.

The output of an event generator is typically processed through a 
\progname{GEANT}~\cite{geant} simulation of the detector
(\progname{d\o geant}). 
However, such a detailed simulation is extremely computationally 
intensive and does not
allow for generation of the necessary high-statistics samples.
As a compromise, the full \progname{d\o geant}
simulation was run on a large sample of electrons and hadrons, and 
the resultant calorimeter showers were stored in a library~\cite{ewv}.
These showers are binned in five quantities representing the input particle:
\begin{itemize}
\item $z$ vertex position (6 bins)
\item $\eta$ (37 bins matching calorimeter segmentation)
\item momentum (7 bins)
\item $\phi$ region: The calorimeter is largely symmetric in $\phi$, the
      exceptions being the cracks between modules in the central 
      electromagnetic calorimeter and the region through which the Main Ring
      passes in the hadronic calorimeter. Hence, there are only two bins
      in $\phi$, representing the ``good'' and ``bad'' regions.
\item particle type: Energy depositions in the calorimeter for 
      electrons/photons and hadrons are stored separately in the 
      shower library.
\end{itemize}
A total of 1.2 million events was used to populate the library. 
As events are sent through the library version of the simulation, a shower
from the library is selected to model the calorimeter response of each 
individual particle. The total energy of the shower is scaled by the ratio 
of the energy of the particle to be simulated to that of the library particle 
which created the shower. Since the full \progname{d\o geant} simulation for
muons is not as time-consuming (owing to their minimum-ionizing nature), muons
are not included in the shower library but are instead tracked through the
detector just as in non-library version of the simulation.

For the muon system, the efficiency is overestimated and the resolution 
is underestimated
by \progname{d{\o}geant}. The next step in the simulation procedure 
therefore smears the muon hit timing information so that the Monte Carlo
hit position resolution matches that in $Z\ra\mu\mu$ data, and 
randomly discards hits to model the chamber inefficiency more accurately. 
In addition,
the muon-system geometry in the Monte Carlo is misaligned in order to 
reproduce the correct overall momentum resolution.

For several of the analyses, a final step in the simulation 
models the level 1 and level 2 triggers (trigger simulator). As discussed 
in Sec.~\ref{triggers}, the level 1 trigger is a collection of hardware 
elements interfaced to an AND-OR network. The level 1 simulation therefore 
consists
of simulated trigger elements and a simulated AND-OR network. Level 2 is
a software trigger that runs in the online data acquisition environment.
The level 2 simulation consists of exactly the same code but has been
ported to the offline environment. The level 1 and level 2 simulations are
typically used as a single entity, referred to simply as the 
trigger simulator.

\section{Analysis of dilepton events}
\label{dilep}

As discussed in Sec.~\ref{sec:intro},
the $ee$, $e\mu$, and $\mu\mu$ dilepton signatures are 
characterized by two isolated high-$p_T$  charged leptons, $\met$, and two 
or more jets (from the $b$ quarks and initial and final state radiation). 
Figures~\ref{fig:int_dil_lep} and \ref{fig:int_dil_jet} show Monte Carlo 
distributions for the lepton and jet $E_T/p_T$ and $|\eta|$, and the $\MEt$ 
expected in $t\bar t \ra e\mu$ events with $m_t = 170~\gevcc$.
Background events with a similar topology are relatively rare and arise 
primarily from Drell-Yan production of ($Z/\gamma$)+jets, $WW$+jets, 
and leptonic $W$+jets events in which the second 
lepton arises from the misidentification of one of the jets. Therefore,
requirements based on the above characteristics 
form the initial selection for all three channels 
(see Tables~\ref{tab:eecuts}, \ref{tab:emucuts}, 
and \ref{tab:mumucuts}). Additionally, for the $ee$ and $\mu\mu$
channels there are cuts designed to reject $Z \ra ee,\mu\mu$ events.

To attack the remaining background, variables were selected based on a 
series of cut optimization studies.
These are designed to maximize the significance, defined as 
$ {\cal S} \equiv {\rm signal}/ \sqrt{{\rm background}}$, and result in the
introduction of a new transverse energy variable, defined as
\begin{equation}
H_T^e \equiv  \displaystyle\sum_{\rm jets} E_T  + 
                ({\rm leading \ electron} \ E_T)
\label{eq:htedef}
\end{equation}
for the $ee$ and $e\mu$ channels and as
\begin{equation}
H_T \equiv  \displaystyle\sum_{\rm jets} E_T 
\label{eq:htmumudef}
\end{equation}
for the $\mu\mu$ channel.
The sums are over all jets with $E_T \geq 15$ GeV and $|\eta| 
\leq 2.5$. The optimized $H_T^e$ and $H_T$ cut values are given in 
the event selection tables in Secs.~\ref{ee}, \ref{emu}, and \ref{mumu}.
An additional result of the optimization studies was the requirement that,
for the $ee$, $e\mu$, and $\mu\mu$ channels, there should be at least two 
jets with $E_T \geq 20$ GeV.
As discussed below, both of these requirements are very effective
in distinguishing $t \bar t$ events from background.

In addition to the $ee$, $e\mu$, and $\mu\mu$ channels, a new
channel was introduced that is designed to catch dilepton events
in which one of the leptons either fails the $p_T$ requirement or 
escapes detection (perhaps by passing through
an uninstrumented region of the detector). This ``$e\nu$'' channel selects
events that contain one high-$p_T$ electron, significant missing transverse 
energy, and two or more jets. The analogous $\mu\nu$ channel has not been
considered.

Acceptances for all four dilepton channels were computed from Monte Carlo 
events generated by the \progname{herwig} program 
for 24 top quark mass 
values ($m_t$ = 90 -- 230 GeV/$c^2$)  and then passed through the full D\O\ 
detector simulation (see Sec.~\ref{mc}). The expected number of 
$t \bar t$ events passing the selection for a given channel is 
\begin{equation}
N = \sigma_{t \bar t}(m_t) \sum_{i = {\rm runs}} \sum_{j = {\rm det}}  
A(i, j, m_t) \cdot {\mathcal L}_{i,j} 
\label{eq:numllevts}
\end{equation}
where $\sigma_{t \bar t}$ is the theoretical $t\bar t$ cross section at 
a top quark mass of $m_t$ \cite{laenen}, ${\mathcal L}_{i,j}$ is the 
integrated luminosity for run $i$ and pair of lepton detector regions $j$ 
(for $ee$ $j$=CC+CC,CC+EC,EC+EC, for $e\mu$ $j$=CC+CF,CC+EF,EC+CF,EC+EF, and
for $\mu\mu$ $j$=CF+CF,CF+EF,EF+EF),
and the acceptance, $A$, is 
\begin{equation}
A = \varepsilon_{\rm trig} \cdot \varepsilon_{\rm pid} \cdot 
             \varepsilon_{\rm sel} \cdot G \cdot {\cal B},
\label{eq:llaccept}
\end{equation}
where $\varepsilon_{\rm trig}(i,j,m_t)$ is the trigger efficiency, 
$\varepsilon_{\rm pid}(i,j)$ is the efficiency for
identifying the two leptons, $\varepsilon_{\rm sel}(i,j,m_t)$ is
the efficiency of the selection criteria, $G(i,j)$ is the geometric
acceptance, and ${\cal B}$ is the branching fraction for the sample 
being studied.
Trigger efficiencies are obtained from data or Monte Carlo, depending
on the channel, and are discussed in greater detail below. 
Particle identification
efficiencies are obtained from data in the case of electrons (as discussed
in Sec.~\ref{e-id}), and from a combination of data and Monte Carlo in
the case of muons (as discussed in Sec.~\ref{muid}).
The selection efficiencies $\varepsilon_{\rm sel}$ and geometrical acceptances
$G$ are calculated from Monte Carlo.
As will be discussed in Sec.~\ref{crsec}, it is the acceptance, rather than
the expected number of $t\bar t$ events, that is used to calculate
the $t\bar t$ cross section.
Typical values for acceptance, often denoted as the 
``efficiency times branching fraction'' ($\varepsilon \times {\cal B}$),
for all eight leptonic channels, are tabulated in Sec.~\ref{crsec} 
for seven values of top quark mass. The numbers of 
$t \bar t$ events expected in the four dilepton channels are tabulated
in Secs.~\ref{ee}, ~\ref{emu}, ~\ref{mumu}, and ~\ref{enu}, 
for the same set of top quark masses. Systematic 
uncertainties on the acceptances are discussed in Sec.~\ref{syserr}.

Whenever possible, backgrounds are 
measured directly from data. If not, then the backgrounds 
are determined from Monte Carlo events in which the initial cross sections
are normalized either to measured or theoretical values:
\begin{equation}
B = \sigma_{\rm bkg} \sum_{i = {\rm runs}} \sum_{j = {\rm det}}  
         \Bigl( \varepsilon_{\rm trig} \cdot \varepsilon_{\rm pid} \cdot 
         \varepsilon_{\rm sel} \cdot G \Bigr)_{i,j} \cdot {\mathcal L}_{i,j} 
\label{eq:numllbkevts}
\end{equation}
where $\sigma_{\rm bkg}$ is the measured or theoretical cross section 
for the background under consideration.

\subsection{The $ee$ channel}
\label{ee}


\begin{table*}
\caption{Number of observed and expected $ee$ events passing at
each cut level of the offline analysis. Expected number of $\ttb$ events are
for $m_t = 172.1 \gevcc$. Uncertainties correspond to 
statistical and systematic contributions added in quadrature. 
\label{tab:eecuts}}
\vskip 0.5cm
\begin{tabular}{l|c|c|c|c|c}
\multicolumn{6}{c}{Number of $ee$ events at each cut level } \\
\multicolumn{1}{l|}{}&       & Total       &mis-id& physics &      \\
\multicolumn{1}{l|}{}& Data  & sig + bkg   & bkg  & bkg     & $\ttb$ \\
\hline 
$2e$, $E_{T}^{e} > 20$ GeV, + $e$ id + trig 
 &     &               &               & $4168 \pm 1243$ & $1.9 \pm 0.3$ \\
+ 2 jets, $E_{T}^{\rm jet} > 15$ GeV  
 & 112 & $125 \pm 36$ & $9.0 \pm 0.08$ & $114 \pm 35$ & $1.8 \pm 0.3$ \\
+ $\metcal > 25$ GeV                  
 & 3   & $3.2 \pm 1.9$ & $0.23 \pm 0.06$ & $1.5 \pm 1.9$ & $1.5 \pm 0.3$ \\
+ $\metcal > 40$ GeV or         
 &     &               &               &       &  \\
 \ \ \ $M_{ee} < 79$ \gevcc or $M_{ee} > 103$ \gevcc  
                                      
 & 2   & $2.3 \pm 0.5$ & $0.22 \pm 0.06$ & $0.62 \pm 0.21$ & $1.5 \pm 0.3$ \\
+ 2 jets, $E_{T}^{\rm jet} > 20$ GeV  
 & 2   & $1.9 \pm 0.4$ & $0.20 \pm 0.05$ & $0.39 \pm 0.12$ & $1.4 \pm 0.3$ \\
+ $H_{T}^e > 120$ GeV                   
 & 1   & $1.7 \pm 0.2$ & $0.20 \pm 0.05$ & $0.28 \pm 0.09$ & $1.2 \pm 0.2$ \\
\end{tabular}
\end{table*}


The signature for an event in the $ee$ channel consists of two 
isolated high-$E_T$ electrons, two or 
more jets (from the $b$ quarks and initial and final state radiation),
and significant $\MEt$ (from the neutrinos). The trigger for this 
channel was (depending on run period) \progname{ele-jet}(1a), 
\progname{ele-jet-high}(1b), or 
\progname{ele-jet-higha}(1c), requiring an electron, 2 jets, and
$\MEt$ at level 2 (see Sec.~\ref{triggers} for details).
As discussed in Appendix~\ref{mrrecov}, for this analysis Main-Ring events were
corrected and not rejected. 
Over the complete Run 1 data set, these triggers provided a total integrated
luminosity of $130.2 \pm 5.6 \ {\rm pb}^{-1}$. 
The event sample passing these triggers consists primarily of misidentified
multijet and heavy flavor events.

The backgrounds to this signature arise from Drell-Yan
($Z/\gamma^*$) production that
results in a dielectron final state ($Z \rightarrow ee$, 
$Z \rightarrow \tau\tau \rightarrow ee$, and $\gamma^* \rightarrow ee$), 
$WW \rightarrow ee$, and multijet events containing one or more misidentified
electrons. The latter background consists primarily of 
$W(\rightarrow e\nu)$+3 jet events in which one of the jets is misidentified 
as an electron.

\begin{figure}
\vbox{
\vskip -0.9cm
\centerline{\psfig{figure=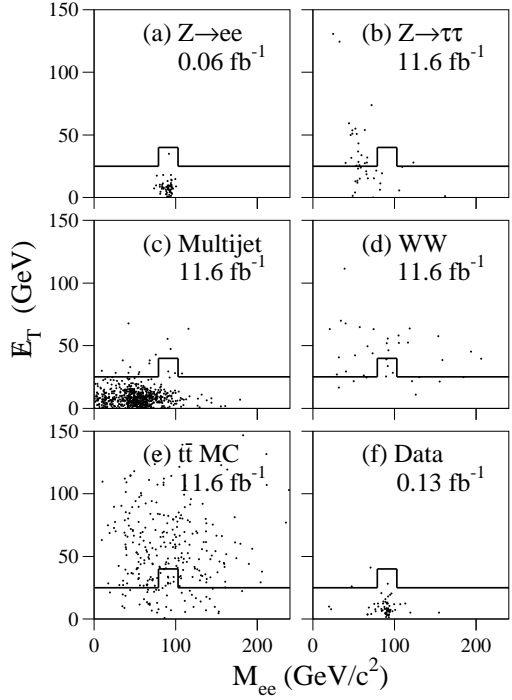,width=3.0in}}
\hskip 0.1cm
\caption{Scatter plots of $\MEt$ vs $M_{ee}$ for the $ee$ channel:
(a) $Z \ra ee$ events,(b) $Z \ra \tau\tau \ra ee$ MC events,
(c) QCD multijet events, (d) $WW \ra ee$ MC events, (e)
$t{\bar t} \ra ee$ MC signal ($m_t = 172.1$ GeV/$c^2$), and (f) data. 
The signal region is defined as being above the solid line in each plot.
\label{fig:ee}}}
\end{figure}

The offline selection cuts and their cumulative effect are summarized in 
Table~\ref{tab:eecuts}. After passing the trigger requirement, events are 
required to have 2 electrons (loose electron identification, see
Sec.~\ref{e-id}) with $E_T > 20$ GeV and $|\eta| \leq 2.5$. 
This initial selection has an acceptance ($\varepsilon \times {\cal B}$) 
of ($0.26 \pm 0.03$)\% (for $m_t = 170$ GeV/$c^2$), and essentially eliminates 
any background from heavy flavor production 
and reduces the QCD multijet background to a small fraction of the 
remaining dominant background from $Z \rightarrow ee$.
The number of $Z+n$ jet events is proportional to $\alpha_s^n$, and a similar
steep falloff in jet multiplicity is observed for the other backgrounds
present at this stage. Requiring 2 jets with 
$E_T > 15$ GeV and $|\eta| \leq 2.5$ significantly reduces backgrounds 
from $Z$ boson, Drell-Yan and $WW$ production, and QCD multijet events. 
Most of these
($Z$, Drell-Yan, and QCD multijet) do not contain
high-$p_T$ neutrinos. Therefore, a hard cut on the 
$\met$ brings these events to an even more manageable level.
At this point the background is still dominated by $Z \rightarrow ee$
events, so the next step requires that the dielectron invariant mass not
be within the mass window of the $Z$ boson (see Table~\ref{tab:eecuts}). 
However, since 
$Z \rightarrow ee$ events have no real $\MEt$, this cut is only
made for events with $\MEt < 40$ GeV, thereby reclaiming a considerable
amount of $t \bar t$ efficiency.
The final two cuts, $H_T^e > 120$ GeV and $N_{\rm jets} \geq 2$ with 
$E_T^{\rm jet} > 20$ GeV and
$|\eta^{\rm jet}| \leq 2.5$, are obtained through the optimization procedure 
discussed in Sec.~\ref{dilep},
and provide rejection against the remaining 
background from $Z \rightarrow \tau\tau$, $WW$, and Drell-Yan production, 
and QCD multijet events.
Table~\ref{tab:eecuts} shows the number of data events, expected signal 
($m_t = 172.1$ GeV/$c^2$), and expected background surviving at each stage 
of the selection. It is clear from this table that the $\met$ requirement
greatly reduces
the background. This is shown in Fig.~\ref{fig:ee}, where $\MEt$ is plotted
vs $M_{ee}$ for all the major backgrounds (a-d), for $t \bar t$
Monte Carlo (e), and for data (f). Because of the presence of 
two neutrinos, the $WW$ background is not reduced much by the 
selection on $\MEt$. 
It is, however, reduced significantly by the jet and $H_T^e$ requirements. 
The effect of the $H_T^e$ cut on $WW$ events can be seen in 
Fig.~\ref{fig:emu}(b),
which gives the $H_T^e$ distribution for $t \bar t \rightarrow e\mu$ events, 
but is very similar to that for $t \bar t \rightarrow ee$ events.
After the above selection, only one $ee$ candidate remains.

\begin{table}
\caption{Expected number of $ee$ signal and background events after all cuts 
in 130.2 ${\rm pb}^{-1}$. Uncertainties are statistical and systematic 
contributions added in
quadrature. The systematic uncertainty on the total background 
includes correlations among the different background sources.
\label{tab:eeexp}}
\vskip 0.5cm
\begin{tabular}{c|c}
\multicolumn{2}{c}{Expected number of $ee$ events in 130.2 ${\rm pb}^{-1}$ } \\
\hline 
\underline{top MC  $m_{t}$ (\gevccn)} &     \\
                   140      &  $2.34 \pm 0.34 $ \\
                   150      &  $1.96 \pm 0.29 $ \\
                   160      &  $1.62 \pm 0.23 $ \\
                   170      &  $1.25 \pm 0.18 $ \\
                   180      &  $1.02 \pm 0.15 $ \\
                   190      &  $0.79 \pm 0.12 $ \\
                   200      &  $0.62 \pm 0.09 $ \\
\hline
$Z\ra \tau\tau \ra ee $    & $0.08 \pm 0.06 $ \\
multijet (mis-id $e$)      & $0.20 \pm 0.05 $ \\ 
$Z\ra ee$                  & $0.06 \pm 0.01 $ \\
$WW\ra ee$                 & $0.09 \pm 0.03 $ \\
${\rm DY}\ra ee$           & $0.06 \pm 0.03 $ \\
\hline
Total background           & $0.48 \pm 0.10 $ \\
\end{tabular}
\end{table}


The $Z\rightarrow ee$ background is determined entirely from data. As noted
above, $Z(\rightarrow ee$)+jets events have no real $\met$, and due to the
excellent electron momentum resolution, any $\met$ observed in the detector
will arise from mismeasurement of jet $E_T$ and other noise in the 
calorimeter. Due to 
the extremely high rejection power of the $\met$ requirement on 
$Z\rightarrow ee$+jet events, a $\met$ mismeasurement rate is 
determined from a sample 
of QCD multijet data selected to closely match the jet 
requirements in this analysis: $\geq 2$ jets, $E_T > 20$ GeV, 
$H_T > 70$ GeV (where the remaining 50 GeV contribution to the 
$H_T^e > 120$ GeV is assumed to originate from the highest-$E_T$ electron). 
The fraction of
events in this sample that passes the $\met > 25$ GeV requirement is 
taken as the $\met$ mismeasurement rate 
(i.e., the fraction of the time that the detector
resolution will result in a false $\met$ signal). Due to a slight dependence 
on jet multiplicity, the $\met$ mismeasurement rate is determined as a 
function of the $\met$ cut and number of jets $n$ in the event and is found 
to be ($1.02 \pm 0.09$)\% for $n=2$, ($0.86 \pm 0.02$)\% for $n=3$, and 
($1.12 \pm 0.02$)\% for $n=4$ for $\met > 25$ GeV; and 
($0.20 \pm 0.04$)\% for $n=2$, ($0.14 \pm 0.01$)\% for $n=3$, and 
($0.17 \pm 0.01$)\% for $n=4$ for $\met > 40$ GeV. These factors are 
then applied to
the number of dielectron events that pass all selection requirements 
(including the $Z$ boson mass window cut), except for that on 
$\met$, to obtain the
total expected $Z\rightarrow ee$ background of $0.058 \pm 0.013$ events.
The systematic uncertainty on this determination is discussed in 
Sec.~\ref{syserr}.

The background from multijet events is also obtained entirely from data.
The probability for an extra-loose electron to pass the loose electron 
identification criteria (see electron misidentification rate discussion 
in Sec.~\ref{e-id})
is applied to both the full Run 1 sample (not including Main-Ring, 
MR, events) of dielectron events in which one electron
candidate passes the loose identification and the other fails the 
loose identification but passes the extra-loose identification, and to 
that where both electron candidates fail the
loose identification but pass the extra-loose identification. 
The resultant misidentification background is then 
scaled up by the (nonMR + MR)/nonMR luminosity ratio to account for the
misidentification background expected in the MR data.

Backgrounds from $Z\rightarrow\tau\tau\rightarrow ee$, $WW\rightarrow ee$,
and $\gamma^* \rightarrow ee$ are obtained from \progname{pythia} and 
\progname{isajet} Monte Carlo samples via Eq.~\ref{eq:numllbkevts},
and are normalized either to experimental or theoretical values. 

The $Z\rightarrow\tau\tau \rightarrow ee$ Monte Carlo
samples are normalized to D\O's $Z$ boson cross section measurement and its
measurement of $p_T^Z$ (to obtain more $Z$+jets events and thus enhance the 
final statistics, generator-level cuts are placed on $p_T^Z$)
\cite{d0zcsec,d0ptz} and corrected for the $Z\rightarrow\tau\tau$ and 
$\tau\rightarrow e{\bar \nu_e} \nu_\tau$ branching fractions \cite{pdg}.
The $\gamma^* \rightarrow ee$ Monte Carlo sample is likewise normalized to 
D\O's measurement of the Drell-Yan ($\gamma^* \rightarrow ee$) cross section
in the dielectron mass range 
30 \gevcc $\leq M_{ee} \leq$ 60 \gevcc \cite{jmthesis}.
The $WW\rightarrow ee$ Monte Carlo samples are normalized to 
theory~\cite{wwwz}, and a 10\% uncertainty is
assigned~\cite{pbthesis}.

For the $Z\rightarrow\tau\tau\rightarrow ee$ background, 
the associated jet spectrum in \progname{pythia}, 
\progname{herwig}, and \progname{isajet} does not agree 
with that found in the $Z\rightarrow ee$ data. This is corrected by 
incorporating the jet cut survival probabilities from the 
$Z(\rightarrow ee)$+jet data 
(where the $H_T$ cut is taken as 70 GeV, as in the mismeasured
$\met$ calculation) rather than from Monte Carlo.

As described in the previous section, the $t\bar t$ 
acceptances are computed via Eq.~\ref{eq:llaccept}
using Monte Carlo events generated with \progname{herwig} and 
passed through the D\O\ detector simulation (see Sec.~\ref{mc}).
The trigger efficiency is obtained from $Z\rightarrow ee$ data but  
cross checked with the trigger simulator (see Sec.~\ref{mc}). 
Both approaches result in a trigger efficiency of $99 \pm 1$\%~\cite{bkthesis}.

The acceptance values after all cuts for seven top quark masses 
(for all channels) are given in Sec.~\ref{crsec}. 
The expected numbers of $t \bar t$ events, determined via 
Eq.~\ref{eq:numllevts}, are given in 
Table~\ref{tab:eeexp} for each of these seven masses. Finally, a cross 
section of $ 2.4 \pm 4.6 $ pb is obtained for the $ee$ channel.

To test the robustness of the background predictions,
comparison is made of data and expectations in regions dominated
by background (i.e., at earlier steps along the selection chain).
Making use of Eqs.~\ref{eq:numllevts} -- \ref{eq:numllbkevts} for the
different stages of the selection, Table~\ref{tab:eecuts} shows that the 
expectation from background and $t\bar t$ compare well with what is observed 
in the data at the various stages of the selection procedure.

\subsection{The e$\mu$ channel}
\label{emu}

\begin{table*}
\caption{Number of observed and expected $e\mu$ events passing at each 
cut level of the conventional analysis. Expected number of $\ttb$ events 
are for $m_t = 172.1 \gevcc$. Uncertainties correspond to 
statistical and systematic contributions added in quadrature.
\label{tab:emucuts}}
\vskip 0.5cm
\begin{tabular}{l|c|c|c|c|c}
\multicolumn{6}{c}{Number of $e\mu$ events passing cuts} \\
\multicolumn{1}{l|}{}&       & Total       & Mis-id & Physics &  \\
\multicolumn{1}{l|}{}& Data  & sig + bkg   & bkg  &  bkg    &  $\ttb$ \\
\hline 
$E_{T}^{e} > 15$ GeV, $p_{T}^{\mu} > 15$ GeV       &   &   &   &   &   \\

+ $e$ id + $\mu$ id + trig                  & 130 &  $93  \pm 7$     &   $50 \pm 2$      &  $39  \pm 6$      & $4.3 \pm 0.9$ \\

+ $\Delta R({\mu,{\rm jet}}) > 0.5 $     &  60 &  $59  \pm 6$     & $17.8 \pm 0.9$    &  $38  \pm 6$      & $3.4 \pm 0.7$ \\

+ \MEt $> 10$ GeV                       &  41 &  $38  \pm 3$     & $13.5 \pm 0.7$    & $21.4 \pm 3.3$    & $3.4 \pm 0.7$ \\

+ $\metcal > 20$ GeV                    &  22 & $21.8 \pm 2.2$   &  $4.5 \pm 0.4$    & $14.0 \pm 2.1$    & $3.2 \pm 0.6$ \\

+ $\Delta R({e,\mu}) > 0.25 $            &  20 & $19.5 \pm 2.2$   &  $2.3 \pm 0.3$    & $14.0 \pm 2.0$    & $3.2 \pm 0.6$ \\

+ 2 jets, $E_{T}^{\rm jet} > 15$ GeV    &   4 &  $3.4 \pm 0.6$   & $0.32 \pm 0.14$   & $0.34 \pm 0.09$   & $2.7 \pm 0.6$ \\

+ $H_{T}^e > 100$ GeV                     &   4 &  $2.8 \pm 0.5$   & $0.11 \pm 0.12$   & $0.24 \pm 0.08$   & $2.5 \pm 0.5$ \\

+ $H_{T}^e > 120$ GeV                     &   3 &  $2.6 \pm 0.5$   & $0.08 \pm 0.12$   & $0.20 \pm 0.08$   & $2.3 \pm 0.5$ \\

+ 2 jets, $E_{T}^{\rm jet} > 20$ GeV    &   3 &  $2.5 \pm 0.5$   & $0.08 \pm 0.12$   & $0.19 \pm 0.10$   & $2.2 \pm 0.5$ \\
\end{tabular}
\end{table*}
 
The signature for an event in the $e\mu$ channel consists of 
one high-$E_T$ isolated electron, one high-$p_T$ isolated muon,
two or more jets (from the $b$ quarks and initial and final state radiation),
and significant $\MEt$ (from the neutrinos). The trigger for this 
channel required one of the following level 2 terms to be satisfied:
\begin{itemize} 
\item \progname{ele-jet}(1a),\progname{ele-jet-high}(1b), or
      \progname{ele-jet-higha} (1c), which required an electron, 2 jets,
      and $\MEt$.
\item \progname{mu-ele}(1a and b) or \progname{mu-ele-high}(1c), which required an 
      electron and a muon.
\item \progname{mu-jet-high}(1a and b) or \progname{mu-jet-cent}(1c), which required
      a muon and a jet.
\end{itemize}
Details of these triggers are discussed in Sec.~\ref{triggers}.
Main-Ring events are not included in this analysis.
Over the complete Run 1 data set, these triggers provided a total integrated
luminosity of $112.6 \pm 4.8 \ {\rm pb}^{-1}$.

The backgrounds to this signature arise from Drell-Yan
production of $\tau\tau$ which
can lead to $e\mu$ final states ($Z \rightarrow \tau\tau \rightarrow e\mu$ 
and $\gamma^* \rightarrow \tau\tau \rightarrow e\mu$), 
$WW \rightarrow e\mu$, and multijet events containing an isolated muon 
and a misidentified electron. The latter background consists primarily of 
$W(\rightarrow \mu\nu)$+3 jet events, where one of the jets is misidentified 
as an electron. Backgrounds containing a real electron and a 
misidentified isolated muon, and those containing both a misidentified 
electron and a misidentified isolated muon 
were discussed in Ref.~\cite{prd1} and found to be negligible.

The offline selection cuts and their cumulative effect are summarized in 
Table~\ref{tab:emucuts}. After passing the trigger requirement, events are 
required to have $\geq 1$ electron (loose electron identification, see
Sec.~\ref{e-id}) with $E_T > 15$ GeV, $|\eta| \leq 2.5$
and $\geq 1$ muon (loose muon identification, see Sec.~\ref{muid}) with
$p_T > 15$ \gevcn. This initial selection has an acceptance 
($\varepsilon \times {\cal B}$) of $0.68 \pm 0.15$\% 
for $m_t = 170$ GeV/$c^2$. At this stage, the background is dominated by 
QCD multijet events containing a jet misidentified as an electron and a 
non-isolated muon from
the semi-leptonic decay of a $b$ or $c$ quark. This background is reduced
significantly by requiring the muon to be isolated, 
$\Delta R(\mu,{\rm jet}) > 0.5$. To further reduce the misidentification
background, the next two steps require $\MEt > 10$ GeV and 
$\metcal > 20$ GeV.
The cut on $\metcal$ is particularly effective against background
from $W(\rightarrow \mu\nu)$+jets events (where one of the jets is 
misidentified as an electron) due to the fact that $\metcal$ provides a 
measure of the transverse momentum of the $W$ boson since both of its decay 
products deposit little or no energy in the calorimeter. 
Studies also show that QCD multijet events 
that contain a highly electromagnetic jet (misidentified as an electron)
which gives rise to an isolated muon from the semi-leptonic decay 
of a $b$ or $c$ quark, can easily enter this analysis (as can 
$W(\ra\mu\nu)+$jets events where there is significant bremsstrahlung
from the muon as it passes through the EM calorimeter). Such events 
typically have the $e$ and $\mu$ very close in ($\eta,\phi$) space, 
and a requirement
of $\Delta R({e,\mu}) > 0.25$ effectively eliminates this class of 
misidentification background. 

After the above requirements, the background is primarily from 
$Z \ra \tau\tau \ra e\mu$ events and, to a lesser extent, from 
$WW \rightarrow e\mu$ events. The jets associated with these processes 
arise from 
initial state radiation (recoil) and are therefore 
softer in $E_T$ than the $b$ jets in a $t\bar t$ event. In addition, as 
noted above (see Sec.~\ref{ee}), the number of $Z+n$ jet events is 
proportional to $\alpha_s^n$, and a similar
steep falloff in jet multiplicity is observed for the 
Drell-Yan (and presumably $WW$) backgrounds. 
Requiring two jets with 
$E_T^{\rm jet} > 15$ GeV and $|\eta^{\rm jet}| \leq 2.5$ significantly 
reduces these backgrounds and that from QCD multijet production. 
The final cuts on $H_T^e > 120$ GeV and 
$N_{\rm jets} \geq 2$ for $E_T^{\rm jet} > 20$ GeV 
and $|\eta^{\rm jet}| \leq 2.0$ are obtained through the 
optimization procedure 
discussed in Sec.~\ref{dilep}
and provide further rejection against the remaining 
backgrounds. After the above selection, three $e\mu$ candidates remain 
in the data.

\begin{figure}
\vbox{
\centerline{\psfig{figure=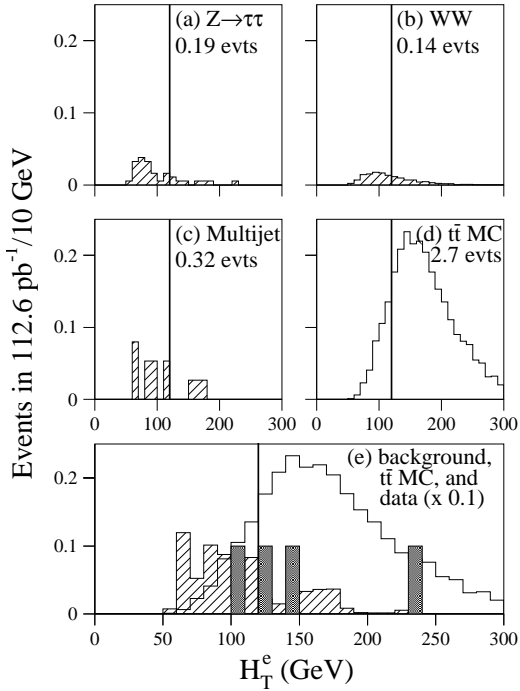,width=3.0in}}
\hskip 0.15cm
\caption{$H_T^e$ distributions for the $e\mu$ channel for expected background 
(hatched), expected signal (open), and data (solid) after all cuts except
$H_T^e > 120$ GeV (shown by solid vertical line) and 2 jets with 
$E_T^{\rm jet} > 20$ GeV (corresponding to line 6 of Table~\ref{tab:emucuts}).
Plots (a) -- (c) show the
individual contributions of the three leading backgrounds and give the 
expected number of events. Plot (d) gives the
expected $t\bar t$ contribution ($m_t = 170$ \gevccn), and plot (e) overlays
the total expected background, expected signal, and data ($\times 0.1$).
\label{fig:emu}}}
\end{figure}

Table~\ref{tab:emucuts} shows the number of data events, 
expected signal ($m_t = 172.1$ GeV/$c^2$), and expected background surviving 
at each stage of the selection. It is clear from this table that the $H_T^e$ 
cut is the most effective cut during the final stages of the analysis.
This is also shown in Fig.~\ref{fig:emu}, where the $H_T^e$
distributions are given for the three major backgrounds (a-c), for $t \bar t$
Monte Carlo (d), and for data superimposed on the total background and 
expected $t\bar t$ signal (e). 

As in the case of the $ee$ channel, the background from multijet events is 
obtained entirely from data.
The probability for an extra-loose electron to pass the loose electron 
identification criteria (see misidentification rate discussion 
in Sec.~\ref{e-id})
is applied to the full Run 1 sample of $e\mu$ events, where the electron 
candidate passes the extra-loose electron identification but fails the
loose electron identification, with all the other kinematic cuts 
applied. As shown in Table~\ref{tab:emuexp}, the QCD multijet 
(misidentified $e$) background is determined to be $0.08 \pm 0.12$ events.

Background estimates for 
$Z\rightarrow\tau\tau\rightarrow e\mu$, $WW\rightarrow e\mu$,
and $\gamma^* \rightarrow e\mu$ events are obtained 
via Eq.~\ref{eq:numllbkevts} 
using normalized \progname{pythia} and \progname{isajet} Monte Carlo samples.
The $Z\rightarrow\tau\tau \rightarrow e\mu$ Monte Carlo
samples are normalized to D\O's measurement of
$\sigma(p\bar{p}\to Z+X)B(Z\to ee)$ and the associated measurement of $p_T^Z$ 
\cite{d0zcsec,d0ptz}, and incorporate the $Z\rightarrow\tau\tau$,
$\tau\rightarrow e{\bar \nu_e} \nu_\tau$, and 
$\tau\rightarrow \mu{\bar \nu_{\mu}} \nu_\tau$ branching fractions \cite{pdg}.
The $\gamma^* \rightarrow \tau\tau$ Monte Carlo sample is likewise 
normalized to D\O's measurement of the Drell-Yan ($\gamma^* \rightarrow ee$) 
cross section
in the dielectron mass range 
30 \gevcc $\leq M_{ee} \leq$ 60 \gevcc \cite{jmthesis}
also incorporating the $\tau\rightarrow e{\bar \nu_e} \nu_\tau$ and 
$\tau\rightarrow \mu{\bar \nu_{\mu}} \nu_\tau$ 
branching fractions \cite{pdg}.
The $WW\rightarrow e\mu$ Monte Carlo samples are normalized to 
theory~\cite{wwwz}, and a 10\% uncertainty assigned \cite{pbthesis}. 

As for the $ee$ channel, the 
$Z\rightarrow\tau\tau\rightarrow e\mu$ Monte Carlo samples are not used
to model the jet and $H_T^e$ requirements.  
Instead, survival probabilities for 
these cuts are obtained from $Z(\rightarrow ee)$+jet data.

\begin{table}
\caption{Expected number of $e\mu$ signal and background events in 
112.6 ${\rm pb}^{-1}$ after all cuts in the conventional analysis. 
Uncertainties are statistical and systematic contributions added in
quadrature. The systematic uncertainty on the total background 
includes correlations among the different background sources.
\label{tab:emuexp}}
\vskip 0.5cm
\begin{tabular}{c|c}
\multicolumn{2}{c}{Expected number of $e\mu$ events in 112.6 ${\rm pb}^{-1}$ } \\
\hline 
\underline{$\ttb$ MC  $m_{t}$ (\gevccn)} &     \\
                   140       &  $4.07 \pm 0.88 $ \\
                   150       &  $3.32 \pm 0.72 $ \\
                   160       &  $2.77 \pm 0.60 $ \\
                   170       &  $2.29 \pm 0.49 $ \\
                   180       &  $1.84 \pm 0.40 $ \\
                   190       &  $1.48 \pm 0.32 $ \\
                   200       &  $1.12 \pm 0.24 $ \\
\hline
$Z\ra \tau\tau \ra e\mu$          & $0.10 \pm 0.09 $ \\
QCD multijet (mis-id $e$)         & $0.08 \pm 0.12 $ \\ 
$WW\ra e\mu$                      & $0.08 \pm 0.02 $ \\
${\rm DY}\ra \tau\tau \ra e\mu$   & $0.006 \pm 0.004 $ \\
\hline
Total background   & $0.26 \pm 0.16 $ \\
\end{tabular}
\end{table}

The $t\bar t$ acceptances are computed via Eq.~\ref{eq:llaccept}
using Monte Carlo events that are generated with \progname{herwig} and 
passed through the D\O\ detector simulation (see Sec.~\ref{mc}).
The trigger efficiency is obtained from the trigger simulator and 
is dependent on the detector region of the electron and muon, giving
($95 \pm 5)\%$ for CC($e$)CF($\mu$), ($93 \pm 5)\%$ for EC($e$)CF($\mu$),
($90 \pm 4)\%$ for CC($e$)EF($\mu$), and ($93 \pm 5)\%$ for EC($e$)EF($\mu$).
The acceptance values after all cuts for seven top quark masses 
(and for all channels) are given in Sec.~\ref{crsec}.
The expected number of $t \bar t$ events passing this selection is
determined via Eq.~\ref{eq:numllevts} and are given in 
Table~\ref{tab:emuexp} for these same seven masses. Finally, a 
cross section of $ 6.8 \pm 4.6 $ pb is obtained for the $e\mu$ channel.

\subsection{The $\mu \mu$ channel}
\label{mumu}

\def\mexc{\mbox{$\rlap{\kern0.25em/}E_x^{\rm cal}$}}
\def\meyc{\mbox{$\rlap{\kern0.25em/}E_y^{\rm cal}$}}

The signature for an event in the $\mu\mu$ channel consists of two isolated 
high-$p_T$ muons, 
two or more jets (from the $b$ quarks and initial and final state radiation),
and significant $\MEt$ (from the neutrinos). 
The trigger for this 
channel required one of the following level 2 terms to be satisfied:
\progname{mu-jet-high}(1a and 1b), \progname{mu-jet-cal}(1b), 
\progname{mu-jet-cent}(1b and 1c), or \progname{mu-jet-cencal}(1b and 1c). 
Each of these required a muon and one jet at level 2 
(see Sec.~\ref{triggers} for details).
Main-Ring events are not included in this analysis.
Over the complete Run 1 data set, these triggers provided a total integrated
luminosity of $108.5 \pm 4.7 \ {\rm pb}^{-1}$. 

The backgrounds to this signature arise from Drell-Yan production with
dimuon final states ($Z \rightarrow \mu\mu$,
$Z \rightarrow \tau\tau \rightarrow \mu\mu$, and 
$\gamma^* \rightarrow \mu\mu$), 
$WW \rightarrow \mu\mu$, and multijet events containing misidentified isolated 
muons. The latter background consists primarily of 
four-jet events where the semi-leptonic decay of $b$ and/or $c$ quarks 
result in two muons that pass the isolation requirement, and of
$W(\rightarrow \mu\nu)$+3 jet events where one of the jets gives rise
(through the semi-leptonic decay of a $b$ or $c$ quark) to a muon that
passes the isolation requirement.

\begin{table*}
\caption{Number of observed and expected $\mu\mu$ events passing at each 
cut level of the offline analysis. 
Shown are results for Run 1b+1c (CF-CF) only. 
Expected number of $\ttb$ events are for $m_t = 172.1 \gevcc$.
Uncertainties correspond to statistical and systematic contributions 
added in quadrature. 
\label{tab:mumucuts}}
\vskip 0.5cm
\begin{tabular}{l|c|c|c|c|c}
\multicolumn{6}{c}{number of $\mu\mu$ events passing cuts} \\
\multicolumn{1}{l|}{}&       & Total       & Mis-id & Physics &  \\
\multicolumn{1}{l|}{}& Data  & sig + bkg   & bkg  & bkg     & $\ttb$ \\
\hline 
$2\mu$, $p_{T}^{\mu} > 15$ GeV/$c$, + $\mu$ id &     &               &               &               &               \\
+ trig + 1 jet, $E_{T}^{\rm jet} > 20$ GeV & 606 &      --       
                        &      --       & $174 \pm 50 $ & $1.6 \pm 0.2$ \\

+ $\Delta \phi(\vec{\mu}_1,\vec{\mu}_2) < 165^{\circ} $ for 
  $|\eta_{\mu_1} + \eta_{\mu_2}| < 0.3$  
  & 207 &      --       &      --       & $146 \pm 42 $ 
  & $1.5 \pm 0.2$ \\

+ $M_{\mu\mu} > 10$ \gevcc (J$/\psi$ rej)   & 165  & $187 \pm 43 $  & $40  \pm 9  $   & $146 \pm 42 $      & $1.5 \pm 0.2$ \\

+ $\Delta R({\mu,{\rm jet}}) >0.5$  & 105 & $136 \pm 39$ & $0.70 \pm 0.33$ & $134 \pm 39 $ & $0.9 \pm 0.1$ \\

+ 2nd jet, $E_{T}^{\rm jet} > 20$ GeV   &  19 & $13.6 \pm 8.0$ & $0.22 \pm 
0.10 $ & $12.7 \pm 8.0$ & $0.72 \pm 0.09$ \\

+ $H_{T} > 100$ GeV  & 6  & $5.1 \pm 3.3$ & $0.03 \pm 0.02$ & $4.5 \pm 3.3$
& $0.53 \pm 0.07$ \\
+$Z$ fit prob$(\chi^{2})<1$ \% &  1   & $0.9 \pm 0.3$ & $0.03 \pm 0.02$ 
& $0.42 \pm 0.16$ & $0.48 \pm 0.06$ \\
\end{tabular}
\end{table*}

The offline selection cuts and their cumulative effect are summarized in 
Table~\ref{tab:mumucuts}. After passing the trigger requirement, events 
are required to have two muons (loose muon identification, see
Sec.~\ref{muid}) with $p_T > 15$ \gevc and $|\eta| \leq 1.0$ 
($|\eta| \leq 1.7$ in Run 1bc postclean) and one jet with 
$E_T^{\rm jet} > 20$ GeV and $|\eta| \leq 2.5$. 
This initial selection has an acceptance ($\varepsilon \times {\cal B}$) 
of 0.35\% ($m_t = 170$ GeV/$c^2$). At this stage,  
the dominant background is from cosmic rays. This
is minimized by rejecting tracks that are back-to-back in both
$\eta$ and $\phi$:
\begin{equation}
\Delta \phi (\vec{\mu}_1,\vec{\mu}_2)<165^{\circ} 
{\rm \ for \ }  | \eta(\vec{\mu}_1)+\eta(\vec{\mu}_2)|<0.3.
\label{eq:mumucos}
\end{equation}
It is necessary to exclude background from 
J/$\psi \rightarrow \mu\mu$. As discussed below, the muon momentum
resolution prohibits an efficient cut on $M_{\mu\mu}$ at the $Z$ boson 
mass peak.
However, at lower muon $p_T$, it is an effective quantity and is used to 
reject low-mass pairs resulting from high-\pt\ J/$\psi$ 
production with recoil jets: $M_{\mu \mu} > 10 \ \gevcc$ is required.
At this stage, the background is dominated by QCD multijet events rich in 
heavy flavor with muons originating from semi-leptonic decays of  
$b$ or $c$ quarks. By requiring both muons to be isolated 
($\Delta R(\mu,{\rm jet}) > 0.5$), this background is reduced to a 
negligible level. The remaining background is mainly from events
containing isolated dimuons from $Z/\gamma^*$ and $WW$ production.
The jets associated with these processes arise from recoil and are thus 
softer in $E_T$ than the $b$ jets in a $t\bar t$ event. Also, as 
noted in Sec.~\ref{ee}, the number of $Z+\geq n$ jet events is 
proportional to $\alpha_s^n$, and a similar
steep falloff in jet multiplicity is observed for the Drell-Yan and $WW$
backgrounds. The next step in the analysis therefore requires a 
second jet with 
$E_T > 20$ GeV and $|\eta| \leq 2.5$, reducing 
the dimuon background from these sources.
The requirement of $H_T > 100$ GeV is obtained through the optimization 
procedure, as discussed in Sec.~\ref{dilep},
and provides further rejection against the remaining 
background, leaving only the contribution from $Z \rightarrow \mu\mu$ at 
a non-negligible level.

As noted above, because of limitations on the momentum resolution 
of the D\O\ muon system, the invariant mass peak of the $Z$ boson is smeared 
and a simple cut on $M_{\mu\mu}$ is ineffective in reducing this background. 
Instead, rejection is achieved using the result of a $\chi^2$ minimization
procedure that involves a refitting of the muon
momenta with a constraint that the transverse momentum of the dimuon system
balance the remaining transverse energy in the event:
\begin{eqnarray}
\chi^2 = & \frac{\bigl(\frac{1}{p_{\mu 1}} - \frac{1}{p^0_{\mu 1}}\bigr)^2}
         {\sigma^2\bigl(\frac{1}{p_{\mu 1}}\bigr)} +
         \frac{\bigl(\frac{1}{p_{\mu 2}} - \frac{1}{p^0_{\mu 2}}\bigr)^2}
         {\sigma^2\bigl(\frac{1}{p_{\mu 2}}\bigr)} \nonumber \\
         & + \frac{\bigl(\mexc - \bigl(p^0_{\mu 1}\bigr)_x - \bigl(p^0_{\mu 2}\bigr)_x\bigr)^2}
         {\sigma^2\bigl(\mexc\bigr)} 
      \label{eq:zfit}  \\
         & + \frac{\bigl(\meyc - \bigl(p^0_{\mu 1}\bigr)_y - \bigl(p^0_{\mu 2}\bigr)_y\bigr)^2}
         {\sigma^2\bigl(\meyc\bigr)},  \nonumber
\end{eqnarray}
with the constraint that $M_{\mu\mu} = M_{Z}$: 
\begin{equation}
M_Z^2 = 2(p^0_{\mu1}p^0_{\mu2} - {\vec{p}}^0_{\mu1}\cdot{\vec{p}}^0_{\mu2})
\label{eq:zfitcon}
\end{equation}
where $p_{\mu i}$ is the measured momentum for the $i$-th muon, $p_{\mu i}^0$
is the fitted value of $p_{\mu i}$, $\sigma(\frac{1}{p_{\mu i}})$ is the
measured muon momentum resolution (see Eq.~\ref{eq:mumomres}), 
$\mexc$ and $\meyc$ are the $x$ and $y$ components of $\metc$, and 
$\sigma(\mexc)$ and $\sigma(\meyc)$
are their measured resolutions (see Eq.~\ref{eq:metres}).
This $\chi^2$ is minimized as a function of $p^0_{\mu1}$ and $p^0_{\mu2}$.
An event is considered to be a $Z\rightarrow \mu\mu$ candidate, and is thus
rejected, if ${\rm Prob}(\chi^2) > 0.01$. This procedure is also used to 
remove $Z\ra\mu\mu$ background from the 
$t\bar t \ra \mu + {\rm jets} + \mu$ tag channel (see Sec.~\ref{ljets_mutag}).

\begin{figure}[h]
\vbox{
\vskip -0.8cm
\centerline{\psfig{figure=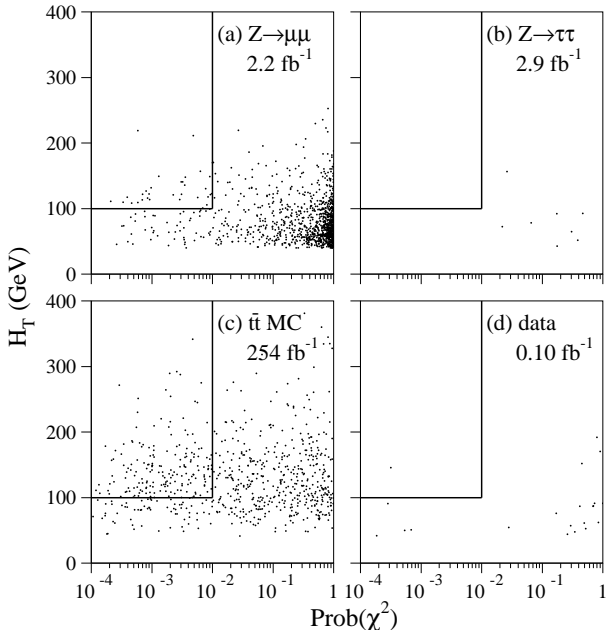,width=3.5in}}
\hskip 0.1cm
\caption{Scatter plots of $H_T$ vs Prob($\chi^2$) for the $\mu\mu$ channel:
(a) $Z \ra \mu\mu$ background, (b) $Z \ra \tau\tau \ra \mu\mu$ background,
(c) $t{\bar t} \ra \mu\mu$ signal, and (d) data. The signal region is shown
in the upper left corner of each plot (Prob($\chi^2$) $<$ 1\%, 
$H_T > 100$ GeV).
\label{fig:mumu}}}
\end{figure}

Table~\ref{tab:mumucuts} shows the number of observed events, expected signal 
(for $m_t = 172.1$ GeV/$c^2$), and expected background surviving at each stage 
of the selection. It is clear from this table that the $H_T$ and 
${\rm Prob}(\chi^2)$ cuts provide significant background rejection
in the final stages of the analysis. 
This is shown in Fig.~\ref{fig:mumu}, where $H_T$
vs ${\rm Prob}(\chi^2)$ is plotted for $Z\rightarrow \mu\mu$ and
$Z\rightarrow \tau\tau \rightarrow \mu\mu$ MC events (a,b), for $t \bar t$
MC events (c), and for data (d). 

One ${t\bar t}\rightarrow \mu\mu$ candidate survives the above selection.
Both muons in the event are central, and each 
track has the maximum of ten hits in the muon chambers, the case where the 
momentum resolution is best modeled and understood. An interesting 
feature of this event is that all the muons and jets are in one hemisphere 
in $\phi$ in the detector, leaving only $\MEt$ in the other half; 
this topology is highly unlikely to come from the main background of 
$Z\rightarrow \mu\mu$ production.  

The background from multijet events is determined entirely from data.
The probability for a jet to give rise to an isolated muon is determined
separately for the CF and EF regions of the muon system using a sample of 
of multijet events. These probabilities are then applied to the jets 
in a sample of muon (loose identification, see Sec.~\ref{muid}) + jet
events to obtain the background expected from 
$W(\rightarrow \mu\nu)$+jets, QCD multijet production, and
$Z\rightarrow \tau\tau \rightarrow \mu$ + hadrons where the second muon
originates from the semi-leptonic decay of a $b$ or $c$ quark 
from initial or final state radiation.

In a manner analogous to the background calculations used for the 
$ee$ and $e\mu$ channels, backgrounds from $Z\rightarrow\mu\mu$,
$Z\rightarrow\tau\tau\rightarrow \mu\mu$, $WW\rightarrow \mu\mu$,
and $\gamma^* \rightarrow \mu\mu$ are obtained 
via Eq.~\ref{eq:numllbkevts} 
from \progname{pythia} and \progname{isajet} 
Monte Carlo samples which are normalized to experimental or theoretical 
values. In particular, the $Z\rightarrow\tau\tau \rightarrow \mu\mu$ MC
samples are normalized to the D\O\ $Z$ boson cross section measurement
but incorporate $Z\rightarrow\tau\tau$ and 
$\tau\rightarrow \mu{\bar \nu_{\mu}} \nu_\tau$ 
branching fractions from elsewhere~\cite{pdg}.
Similarly, the $\gamma^* \rightarrow \mu\mu$ 
Monte Carlo sample is 
normalized to D\O's measurement of the Drell-Yan ($\gamma^* \rightarrow ee$) 
cross section in the dielectron mass 
range 30 \gevcc $\leq M_{ee} \leq$ 60 \gevcc \cite{jmthesis}.
The $WW\rightarrow \mu\mu$ Monte Carlo sample
is normalized to theory~\cite{wwwz} and a 10\% 
uncertainty assigned~\cite{pbthesis}.

As for the $ee$ and $e\mu$ channels, the 
$Z\rightarrow\tau\tau\rightarrow \mu\mu$ Monte Carlo samples are not used
to model the jet and $H_T$ requirments.  Instead, survival probabilities for
these cuts are obtained from $Z(\rightarrow ee)$+jet data.

As described in Sec.~\ref{dilep}, the $t\bar t$ 
acceptances are computed via Eq.~\ref{eq:llaccept}
using Monte Carlo events that are generated with \progname{herwig} and 
passed through the D\O\ detector simulation (Sec.~\ref{mc}).
The trigger efficiency is computed using data-derived trigger
turn-on curves applied to $t\bar t$ Monte Carlo and is determined 
to be ($95\pm5$)\%. The acceptance values after all cuts for seven 
top quark masses (and for all channels) are given in Sec.~\ref{crsec}.
The expected numbers of $t \bar t$ events passing 
this selection are determined via Eq.~\ref{eq:numllevts} and are given in 
Table~\ref{tab:mumuexp} for these same seven masses. 
Finally, a cross section of $ 2.1 \pm 8.8 $ pb is obtained 
for the $\mu\mu$ channel.

To test the robustness of the background predictions,
comparisons are made between the data and expectations in regions dominated
by background (i.e., at earlier steps along the selection chain).
Equations.~\ref{eq:numllevts} -- \ref{eq:numllbkevts} give, for the different 
stages of the selection, the results in Table~\ref{tab:mumucuts}, which show 
that that the expectation from background and $t\bar t$ compare well with 
what is observed in the data at the various stages of the selection procedure.


\begin{table}
\caption{Expected number of $\mu\mu$ signal and background events after all 
cuts in 108.5 ${\rm pb}^{-1}$. Uncertainties are statistical and systematic 
contributions added in quadrature. The systematic uncertainty on the total 
background includes correlations among the different background sources.
\label{tab:mumuexp}}
\vskip 0.5cm
\begin{tabular}{c|c}
\multicolumn{2}{c}{Expected number of $\mu\mu$ events in 108.5 ${\rm pb}^{-1}$ } \\
\hline 
\underline{$\ttb$ MC  $m_{t}$ (\gevccn)} &     \\
 140 &  $ 1.02 \pm 0.15 $ \\
 150 &  $ 0.88 \pm 0.13 $ \\
 160 &  $ 0.78 \pm 0.11 $ \\
 170 &  $ 0.67 \pm 0.09 $ \\
 180 &  $ 0.54 \pm 0.08 $ \\
 190 &  $ 0.44 \pm 0.06 $ \\
 200 &  $ 0.33 \pm 0.05 $ \\
\hline
$Z\ra \tau\tau \ra \mu\mu$  & $ 0.03 \pm 0.03 $ \\
QCD multijet (mis-id $\mu$) & $ 0.07 \pm 0.01 $ \\ 
$Z\ra \mu\mu$               & $ 0.58 \pm 0.22 $ \\
$WW\ra \mu\mu$              & $ 0.007 \pm 0.004 $ \\
${\rm DY}\ra \mu\mu$        & $ 0.07 \pm 0.04 $ \\
\hline
Total background   & $ 0.75 \pm 0.24 $ \\
\end{tabular}
\end{table}

\subsection{The $e \nu$ channel}
\label{enu}

The $e\nu$ channel is based on the assumption that
one of the $W$ bosons decays to $e\nu$ and that the remaining $t\bar t$
decay products conspire to give rise to significant $\metcal$
($> 50$ GeV).
As can be inferred from Figs.~\ref{fig:int_dil_lep} and 
\ref{fig:int_lj_lep}, this is most probable 
for $ee$ and $e\mu$ events but will also occur in some fraction of the $e$+jets
events. To eliminate overlap with the dilepton channels, it is further
assumed that for $e\mu$($ee$) events, the muon (second electron) is either 
too low in $p_T$($E_T$) to pass the selection or escapes detection.
The signature for an event in the $e\nu$ channel is therefore one, and only
one, high-$E_T$ electron, 
two or more jets (from the $b$ quarks and initial and final-state radiation),
and very large $\MEt$ (from the neutrinos and possibly a lost lepton). 
The virtue of this channel is that it can recover some of the
$\ttb$ cross section not seen by the other channels. Indeed, investigating 
\progname{herwig} $t\bar t$ Monte Carlo events (at $m_t = 170$ \gevccn),
the final $e\nu$ sample is found to consist of one-half dilepton
($ee$ and $e\mu$) events, one-third $e$+jets events, and one-sixth 
$e$ + hadronic-tau events.

The trigger for the $e\nu$ channel required one of the following 
level 2 terms to be satisfied (see Sec.~\ref{triggers}):
\begin{itemize}
\item \progname{ele-jet}(1a), \progname{ele-jet-high}(1b), 
      \progname{em1-eistrkcc-ms}(1b) or \progname{ele-jet-higha}(1c), all of
      which required an electron, 2 jets, and $\MEt$.
\item \progname{missing-et}(1ab) or \progname{missing-et-high}(1c), both of 
      which required only very large $\metcal$.
\end{itemize}
Note that Main-Ring events were not included in this analysis.
Over the complete Run 1 data set, these triggers provided a total integrated
luminosity of $112.3 \pm 4.8 \ {\rm pb}^{-1}$. 

\begin{table*}
\caption{Number of observed and expected $e\nu$ events passing at each 
cut level of the offline analysis. Expected number of $\ttb$ events 
are for $m_t = 172.1 \gevcc$. Uncertainties correspond to statistical 
and systematic contributions added in quadrature.
\label{tab:enucuts}}
\vskip 0.5cm
\begin{tabular}{l|c|c|c|c|c}
\multicolumn{6}{c}{Number of $e\nu$ events passing cuts} \\
\multicolumn{1}{l|}{}&       & Total       & Mis-id & Physics &  \\
\multicolumn{1}{l|}{}& Data  & sig + bkg   & bkg  & bkg     & $\ttb$ \\
\hline 
$1e$, $E_{T}^{e} > 20$ GeV, + min $e$ id + trig & 
119,263 &      --         &          --           &       --       & $71.5 \pm 20.2$ \\
+ $\metcal > 50$ GeV         & 
   3941 &      --         & $434 \pm 74$ &       --       & $36.0 \pm 10.2$ \\
+ 1 jet, $E_{T}^{\rm jet} > 30$ GeV  & 
   1422 &      --         & $357 \pm 61$ &       --        & $35.5 \pm 10.1$ \\
+ $2^{\rm nd}$ jet, $E_{T}^{\rm jet} > 30$ GeV  & 
    192 & $244.4 \pm 39.0$ & $92.9 \pm 16.0$ & $121.2 \pm 35.6$ & $30.3 \pm 8.6$ \\
+ $M_T^W > 115$ \gevcc  & 
     25 & $29.3 \pm 4.8$ & $24.4 \pm 4.7$ & $1.0 \pm 0.4$ & $3.9 \pm 1.1$ \\
+ $\Delta\phi(\vec{\MEt},2^{\rm nd} E_T {\rm \ object}) \geq 0.5 $ & 
     12 & $18.1 \pm 3.0$ & $13.7 \pm 2.9$ & $0.9 \pm 0.4$ & $3.6 \pm 1.0$ \\
+ loose $e$ id  & 
      5 & $4.1 \pm 0.8$  & $0.69 \pm 0.12$ & $0.75 \pm 0.35$ & $2.7 \pm 0.8$ \\
+ orthogonality to other channels & 
      4 & $2.9 \pm 0.7$  & $0.47 \pm 0.15$ & $0.72 \pm 0.34$ & $1.7 \pm 0.5$ \\
\end{tabular}
\end{table*}

The primary backgrounds to this signature arise from 
$W(\rightarrow e\nu)$+2 jet events and QCD production of
three-jet events where one jet is misidentified as an 
electron and the $\MEt$ is an
artifact of jet $E_T$ mismeasurement. An additional source of background
is $WW$ + $n$ jets production where one of the $W$ bosons decays 
to $e\nu$ and, in the case of $n=0$ or 1, the other $W$ decays hadronically. 
Similarly, backgrounds from $WZ$ + $n$ jets also contribute, but to a lesser
extent.

The offline selection cuts and their cumulative effects are summarized in 
Table~\ref{tab:enucuts}. After passing the trigger requirement, events are 
required to have one electron (minimal electron identification, see 
Sec.~\ref{e-id}) with $E_T > 20$ GeV and $|\eta| \leq 1.2$. 
This channel differs from the other $t\bar t$ channels both in 
choosing its initial 
electron identification to be minimal (loose electron identification
is required at a later stage) and in the restriction of electrons to the
CC region of the calorimeter (to suppress QCD multijet background, which 
increases in the forward region). This initial selection has an acceptance 
($\varepsilon \times {\cal B}$) of ($11.1 \pm 3.2$)\% 
(for $m_t = 170$ GeV/$c^2$).
The next step requires $\MEt >50$ GeV to select high-$\MEt$ 
$t\bar t$ events, reject QCD multijet background, and decrease the number of 
$W(\rightarrow e\nu)$ and $WW$ events. To further decrease these backgrounds,
two jets with $E_T > 30$ GeV and $|\eta| \leq 2.0$ are required.
At this stage the background is dominated by $W(\rightarrow e\nu)$+2 jet events
and a cut on the $e,\MEt$ transverse mass, $M_T^W > 115$ GeV, brings it
down to approximately one event. The transverse mass is defined by
\begin{equation}
M_T^W(e,\MEt) = 
\sqrt{(|\vec{E_T^e}| + |\vec{\MEt}|)^2 - (\vec{E_T^e} + \vec{\MEt})^2}.
\end{equation}
This cut is also effective against QCD multijet background, 
being similar to the 
$E_T^L$ (= $E_T^e + \MEt$) cut which will be described in 
Sec.~\ref{ljets_topo}, and tends to reject events where the electron is 
parallel to the $\MEt$ in $\phi$. The background that remains is dominated 
by 3-jet events, where one of the jets is misidentified as 
an electron and the $\MEt$ is 
an artifact of jet $E_T$ mismeasurement. A topological cut,
$\Delta\phi(\rlap{\kern0.25em/}E_T$, {$2^{\rm nd}$} $E_T$
object$)>0.5$~rad, rejects two-jet-like events where the $\MEt$ is
aligned with one of the jets due to an upward fluctuation of the highest
$E_T$ jet or a downward fluctuation of the second-highest $E_T$ jet. 
Note that the electron is treated as a jet in this $E_T$ ordering.

The next step requires that the loose electron identification criteria 
be applied to all electron candidates and brings the remaining QCD 
multijet background
down to an acceptable level. The final step in the selection requires, for the
purpose of obtaining a combined cross section, that this channel be
orthogonal with the other top channels with which it overlaps:
$ee$, $e\mu$, and $e$+jets. This is accomplished by vetoing any event that
passes the selection requirements of any one of these channels.
As shown in Table~\ref{tab:enucuts}, four events pass all $e\nu$ selection
requirements. One of the events has four jets with $E_T > 15$ GeV,
as would be expected for a $\ell+$jets event, and the
remaining three events have only two jets, which is more characteristic of 
dilepton events.

The background from $W$+jets is modeled with \progname{vecbos} Monte
Carlo distributions that are scaled to match the jet $E_T$, $\MEt$, and 
$M_T^W$ spectra found in data. 
The Monte Carlo sample is 
normalized to the number of $W(\rightarrow e\nu)+$2 jet events found in
data and Eq.~\ref{eq:numllbkevts} is used to compute the expected
background of $0.5 \pm 0.3$ events, as shown in Table~\ref{tab:enuexp}.

The QCD multijet background estimate is obtained from data
and is defined as the mean of the results from
two independent methods. In the first method, the probability for a jet
to be misidentified as a loose electron is determined from a sample
of multijet data to be ($0.0091 \pm 0.0012$)\% in the CC region of the
calorimeter. This probability is then applied to the number of jets
with $E_T > 20$ GeV in a sample
of three or more jet events where all requirements except
that of electron identification have been applied. This method results
in an estimate of the QCD multijet background of 
$0.576 \pm 0.077 {\rm \ (stat)} \pm 0.076 {\rm \ (sys)}$ events.
In the second method, the standard rate for an 
extra-loose candidate to be misidentified
as a loose candidate (see Table~\ref{tab:eid}) is applied to a sample
of electron + jet events (extra-loose electron identification) to which
all other other kinematic cuts have been applied. This method results
in an estimate of the QCD multijet background of 
$0.367 \pm 0.129 {\rm \ (stat)} \pm 0.005 {\rm \ (sys)}$ events.
The mean of these two approaches yields an expected QCD multijet background 
of $0.47 \pm 0.15$ events, as shown in Table~\ref{tab:enuexp}.

The backgrounds from $WW$ and $WZ$ events are obtained via 
Eq.~\ref{eq:numllbkevts} from \progname{pythia} Monte Carlo 
normalized to the theoretical cross section~\cite{wwwz}, and
are given in Table~\ref{tab:enuexp}.

\begin{figure}
\vbox{
\centerline{\psfig{figure=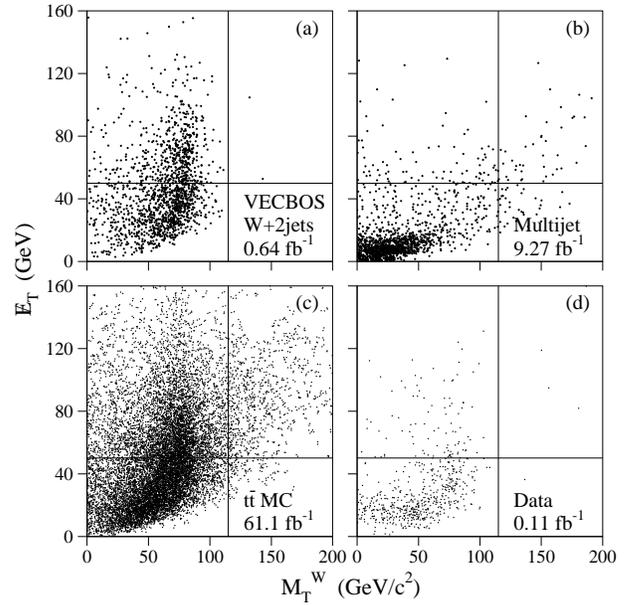,width=3.2in}}
\caption{Scatter plots of $\MEt$ vs $M_T^W$ for the $e\nu$ channel:
(a) $W$+jets background, 
(b) background from multijet events with a misidentified electron,
(c) $t\bar t$ signal ($m_t = 170$ \gevccn), 
and (d) data. The signal region is shown
in the upper right corner of each plot ($M_T^W \geq 115$ \gevccn, 
$\MEt \ge 50$ GeV).
\label{fig:enu}}}
\end{figure}

As shown in Table~\ref{tab:enucuts}, the cuts on $\metcal$ and 
$M_T^W$ are most effective in reducing the background. This is shown 
in Fig.~\ref{fig:enu}, where $\metcal$ vs $M_T^W$ is plotted
for the $W$+jets and QCD multijet 
backgrounds (a,b), for $t \bar t$ Monte Carlo events (c), 
and for data (d). It can be seen that the four candidate 
events are well inside the signal region and far from the cut boundaries.

\begin{table}[h]
\caption{Expected number of $e\nu$ signal and background events after 
all cuts in 112.3 ${\rm pb}^{-1}$. Uncertainties are statistical and 
systematic contributions added in quadrature. The systematic uncertainty
on the total background includes correlations among the background sources.
\label{tab:enuexp}}
\vskip 0.5cm
\begin{tabular}{c|c}
\multicolumn{2}{c}{Expected number of $e\nu$ events in 112.3 ${\rm pb}^{-1}$ } \\
\hline 
\underline{$\ttb$ MC  $m_{t}$ (GeV/$c^2$)} &     \\
 140     &  $ 2.96 \pm 0.88 $ \\
 150     &  $ 2.64 \pm 0.77 $ \\
 160     &  $ 2.06 \pm 0.60 $ \\
 170     &  $ 1.72 \pm 0.50 $ \\
 180     &  $ 1.49 \pm 0.43 $ \\
 190     &  $ 1.15 \pm 0.33 $ \\
 200     &  $ 0.91 \pm 0.27 $ \\
\hline
$WW$                 & $0.16 \pm 0.05 $ \\
$WZ$                 & $0.017 \pm 0.005 $ \\
$W$+jets             & $0.54 \pm 0.32 $ \\
QCD multijet         & $0.47 \pm 0.15 $ \\
\hline
Total background    & $1.19 \pm 0.38 $ \\
\end{tabular}
\end{table}

As described in Sec.~\ref{dilep}, $t\bar t$ 
acceptances are computed via Eq.~\ref{eq:llaccept}
using Monte Carlo events generated with \progname{herwig} and 
passed through the D\O\ detector simulation (see Sec.~\ref{mc}).
The trigger efficiency is obtained from the Trigger Simulator 
(see Sec.~\ref{mc}) and found to be $ 99.4^{+0.6}_{-3.1}$\%.
The final acceptances for seven 
top quark masses (and for all channels) are given in Sec.~\ref{crsec}.
The expected numbers of $t \bar t$ events 
passing this selection are
determined via Eq.~\ref{eq:numllevts} and are given in 
Table~\ref{tab:enuexp} for these same seven masses. Finally, a
cross section of $ 9.1 \pm 7.2 $ pb is obtained for the $e\nu$ channel.

To test the robustness of the background predictions,
a comparison is made between the data and expectations in regions dominated
by background (i.e., at earlier steps along the selection chain).
Making use of Eqs.~\ref{eq:numllevts} -- \ref{eq:numllbkevts} for the
different stages of the selection, Table~\ref{tab:enucuts} shows that the 
expectation from background and $t\bar t$ compare well with what is observed 
in the data at the various stages of the selection procedure.

\section{Analysis of lepton + jets events}
\label{ljets_intro}

As discussed in Sec.~\ref{sec:intro}, the lepton+jets signatures are 
characterized by one isolated, high-$p_T$ charged lepton, $\MEt$, and four 
or more jets. This signature is similar to that of $W$+jets production. 
Figures~\ref{fig:int_lj_lep} and \ref{fig:int_lj_jet}
include Monte Carlo distributions for the
lepton and jet $E_T/p_T$ and $|\eta|$, and $\met$ expected in $t\bar t$ 
lepton+jets events. 
As shown in Table~\ref{tab:ljini}, requirements based on these 
characteristics form the initial selection for all four channels.

\begin{table}[htp]
\caption{Initial selection for $\ell$+jets analyses. 
The $|\eta(W)|$ cut is introduced and described in Sec.~\ref{ljets_topo}.
\label{tab:ljini}}
\vskip 0.5cm
\begin{tabular}{l|cc|cc}
     & \multicolumn{2}{c|}{Topological} & \multicolumn{2}{c}{Muon tag} \\
 Selection cut  & $e$+jets & $\mu$+jets & $e$+jets & $\mu$+jets \\
\hline 
1 isol $e$, $E_T^e \geq 20$ GeV,   &  &  &  & \\
$|\eta^e| \leq 2.0$ + tight $e$ id    &   yes    &   no    &  yes   &  no  \\ 
1 isol $\mu$, $p_T^\mu \geq 20$ GeV/$c$,  &  &  &  & \\
$|\eta^{\mu}| \leq 1.7(1.0) $  + tight $\mu$ id  &     no     &    yes     &    no     &  yes      \\ 
 $\mu$ tag          & \multicolumn{2}{c|}{veto}& \multicolumn{2}{c}{yes} \\ 
\metc (GeV)              & $\geq 25$  & $\geq 20$  &     --     & $\geq 20$ \\
\met (GeV)               &    --       & $\geq 20$  & \multicolumn{2}{c}
{$\geq 20$} \\
$|\eta(W)|$      & \multicolumn{2}{c|}{$\leq 2.0$} & \multicolumn{2}{c}{--} \\
$N_{\rm jets}$ & \multicolumn{2}{c|}{$\geq 4$} & 
\multicolumn{2}{c}{$\geq 3$} \\
$E_T^{\rm jet}$ (GeV)  & \multicolumn{2}{c|}{$\geq 15$} & 
\multicolumn{2}{c}{$\geq 20$} \\
$|\eta({\rm jet})|$            & \multicolumn{2}{c|}{$\leq 2.0$} & 
\multicolumn{2}{c}{$\leq 2.0$} \\
\end{tabular}
\end{table}

The triggers used to select the candidate events require at least one
high-$p_T$ lepton and some combination of $\MEt$ and jet requirements
(see Sec.~\ref{triggers} for details). The run ranges and luminosities 
for the four channels are given in Table~\ref{tab:ljrrl}.

\begin{table}[htp]
\caption{$\ell$+jets run ranges and luminosities. Channel names are as 
defined in the text.
\label{tab:ljrrl}}
\vskip 0.5cm
\begin{tabular}{l|c|c|c|c}
    & $e$+jets/topo & $\mu$+jets/topo & $e$+jets/$\mu$ & $\mu$+jets/$\mu$ \\
\hline 
Run range                      & 1a,1b  & 1a,1b   & 1a,1b,1c  & 1a,1b \\ 
Lum.  (${\rm pb}^{-1}$)   & 119.5  & 107.7   & 112.6     & 108.0 \\ 
\end{tabular}
\end{table}

The primary background sources are $W$+multijet production 
and QCD multijet events with a misidentified isolated lepton and 
mismeasured $\MEt$. 
As indicated in Table~\ref{tab:ljini}, the initial 
selection requires a high-$p_T$ tight lepton 
(which dramatically reduces the QCD multijet background), large $\MEt$, and
several jets.

\begin{figure}
\vbox{
\vskip -0.7cm
\centerline{\psfig{figure=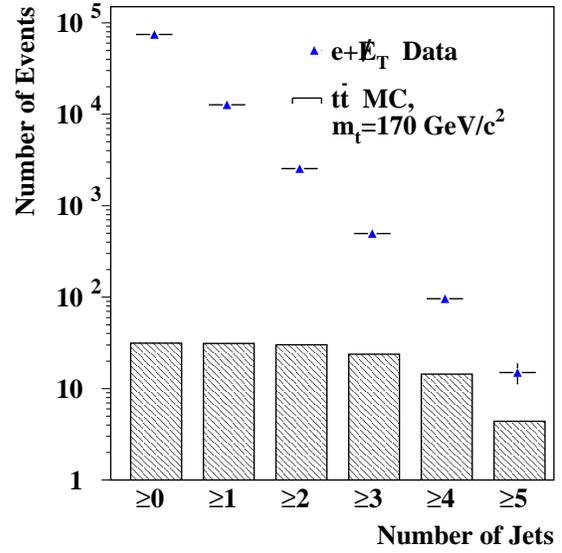,width=3.2in}}
\caption{Jet multiplicity distribution for $e+\MEt$+jets data (triangle 
points) and $\ttb$ Monte Carlo (hatched histogram) after initial selection. 
Trigger inefficiency is not included in the Monte Carlo samples.
\label{fig:jetmultini}}}
\end{figure}

Figure~\ref{fig:jetmultini} shows the number of events as a function of
the number of jets in the event for $e+$jets inclusive data and for 
$t\bar t$ MC events after the initial selection. As can be seen, the 
signal to background ratio 
is still very low. It is, therefore, necessary to further exploit 
the differences between signal and background. The most obvious differences 
are in the event topology and the presence or absence of a $b$ quark jet. 
The $b$ quark is inferred in the D\O\ detector by the presence of a 
non-isolated muon ({\sl muon tag}).
Therefore, two orthogonal analyses are employed beyond this point:
(1) a purely {\sl topological} analysis, which by construction does not 
contain a muon tag, and (2) an analysis that relies primarily on the 
presence of a {\sl muon tag}, but also makes use of some topological cuts.
These channels are denoted respectively as $\ell$+jets/topo and 
$\ell$+jets/$\mu$. The initial selection for these channels is given in 
Table~\ref{tab:ljini}.

\begin{figure}
\vbox{
\centerline{\hskip -0.3cm \psfig{figure=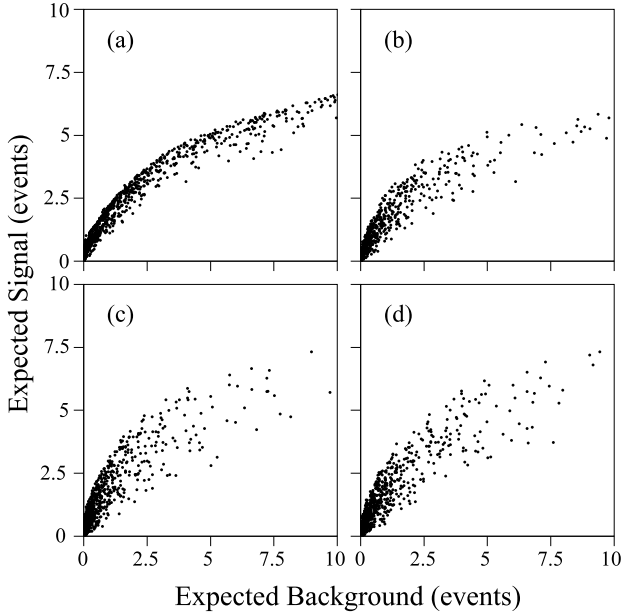,width=3.3in}}
\vskip -1.2cm
\caption{Results of the random grid search in terms of expected signal 
vs expected background for the $e$+jets 
topological analysis for four possible variable sets: 
(a) ${\cal A}(W+{\rm jets})$ and $h$, (b) ${\cal A}(W+{\rm jets})$, $h$,
and $\met$, (c) ${\cal A}(W+{\rm jets})$, $H_T$, and $\met$, (d)
${\cal A}(W+{\rm jets})$, $H_T$, and $p_T(W)$. See text for definitions
of these variables.
\label{fig:ahtopt}}}
\end{figure}

In order to obtain the most precise 
measurement of the $\ttb$ production cross section possible, an optimization
was performed to find those topological variables that provide the best
separation between signal and background. 
This was accomplished through the use of a 
{\sl random grid search}~\cite{randgrid}
in which many possible cut points were tested on the 
signal and background models. 
Many variables were investigated in this way: 
$p_T(W) \equiv |\vec{p}_T^W|$, \met, $N_{\rm jets}$, $h \equiv 
\frac{E_T({\rm lepton})+E_T\!\!\!\!\!\!\!/}{H_T({\rm jets})+p_T(W)}$,
two types of aplanarity ($\cal A$), and two types of $H_T$.
Aplanarity is essentially a measure of the ``flatness'' of
an event and is defined to be ${3 \over 2}$ of 
the smallest eigenvalue of
the normalized laboratory momentum tensor $({\cal M})$, where this tensor
is defined by~\cite{momentens} 
\begin{equation}
 {\cal M}_{ij} = \Bigl( \sum_o p_{o,i}p_{o,j}\Bigr) 
/ \Bigl( \sum_{o} |\vec{p}_{o}|^2 \Bigr) ,
\end{equation}
where $\vec{p}_o$ is the three momentum of object $o$, $i,j$ correspond to
the $x, y,$ and $z$ coordinates, and the objects 
included in the sum depend on the type of aplanarity under consideration: 
(1) only the jets, ${\cal A}({\rm jets})$, and (2) the jets and the 
reconstructed leptonic $W$, ${\cal A}(W+{\rm jets})$. 
Large values of ${\cal A}$ are indicative of spherical events, whereas 
small values correspond to more planar events. 
Events due to $\ttb$ production are
quite symmetric as is typical for the decay of a heavy object. $W$+jet
and QCD multijet events are more planar, owing primarily to the 
fact that the jets in these events arise from gluon radiation.

Analogous to the transverse-energy variable defined for the $ee$ and $e\mu$
channels, and identical in form to that used for the $\mu\mu$ channel 
(see Eq.~\ref{eq:htmumudef}), $H_T$ 
is defined for the lepton+jets channels as 
\begin{equation}
H_T \equiv  \sum_{\rm jets} E_T . 
\label{eq:htdef}
\end{equation}
The sum is over all jets 
with $E_T \geq 15$ GeV and $|\eta| \leq 2.0$ (recall that the $\mu\mu$
channel uses $|\eta| \leq 2.5$). 
The second transverse-energy variable
is simply the sum of the standard $H_T$ and the magnitude of the $W$ boson
transverse momentum vector, $H_T({\rm all}) \equiv H_T + p_T(W)$.
Events due to $t\bar t$ production tend
to have much higher values of $H_T$ than background. 
This is due to the fact that the jet $E_T$ is typically much harder for jets 
originating from the decay of a heavy object 
than are those from gluon radiation.

The $\ttb$ sample used in the optimization of all four channels is 
generated using 
\progname{herwig} with $m_t = 180$ \gevccn. The appropriate
combination of $W$+jets and QCD multijet events is used for background. 
The \zmumu background to the
$\mu$+jets/$\mu$ channel is not included in the optimization.
For the $\ell$+jets/$\mu$ channels, both the $W$+jets and QCD multijet
background estimates 
are based entirely on data. For the topological channels, the QCD multijet
background is based on data and the $W$+jets contribution is modeled 
using the 
\progname{vecbos}
Monte Carlo.  These background samples are used to investigate
the region of phase space remaining after the initial selection
(see Table~\ref{tab:ljini}), and thus differ somewhat from the samples used
in the full background determination to be discussed in Secs.~\ref{ljets_topo} 
and \ref{ljets_mutag}.

\begin{figure}[t]
\vbox{
\centerline{\hskip -3.8cm \psfig{figure=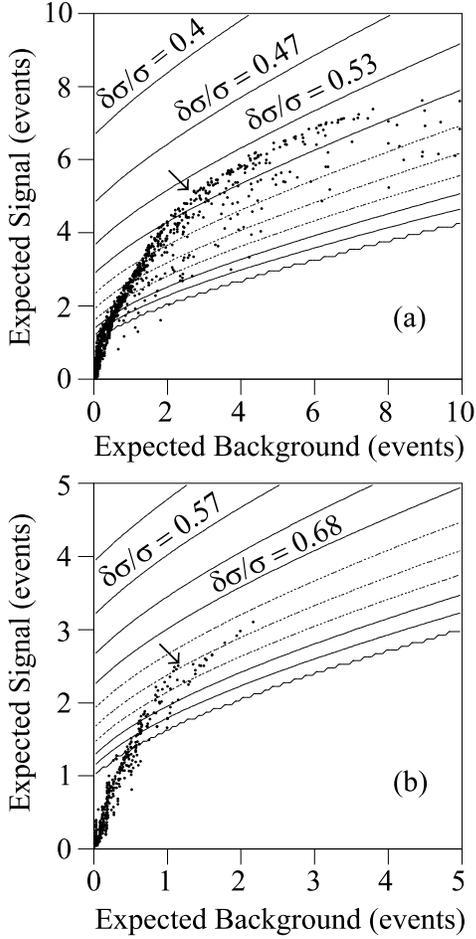,width=4.0in}}
\vskip -2.5cm
\caption{Expected signal vs expected background plots for 
${\cal A}(W+{\rm jets})$ and
$H_T$ optimization variables for (a) $\mu$+jets/topo and 
(b) $e$+jets/$\mu$. The solid curves are contours of constant 
uncertainty on the cross section ($d\sigma/\sigma$). Arrows indicate
chosen cut points.
\label{fig:ahtoptds}}}
\end{figure}

All of these variables are studied in pairs and in different 
combinations, and for
each set of cut points a corresponding point in the 
expected ($S$(signal),$B$(background)) plane 
is found. When all such points are plotted, they define a 
boundary that maximizes
the expected signal for a given background level, which is termed 
the ``optimal boundary'' (see, for example, Fig.~\ref{fig:ahtopt}). 
Comparison of the 
optimal boundaries for the various combinations of variables shows
that the pair ${\cal A}(W+{\rm jets})$ and $H_T$ provides the best 
signal to background ratio for a given signal efficiency.

After determining that ${\cal A}(W+{\rm jets})$ and $H_T$ are 
the best variables, it is necessary to select which cut point 
(on the optimal boundary) results in the most precise 
cross section measurement. Contours of constant uncertainty on
the measured cross section ($d\sigma/\sigma$) can be derived from the relation 
\begin{equation}
 \sigma = \frac{S}{\varepsilon\cdot{\cal L}} 
        = \frac{N - B}{\varepsilon\cdot {\cal L}} 
\label{eq:dsigosig}
\end{equation}
where $N$, $S$, and $B$ are the number of observed, expected signal, 
and expected background events, respectively, $\varepsilon$
is the signal efficiency, ${\cal L}$ is the integrated luminosity, and
$\sigma$ is the measured cross section~\cite{optimize}.
The cut points on the optimal boundary with the
smallest $d\sigma / \sigma$ and best significance ($s/b$) are 
(see Fig.~\ref{fig:ahtoptds}):

\vskip .2cm

$\ell$+jets/topo:
\begin{itemize}
\item   $H_T \geq 180$ GeV \
\item   ${\cal A}(W+{\rm jets}) \geq 0.065$ \
\end{itemize}

$\ell$+jets/$\mu$:
\begin{itemize}
\item   $H_T \geq 110$ GeV \
\item   ${\cal A}(W+{\rm jets}) \geq 0.040$. \
\end{itemize}

Following the initial selection and optimization it is necessary to make
several additional channel-specific requirements. These requirements, along 
with the results and expectations from signal and background, are discussed
in the next two sections (\ref{ljets_topo} and \ref{ljets_mutag}).

Acceptances for all four $\ell+$jets channels are computed from Monte Carlo 
events generated by the \progname{herwig} \cite{herwig} 
program for 24 
top quark mass values ($m_t$ = 90--230 GeV/$c^2$) and then passed 
through the full 
D\O\ detector simulation (see Sec.~\ref{mc}). The expected number of 
$t \bar t$ events passing the selection for a given channel is 
\begin{equation}
N = \sigma_{t \bar t}(m_t) \sum_{i = {\rm runs}} \sum_{j = {\rm det}}  
A(i, j, m_t) \cdot {\mathcal L}_{i,j} 
\label{eq:numljevts}
\end{equation}
where $\sigma_{t \bar t}(m_t)$ is the theoretical $t\bar t$ cross section at 
a top quark mass of $m_t$~\cite{laenen}; ${\mathcal L}_{i,j}$ is the 
integrated luminosity for run $i$ and detector region $j$ 
(CC and EC for electrons, CF and EF for muons); and the acceptance is 
\begin{equation}
A(i,j,m_t) = \varepsilon_{\rm trig} \cdot \varepsilon_{\rm pid} \cdot 
             \varepsilon_{\rm sel} \cdot G \cdot {\cal B},
\label{eq:ljaccept}
\end{equation}
where $\varepsilon_{\rm trig}(i,j,m_t)$ is the trigger efficiency, 
$\varepsilon_{\rm pid}(i,j)$ is the efficiency for lepton 
identification (isolated leptons and muon tag), 
$\varepsilon_{\rm sel}(i,j,m_t)$ is
the efficiency of the selection cuts, $G(i,j)$ is the geometrical
acceptance, and ${\cal B}$ is the branching fraction for the 
sample in question.
Trigger efficiencies are obtained from data or Monte Carlo events, depending
on the channel, and are discussed in more detail below. Particle identification
efficiencies are obtained from data for the case of electrons (as discussed
in Sec.~\ref{e-id}) and from a combination of data and Monte Carlo in
the case of muons (as discussed in Sec.~\ref{muid}).
The selection efficiencies $\varepsilon_{\rm sel}$ 
and the geometrical acceptances $G$ are obtained from Monte Carlo events.
As discussed in Sec.~\ref{crsec}, the acceptance, rather than
the expected number of $t\bar t$ events, is used in the calculation
of the $t\bar t$ cross section.
Typical values for the acceptance, often denoted as the 
``efficiency times branching fraction'' ($\varepsilon \times {\cal B}$),
for all eight leptonic channels, are given in 
Sec~\ref{crsec} for seven top quark masses. The numbers of 
$t \bar t$ events expected in the four $\ell+$jets channels are given in 
Tables~\ref{tab:topoevts} and \ref{tab:tagevts} 
Secs~\ref{ljets_topo} and \ref{ljets_mutag} for the same set of top quark
masses. The systematic uncertainties on the acceptances and backgrounds 
are discussed in Sec.~\ref{syserr}.

\subsection{Topological tag}
\label{ljets_topo}

\begin{figure}[h]
\vbox{
\vskip 0.1cm
\centerline{\psfig{figure=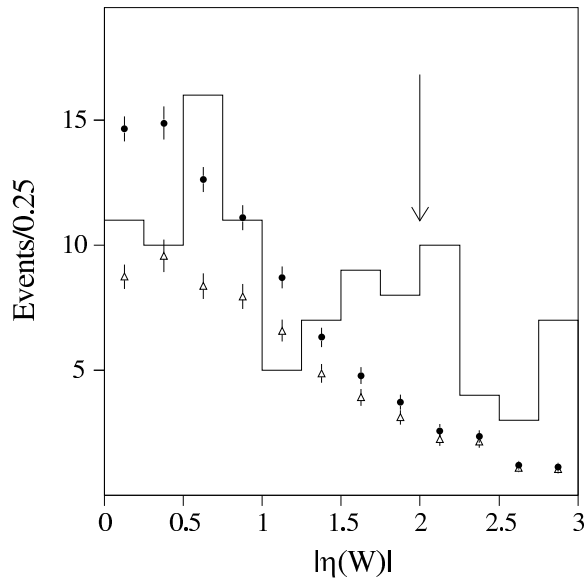,width=3.0in}}
\vskip 0.15cm
\caption{ $|\eta(W)|$ distribution for $\ell$+jets/topo data 
 (histogram) for the sum of 
 predicted signal and background (filled circles), and background alone 
 (open triangles), after application of all selection criteria except 
 the $\eta(W)$ cut.
\label{fig:ejtopo_etaw}}}
\end{figure}

\begin{figure}
\vbox{
\vskip -0.8cm
\centerline{\hskip 0.4in \psfig{figure=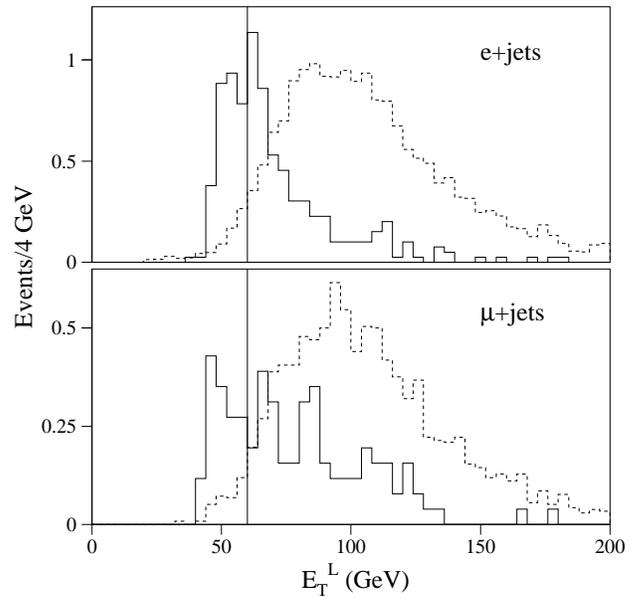,width=3.5in}}
\caption{ $E_T^L$ distributions for $t\bar t$ Monte Carlo 
($m_t = 170$ GeV/$c^2$)  
(dashed histogram), and for QCD multijet data (solid histogram),
after application of all selection criteria except those on
$E_T^L$, ${\cal A}$, and $H_T$. 
The distribution for $W$+jets is similar to that for $t\bar t$.
The solid vertical line at $E_T^L=60$ GeV indicates the cutoff value. 
\label{fig:ljtopo_etl}}}
\end{figure}

As described in the previous section, the first two stages of the 
$\ell$+jets/topo selection require the cuts described 
in Table~\ref{tab:ljini} followed by 
the cuts on ${\cal A}(W+{\rm jets})$ and $H_T$. There is, however, one cut in 
Table~\ref{tab:ljini} which has not yet been discussed. 
This cut on $\eta(W)$, 
the pseudorapidity of the lepton and $\MEt$ fit to a $W$ boson hypothesis,
is designed to remove from consideration those regions of phase space where the
$W$+jets \progname{vecbos} Monte Carlo does not model the 
$W$+jets data very well. As can be seen in Fig.~\ref{fig:ejtopo_etaw}, 
the \progname{vecbos} prediction is considerably below
the data in the forward region~\cite{ljmprd}. Therefore, the initial
selection requires that $|\eta(W)| \leq 2.0$. It should be noted that
only a few percent of $\ttb$ events have $|\eta(W)| > 2.0$, so this cut does 
not represent a serious reduction in acceptance. It should further be
noted that these analyses determine the $W$+jets backgrounds primarily 
from the data. The \progname{vecbos} Monte Carlo is only used to determine 
the survival probability for the cuts on ${\cal A}(W+{\rm jets})$, $H_T$, and
$E_T^L$ which is the scalar sum of the lepton $E_T$ and $\MEt$.
As can be seen in Fig.~\ref{fig:ljtopo_etl}, a requirement of 
$E_T^L \geq 60$ GeV provides significant rejection against QCD multijet 
background while having little effect on the $t\bar t$ signal.

As noted above, the primary backgrounds to the $\ell$+jets/topo 
channels are from $W(\rightarrow \ell\nu)$+jets and QCD
multijet events which contain a misidentified electron or isolated muon 
and mismeasured $\MEt$. The mismeasured $\MEt$ arises primarily
from mismeasurement of jet $E_T$ or vertex $z$ position.

The background calculation proceeds in four steps:
\begin{enumerate}
 \item The QCD multijet background is determined as a function of the
inclusive jet multiplicity from data samples in which the 
${\cal A},H_T,\eta(W)$, and $E_T^L$ 
cuts have not been applied. Due to the different processes that
give rise to a misidentified electron or isolated muon, these backgrounds 
are handled differently:
 \begin{itemize}
  \item Jets that have a large electromagnetic fraction can sometimes pass
   the electron identification criteria and be misidentified 
   as electrons. To determine the background from multijet events
   containing such misidentified electrons and $\MEt$, one begins with the 
   $\MEt$ spectrum from $n+1$ jet ($n \geq 0$) events with 
   $|\eta(W) | \leq 2.0$ which pass an electron trigger but fail the full 
   electron identification cuts (mis-id $e$ + $\MEt$ sample). This sample 
   correctly describes (with sufficient statistics) the $\MEt$ distribution 
   for the QCD multijet background, but the normalization 
   is not correct since the 
   electron identification requirement has not been made. The correct 
   normalization is 
   obtained by matching the number of events at low $\MEt$ ($\MEt \leq 10$
   GeV) to that found in a complementary sample that passes the normal
   electron identification criteria. Requiring $\MEt \geq 25$ GeV 
   then provides the expected number of 
   QCD multijet background events to the $e+n$ jet selection. 
   Uncertainties on this procedure are dominated by the statistics of the 
   samples used and range from 9.5\% (13\%) for the Run 1a (Run 1b) $e+1$ 
   jet selection to 27\% (54\%) for the Run 1a (Run 1b) $e+4$ jet selection.
  \item Muons from the semi-leptonic decay of a $b$ or $c$ quark are normally
   accompanied by an associated jet (non-isolated). However, occasionally 
   the decay kinematics are such that there is
   insufficient hadronic energy to produce a jet. In these cases the muons
   from semi-leptonic $b$ and $c$ decays will appear to be isolated. The
   probability that a muon originating from the decay of a heavy quark
   will appear isolated varies with jet multiplicity, run period, and 
   detector region, and is denoted by $I_{\rm mis-id}({\rm run,det})$.
   Typical CF values are 11\% for $\mu+\geq 1$ jet events and 6\% for
   $\mu+\geq 2,\geq 3, \geq 4$ jet events (the corresponding EF values are
   22\% and 15\% respectively). For a given jet multiplicity, $n$, these 
   probabilities 
   are measured using samples of QCD multijet events with $\MEt \leq 20$ GeV 
   as the ratio of the number of {\sl isolated}-$\mu$ + $\geq n$ jet events to
   the number of {\sl non-isolated}-$\mu$ + ($\geq n+1$) jet events. The QCD
   multijet background is defined by the product of this probability and
   the number of {\sl non-isolated}-$\mu$ + ($\geq n+1$) jet events with
   $\MEt > 20$ GeV. 
   The primary uncertainty in this method stems from the determination of 
   the above misidentified
   muon isolation probabilities. The value of 30\% assigned to this
   uncertainty is dominated by the statistical precision of the control 
   sample used to derive the false isolation fraction for four-jet events.
  \end{itemize}
   These procedures are carried out for each inclusive jet multiplicity, 
   thereby providing the expected QCD multijet 
   contribution to the $\ell+\geq n$ jet selections
   ($n$ = 1,2,3,4), as defined in Table~\ref{tab:ljini}. For the 
   $\ell+\geq4$ jet selection,  
   the expectation is $4.4 \pm 2.2 $ events in the $e+$jets/topo channel 
   and $ 6.44 \pm 2.08$ in the $\mu+$jets/$mu$ channel.
 \item The background from $W(\rightarrow \ell\nu)+$jets is computed
   by performing a fit to the jet-multiplicity spectrum that remains 
   following the subtraction of the QCD multijet background. 
   Inherent in the fit is the assumption of 
   ``Berends ($N_{\rm jets}$) scaling''~\cite{berends1,berends2}  
   which suggests that there is a simple exponential relationship between 
   the number of events and the jet multiplicity:
  \begin{equation}
   \frac{\sigma(W+n {\rm \ jets)}}{\sigma(W+(n-1) {\rm \ jets})} = \alpha,
  \label{eq:berends}
  \end{equation}
   where $\alpha$ is a constant (for any given jet $E_T$ and $\eta$ 
   requirements) and $n$ is the inclusive jet multiplicity.
   For any given inclusive jet multiplicity $i$, the number of events
   which are observed following the QCD multijet subtraction is given by
  \begin{equation}
   N_i^{\rm obs} = N_1^W \cdot \alpha^{i-1} + f_i^{\rm top} \cdot N^{\rm top},
  \label{eq:ni}
  \end{equation}
   where 
   $N_1^W$ is the number of $W+1$ jet events, 
   $N^{\rm top}$ is the number of $t\bar t$ events in the sample, and
   $f_i^{\rm top}$ is the fraction of $t\bar t$ events with jet 
   multiplicity $i$ (obtained from Monte Carlo).
   The values of $N_i^{\rm obs}$ are plotted in 
   Fig.~\ref{fig:ljtopo_jet_mult}. Fits to Eq.~\ref{eq:ni} 
   determine the values of $\alpha$ given in column 2 of 
   Table~\ref{tab:topobknd} ($N_1^W$ and $N^{\rm top}$ are also obtained from
   this fit).
\begin{figure}[t]
\vbox{
\vskip -0.8cm
\centerline{\hskip 0.4in \psfig{figure=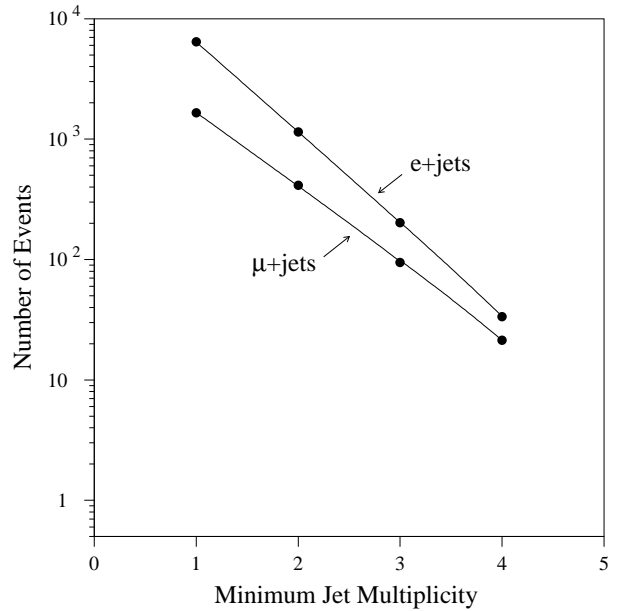,width=3.5in}}
\caption{Number of events as a function of inclusive jet multiplicity for 
the $e$+jets/topo and $\mu$+jets/topo analyses. All cuts have been
applied except ${\cal A}$ and $H_T$. The linear nature of the distributions 
is known as Berends scaling. Note that since the $E_T^L$ cut has been applied,
the values here differ from those in Table~\ref{tab:topocuts}.
\label{fig:ljtopo_jet_mult}}}
\end{figure}
\begin{table}
\caption{Number of $\ell$+jets/topo data events passing at each cut
level. Note that the $e$+jets luminosity of 90.9 ${\rm pb}^{-1}$ 
does not include recovered Main-Ring data (see Appendix~\ref{mrrecov}) -- 
the Main-Ring contribution is given in parentheses. Similarly, the 
luminosity for the $\mu$+jets channel does not include Run 1a or
recovered Main-Ring data. The Main-Ring contribution plus that from 
Run 1a is given in parentheses.
\label{tab:topocuts}}
\vskip 0.5cm
\begin{tabular}{l|r|r}
                        & $e$+jets   & $\mu$+jets \\
\hline
  Lum (${\rm pb}^{-1}$) & 90.9 &  76.6      \\
\hline 
$N_{\rm jets} \geq 1$   &   6604     &  2127 \\
$N_{\rm jets} \geq 2$   &   1225     &  537 \\
$N_{\rm jets} \geq 3$   &    223     &  124 \\
$N_{\rm jets} \geq 4$   &     39     &  28 \\
\hline
$E_T^L \geq 60$ GeV,    &     39     &  22  \\
${\cal A} \geq 0.065$,  &     18     &  10  \\
$H_T \geq 180$ GeV      &    7(2)    &  4(6)
\end{tabular}
\end{table}
\begin{table*}[t]
\squeezetable   
\caption{Steps in $e$+jets/topo and $\mu$+jets/topo background calculation: 
column 2, row 1 gives the expected number of QCD multijet background 
events ($\ell$+4 jets); column 1, row 2 gives the value of $\alpha$ 
determined from the fit to Eq.~\ref{eq:ni}; column 2, row 2 gives the
expected number of $W$+4 jet events; column 3 gives the trigger and Main-Ring 
(MR) correction factors; column 4 gives the result of multiplying column 2
by column 3 (step 3 in the text); column 5 gives the $E_T^L$, ${\cal A}$, 
$H_T$ cut survival probabilities; 
and column 6 gives the final expected background
obtained by multiplying column 4 by column 5. Note that Runs 1a and 1b
are treated separately for the $e$+jets channel whereas they are treated as a
single run for the $\mu$+jets channel.
\label{tab:topobknd}}
\vskip 0.5cm
\begin{tabular}{cr|c|c|c|c|c|c}
            & &          & Exp \# of evts & Trigger \&  &   Exp \# of evts  & $E_T^L, {\cal A}, H_T$ cut & Exp \# evts   \\
            & & $\alpha$ &     Steps 1-2  & MR corr     &    Step 3         & survival prob.  ($f$)     & Step 4        \\
\hline
$e+$jets & QCD multijet 1a & & $0.7\pm0.8$ & $1.09\pm 0.39$ & $0.76\pm0.91$ & 
       $0.071\pm0.040$ & $0.054\pm0.072$ \\
         &          1b & & $3.7\pm2.0$ & $1.71\pm0.12$ & $6.4\pm2.0$   &
       $0.051\pm0.010$ & $0.325\pm0.119$ \\
         &       Total & & $4.4\pm2.2$ &       --       & $7.16\pm2.20$ &
                --      & $0.379\pm0.139$ \\
\hline
         & $W+$jets 1a & $0.17 \pm 0.02$ & $5.45\pm1.53$  & $1.09\pm 0.17$ & $5.9\pm1.9$ &
       $0.092\pm0.061$ & $0.544\pm0.185$ \\
         &          1b & $0.18 \pm 0.01$ & $31.77\pm4.24$ & $1.22\pm0.06$ & $38.9\pm8.3$ &
       $0.092\pm0.061$ & $3.590\pm0.799$ \\
         &       Total & & $37.21\pm4.50$ &    --          & $44.8\pm8.6$ &
                --      & $4.135\pm0.899$ \\
\hline
\hline
$\mu+$jets & QCD multijet & & $6.44\pm2.08$  &  --  & $13.9\pm4.4$ &  --   & $0.993\pm0.498$ \\
\hline
           &    $W+$jets & $0.19 \pm 0.02$ & $18.8\pm3.2$ & $1.37\pm0.07$ & $25.8\pm4.6$ & $0.129\pm0.027$ & $3.324\pm0.911$
\end{tabular}
\end{table*}
   Once $\alpha$ is known, the number of $W+4$ jet events that pass the 
   initial selection can be determined from the equation
\begin{equation}
   N_4^W = N_1^W \cdot \alpha^3. 
\label{eq:n4}
\end{equation}
   The resulting $W$+jets background after the $\ell+4$ jet selection is 
   $37.2 \pm 4.5 $ events for the $e+$jets channel and $18.8 \pm 3.2 $ events 
   for the $\mu+$jets, as indicated in Table~\ref{tab:topobknd}.
   This method, solely based on data, is independent of theoretical 
   calculations of $W+n$ jet
   cross sections which have large uncertainties at high jet multiplicities.
 \item For the $e+$jets channel only, a correction factor of $1.09 \pm 0.39$
  ($1.71 \pm 0.12$) is applied to the Run 1a (Run 1b) QCD multijet 
  background results 
  to account for trigger differences between the background method and the 
  actual data selection and for the increased luminosity from the inclusion 
  of the Main-Ring data (see Sec.~\ref{triggers} and Appendix~\ref{mrrecov}) 
  in the Run 1a and Run 1b data sets. 
  A similar correction factor of $1.09 \pm 0.17$ ($1.22 \pm 0.06$) is applied 
  to the Run 1a (Run 1b) $W+$jets background. Following these corrections, the 
  backgrounds to the $e+4$ jets selection are found to be 
  $7.2 \pm 2.2$ events from QCD multijet and 
  $44.8 \pm 8.6$ events from $W+$jets.
 \item To determine the expected background following the final three
  cuts on $E_T^L$, ${\cal A}$, and $H_T$ (see Table~\ref{tab:topobknd}), 
  a {\it cut survival probability} $f$ 
  is computed for each background. This probability factor 
  is applied to the results obtained after the $\ell+ \geq 4$ jet  
  selections, thus giving the final expected QCD multijet and 
  $W+$jet backgrounds:
 \begin{equation}
  N({\rm total \ bkg}) = N^{\rm QCD}_{\ell+4j} \cdot f_{\rm QCD} + 
  N^W_{\ell+4j} \cdot f_W
 \label{eq:ntot}
 \end{equation}
  where $N^{\rm QCD}_{\ell+4j}$ and $N^W_{\ell+4j}$ are the QCD multijet 
  and $W+$jet background estimates following the $\ell+\geq 4$ jet 
  selections, and $f_{\rm QCD}$
  and $f_W$ are the survival probability factors for the QCD 
  multijet and $W+$jets backgrounds respectively.
  \begin{itemize}
   \item For the $e+$jets channel, $f_{\rm QCD}$ is determined from the
   combined $E_T^L$, ${\cal A}$, and $H_T$ pass rate on a sample of 
   misidentified
   electron+4 jet events that satisfy the $\MEt$ and $\eta(W)$ 
   requirements. 
   \item For the $\mu+$jets channel, the prescription is simply
   an extension of the QCD multijet 
   background computation described above for the
   $\mu+n$ jet selection. Specifically, the selection criteria are applied to
   five-jet events, where the jet associated with the non-isolated muon is
   not included in the ${\cal A}$ and $H_T$ calculations. 
  \end{itemize}
 For both channels, $f_W$ is determined using the \progname{vecbos} 
 Monte Carlo program to measure the final 
 efficiency (including the $\ell+\geq 4$ jet, $E_T^L$, ${\cal A}$, 
 and $H_T$ cuts) relative to that for the $\ell+\geq 4$ jet selection. 
 To investigate the systematic uncertainties associated with this Monte Carlo 
 based procedure,
 samples are generated with two different $Q^2$ scales, $M_W^2$ and 
 $\left< p_T^2({\rm jet})\right>$, and with two different hadronic 
fragmentation 
 prescriptions, \progname{isajet} and \progname{herwig}. 
 Comparison with the background-enriched sample of data indicates
 that \progname{vecbos} generated at
 $Q^2 = \left< p_T^2({\rm jet}) \right>$ and fragmented through 
 \progname{herwig} provides the best match.
 This choice is therefore used to compute the values of $f_W$.
\end{enumerate}
These four steps are summarized in Table~\ref{tab:topobknd}.

\begin{table}
\vbox{
\begin{center}
\caption{Observed and expected number of $\ell$+jets/topo 
signal and background events after all cuts.
Uncertainties shown are statistical and systematic contributions 
added in
quadrature. The total background systematic uncertainty includes 
correlations among the different background sources.
\label{tab:topoevts}}
\vskip 0.5cm
\begin{tabular}{c|c|c}
                        & $e$+jets   & $\mu$+jets \\
  Lum (${\rm pb}^{-1}$) & 119.5 &    107.7      \\
\hline 
Observed &        9        &        10       \\
\hline
\underline{top MC  $m_{t}$ (GeV/$c^2$)} &  &    \\
 140     &  $12.06 \pm 5.20 $ &  $8.22 \pm 3.56 $ \\
 150     &  $11.20 \pm 3.72 $ &  $7.83 \pm 2.98 $ \\
 160     &  $10.11 \pm 2.35 $ &  $7.12 \pm 2.40 $ \\
 170     &  $ 8.97 \pm 1.61 $ &  $5.72 \pm 1.72 $ \\
 180     &  $ 7.44 \pm 1.04 $ &  $4.80 \pm 1.27 $ \\
 190     &  $ 5.70 \pm 0.68 $ &  $3.84 \pm 0.92 $ \\
 200     &  $ 4.60 \pm 0.47 $ &  $3.14 \pm 0.69 $ \\
\hline
$W$+jets            & $4.14 \pm 0.90 $ &  $3.32 \pm 0.91 $ \\
QCD multijet        & $0.38 \pm 0.14 $ &  $0.99 \pm 0.50 $ \\
\hline
Total background    & $4.51 \pm 0.91 $ &  $4.32 \pm 1.04 $ \\
\end{tabular}
\end{center}}
\end{table}

\begin{figure}
\vbox{
\vskip -0.8cm
\centerline{\hskip 0.5cm \psfig{figure=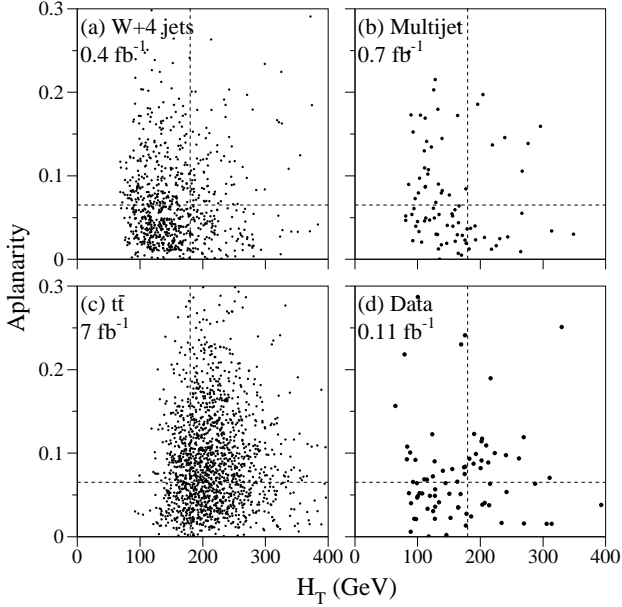,width=3.5in}}
\vskip 0.1cm
\caption{ Scatter plots of ${\cal A}$ vs $H_T$ for $\ell+$jets data (d)
compared to expectations from higher-luminosity samples of $t\bar t$ MC
($m_t$ = 170 \gevccn) (c), and QCD multijet (b) and $W+$4 jet MC (a) 
backgrounds.
The dashed lines represent the threshold values used for selection. 
The effective luminosity given for plot (b) is determined as the product of 
the luminosity of the selected multi-jet sample and the inverse of the 
appropriate misidentification rate.
\label{fig:ljtopo_aht}}}
\end{figure}

Figure~\ref{fig:ljtopo_aht} shows the distribution of 
${\cal A}$ vs $H_T$ for $\ell+$jets
(combined $e+$jets and $\mu+$jets) events for data, \progname{herwig} 
$t \bar t$ Monte Carlo ($m_t$ = 170 \gevccn), QCD multijet, and 
\progname{vecbos} $W+$jets Monte 
Carlo events. From this figure it is clear that ${\cal A}$ and $H_T$ 
provide significant discrimination between signal and background.

As described in Sec.~\ref{ljets_intro}, $t\bar t$ 
acceptances are computed via Eq.~\ref{eq:ljaccept} 
using Monte Carlo events generated with \progname{herwig} and 
passed through the D\O\ detector simulation. 
The trigger efficiency for the $e$+jets channel is obtained from $W$+jets
data and determined to be $ 98.2^{+1.8}_{-4.4}$\%. For the $\mu$+jets
channel, the trigger efficiency is computed using data-derived trigger
turn-on curves applied to $t\bar t$ Monte Carlo and is determined 
to be $89\pm5$\%.
The acceptance values after all cuts for seven different top quark masses 
(and for all channels) are given in Sec.~\ref{crsec}.

Following Eq.~\ref{eq:numljevts}, the expected 
numbers of $t\bar t$ events in the $\ell+$jets/topo channels are 
given in Table~\ref{tab:topoevts} for these same seven masses. 
Also shown are the final numbers of
events observed in the data, 9 in the $e+$jets channel and 10 in the
$\mu+$jets channel. Table~\ref{tab:topocuts} shows the observed number 
of data events passing at the different stages of the selection procedure.
Note that for this table, the $e$+jets luminosity does not include Main-Ring 
data and the $\mu$+jets luminosity does not include Run 1a or Main-Ring data.
Finally, the cross section obtained from
the $e$+jets/topo and $\mu$+jets/topo channels are $2.8 \pm 2.1$ pb and
$5.6 \pm 3.7$ pb, respectively.

\subsection{$\mu$ tag}
\label{ljets_mutag}

The initial selection for $\ell+$jets/$\mu$ events is described 
in Sec.~\ref{ljets_intro} and summarized in Table~\ref{tab:ljini}. 
All events 
are required to have a $\mu$ tag as defined in Sec.~\ref{muid}.

The dominant backgrounds that remain after the initial selection 
arise from $W(\rightarrow \ell\nu)$+jets production, QCD multijet events 
that contain a misidentified electron or isolated muon and mismeasured 
$\MEt$, and also $Z(\ra\mu\mu)+$jets for the $\mu$+jets/$\mu$ channel.

\begin{figure}
\vbox{
\vskip -0.85cm
\centerline{\hskip 0.6cm \psfig{figure=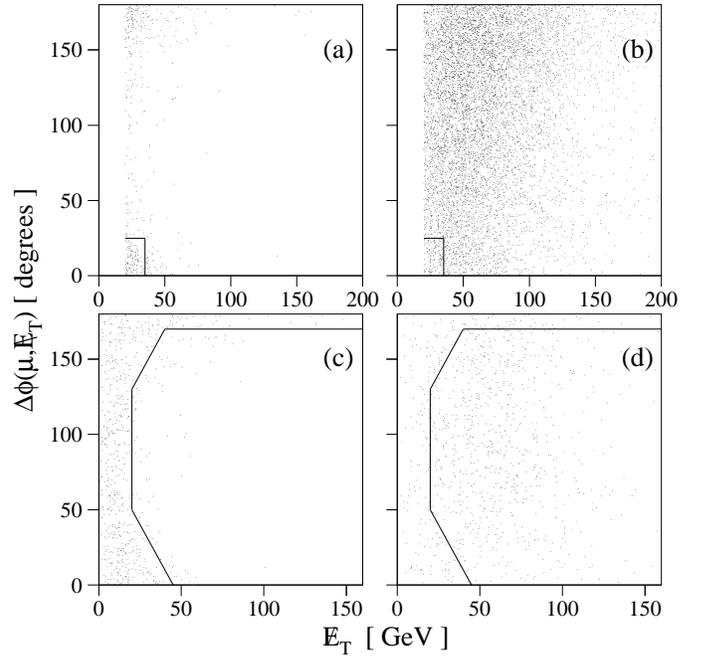,width=3.8in}}
\vskip 0.1cm
\caption{Scatter plots of $\Delta\phi(\mu,\met)$ vs $\met$ for 
(a) $e$+jets/$\mu$ QCD multijet background, 
(b) $t\bar t \ra e+{\rm jets}/\mu$,
(c) $\mu$+jets/$\mu$ QCD multijet background, and
(d) $t\bar t \ra \mu+{\rm jets}/\mu$. The solid lines define the
cut boundaries. 
\label{fig:ljm_metdphi}}}
\end{figure}

For events that have no genuine source of $\MEt$, the presence of a muon, as 
a consequence of the muon system's modest momentum resolution, may lead to 
mismeasured $\MEt$ which is aligned or anti-aligned with the muon $p_T$. 
Indeed, in multijet 
data, the distribution of the angle $\phi$ between the muon momentum 
and the direction of the $\MEt$,
$\Delta\phi(\mu,\MEt)$, peaks at $0^{\rm o}$ and $180^{\rm o}$, whereas for
$t\bar t$ events this distribution rises monotonically from 
$0^{\rm o}$ to $180^{\rm o}$ as indicated in Fig.~\ref{fig:ljm_metdphi}. 
In order to reduce background from QCD multijet events, 
both $\mu$-tag channels make 
a cut on the allowed region in the $\MEt$, $\Delta\phi(\mu,\MEt)$ plane:
\begin{equation}
\met > 35 {\rm \ GeV, \ if \ \ }  |\Delta\phi(\mu,\MEt)| \leq 25^{\rm o}, 
{\rm \ \ \ for \ }e+{\rm jets, } 
\label{eq:dphimetejm}
\end{equation}
and,
\begin{eqnarray}
\Delta\phi(\mu,\MEt) &<& 170^{\rm o} 
{\rm \ \ \ \ \ \ \ \ \ and \ } \nonumber \\
\frac{|\Delta\phi(\mu,\MEt)-90^{\rm o}|}{90^{\rm o}} & \leq & 
\frac{\MEt}{45 {\rm \ GeV}}, {\rm \ \ \ for \ }\mu+{\rm jets}.
\label{eq:dphimetmjm}
\end{eqnarray}
The effectiveness of these cuts 
is displayed in Fig.~\ref{fig:ljm_metdphi}, which shows the distributions in
the $\MEt$,$\Delta\phi(\mu,\MEt)$ plane for QCD multijet events and 
$t\bar t$ Monte Carlo events for both $\mu$-tag channels.

\begin{figure}
\vbox{
\vskip -0.8cm
\centerline{\psfig{figure=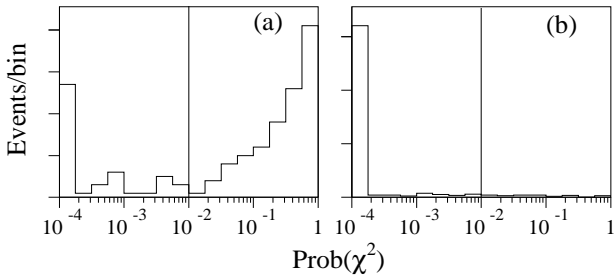,width=3.7in}}
\caption{ $\chi^2$ probability distribution for the $\mu+$jets/$\mu$ channel
after all cuts except $P(\chi^2)$: (a)  $Z(\ra \mu\mu)+$jets MC and 
(b) $t\bar t$ MC.
\label{fig:mjm_zmm_prbchi}}}
\end{figure}

In addition to the QCD multijet and $W+$jets backgrounds noted above, the 
$\mu$+jets/$\mu$ channel, by virtue of the fact that it requires two
muons, has a non-negligible background from $Z(\ra\mu\mu)+$jets production.
Although the muons from $Z$ boson decay are, 
in principle, isolated, there is a 
small probability that one of them will overlap with one of the jets in
the event and thus appear to be non-isolated.
The $\mu$+jets/$\mu$ channel relies therefore on a kinematic
fitting procedure 
to reduce this background. As described in Sec.~\ref{mumu}, a kinematic 
fit to the
$Z\rightarrow\mu\mu$ hypothesis is performed and a $\chi^2$ is obtained 
(see Eqs.~\ref{eq:zfit} and \ref{eq:zfitcon}). Events with a 
$\chi^2$ probability greater than 1\%, $P(\chi^2) > 0.01$, 
are considered likely $Z$ boson candidates and are therefore
rejected. As can be seen in Fig.~\ref{fig:mjm_zmm_prbchi}, this 
procedure provides very good
rejection against the $Z(\ra\mu\mu)+$jets background and has essentially no 
effect on the $t \bar t$ signal.

The general scheme for background calculation proceeds in three steps
which are first outlined and then discussed in detail:
\begin{enumerate}
\item Compute the QCD multijet background:
      \begin{itemize}
      \item For the $e+$jets/$\mu$ channel, the QCD multijet background is 
            computed by applying an electron misidentification rate to a 
            $\mu$-tagged multijet 
            control sample passing all cuts except tight electron 
            identification.
      \item For the $\mu+$jets/$\mu$ channel, the QCD multijet background is 
            computed by applying isolated-muon and muon-tag misidentification 
            rates to an untagged QCD multijet control sample passing all 
            other cuts except the isolated muon requirement.
      \end{itemize}
\item Compute the $W+$jets background: \\
      For both channels, the background from $W+$jets events is computed
      by applying a muon 
      tag rate to the number of untagged multijet-subtracted 
      $\ell$~+~$\geq 3$~jet data 
      events and then subtracting the expectation from $t\bar t$:
      \begin{equation}
      N_W({\rm bkg})  
            = N({\rm data - QCD}) \cdot P_{\rm tag} - N_3^{t\bar t}
      \label{eq:ljmwjb}
      \end{equation}
      where \\
      ``data'' is the number of events passing all cuts except $\mu$-tag; \\
      ``QCD'' is \\ 
        (the number of 
        (extra-loose-$e$+$\geq 3$~jet)/($\geq 4$~jet) events passing 
        $\met,{\cal A}$, and $H_T$ cuts) $\cdot$ ($e/\mu$ mis-id rate); \\
      ``$P_{\rm tag}$'' is the probability (as a function of jet $E_T$ and 
        $\eta$, and run period) for a jet to contain a tagged muon, and is 
        determined from QCD multijet data; \\
      ``$N_3^{t\bar t}$'' is the expected top quark contribution after all 
        cuts and is computed differently for the $e+$jets/$\mu$ and 
        $\mu+$jets/$\mu$ channels:
       \begin{itemize}
       \item for the $e+$jets/$\mu$ channel, the expected top quark 
             contribution is determined from data by fitting the jet 
             spectra of the multijet-subtracted untagged $e+n$~jet data under 
             the assumption of jet scaling and measuring the excess for 
             $n \geq 3$. A tag rate derived from $t\bar t$ MC is applied to
             this excess to obtain $N_3^{t\bar t}$.
       \item for the $\mu+$jets/$\mu$ channel, the expected top quark
             contribution ($N_3^{t\bar t}$) is determined from 
             \progname{HERWIG} MC normalized to the theoretical cross 
             section~\cite{berger2}.
        \end{itemize}
\item For the $\mu$+jets/$\mu$ channel only, determine the background
      from $Z \rightarrow \mu\mu$ using \progname{vecbos} MC events.
\end{enumerate}
The key elements of this procedure, namely the QCD multijet background 
calculations 
and the parameterization of the muon-tagging probability, are motivated and 
developed below.

\vskip 0.5cm

The estimation of the multijet background differs somewhat in the 
$e$+jets/$\mu$ and $\mu$+jets/$\mu$ channels. The calculation for the
$e$+jets/$\mu$ channel is similar to that used for
the $\ell$+jets/topo channels. Namely, the QCD multijet background 
is determined by relaxing the electron identification criteria and 
observing the number of additional events that pass the selection.
It is assumed that the number of events in the extra-loose electron sample, 
$N_l$, consists of both real, $N_e$, and misidentified (often referred to as
``fake''), $N_f$, electrons
\begin{equation}
N_l = N_e + N_f .
\label{eq:ejmfak1}
\end{equation}
The probability for a real electron to pass from the loose sample into the
tight sample, $\varepsilon_t^e$, is determined from 
$Z\rightarrow ee$ 
data. Similarly, the probability for a misidentified electron to make this 
transition,
$\varepsilon_t^f$, is defined as the ratio of tight to loose electron events
in a sample of 
``loose electron+1~jet'' events without \met~\cite{ptthesis}. 
These probabilities are determined separately for the CC and EC regions of 
the calorimeter and are given in Table~\ref{tab:ejmfak}. 
Applying these probabilities to the number of real and misidentified 
electrons in the
loose sample gives the expected number of events in the tight sample:
\begin{equation}
N_t = \varepsilon_t^e N_e + \varepsilon_t^f N_f .
\label{eq:ejmfak2}
\end{equation}
Eqs.~\ref{eq:ejmfak1} and ~\ref{eq:ejmfak2} can be solved for the number of
misidentified electron events in the loose sample:
\begin{equation}
N_f = \frac{\varepsilon_t^e N_l - N_t}{\varepsilon_t^e - \varepsilon_t^f}.
\label{eq:ejmfak3}
\end{equation}
The expected number of misidentified electron events in the final sample is 
the product of the number in the loose sample and the 
probability for a misidentified electron to pass the tight requirement,  
$\varepsilon_t^f N_f$. Values for the CC and EC regions of the 
calorimeter are given in Table~\ref{tab:ejmfak}. The combined (CC+EC) QCD
multijet background for the $e$+jets/$\mu$ channel, including additional 
systematic uncertainties (see Sec.~\ref{syserr}) not given in 
Table~\ref{tab:ejmfak}, is tabulated later in this section.


\begin{table} [htp]
\caption{$e$+jets/$\mu$ QCD multijet background calculation parameters
\label{tab:ejmfak}}
\vskip 0.5cm
\begin{tabular}{c|c|c}
                             &         CC         &         EC         \\
\hline 
$N_t$                        &         4          &         1          \\
$N_l$                        &         8          &         6          \\
$\varepsilon_t^e$            & $0.828 \pm 0.010$  & $0.453 \pm 0.015$  \\
$\varepsilon_t^f$            & $0.027 \pm 0.009$  & $0.053 \pm 0.012$  \\
$N_f$                        & $3.28  \pm 0.11$   & $4.30  \pm 0.31$   \\
$\varepsilon_t^f \cdot N_f$  & $0.088 \pm 0.030$  & $0.228 \pm 0.054$ 
\end{tabular}
\end{table}

The calculation of the QCD multijet background for the $\mu$+jets/$\mu$ 
channel is an extension of that used for the $\mu$+jets/topo channel. 
As described in Sec.~\ref{ljets_topo}, the
QCD multijet background calculation for the $\mu$+jets/topo analysis
applied the probability for a muon from a $b$ or $c$ quark decay to appear 
isolated to the number of non-isolated-$\mu$+jet events to determine the
expected number of misidentified isolated muon events in the signal sample.
The $\mu$+jets/$\mu$ analysis extends this by applying an additional
tag rate function. 
This tag rate function is based on a Monte Carlo sample containing
a high fraction of $b$-quark jets, and is parameterized in terms
of the jet $E_T$ as:
\begin{equation}
h(E_T,{\rm run, det}) = D({\rm run, det}) \cdot {\rm tanh} 
\left( \frac{E_T - 15.0 {\rm \ GeV}}{40.0 {\rm \ GeV}} \right)
\label{eq:mjmqcdtag}
\end{equation}
where $D({\rm run, det})$ is a scale factor that depends on the run period
and detector region under consideration. The QCD multijet background to the
$\mu$+jets/$\mu$ channel is then determined from the product 
\begin{equation}
\label{eq:mjmqcdtot}
N_{\rm QCD} =  \\
\sum_{\rm run,det} \sum_{\rm jets} N_0 \cdot 
I_{\rm mis-id}({\rm run,det}) \cdot h(E_T,{\rm run, det}), \nonumber
\end{equation}
where $N_0$ is the number of events which pass all selection criteria 
except for the isolation requirement on the high-$p_T$ $\mu$ and the
$\mu$-tag requirement, and $I_{\rm mis-id}({\rm run,det})$ is the 
misidentified-isolated-$\mu$ probability discussed in Sec.~\ref{ljets_topo}.
The final value, including systematic uncertainties, is tabulated at the
end of this section.

\vskip 0.5cm

The jets produced in association with $W$ boson production originate primarily
from final state gluon radiation. Therefore, except for a small contribution
from gluon splitting ($g \rightarrow b \bar b$), $W+$jets events are expected
to contain very few $b$ quarks and thus very few muon tags. In order to
estimate this background, it is assumed that the heavy flavor 
($b$ and $c$ quark) content in $W$+jets events is the same as in QCD multijet
events~\cite{prd1}. The expected number of $W$+jets+$\mu$ tag 
events is therefore 
computed from the product of the number of untagged $W$+jet
events and a muon-tag probability ($P_{\rm tag}$)
\begin{equation}
N_W({\rm tagged}) = N_W({\rm not \ tagged}) \cdot P_{\rm tag} .
\label{eq:numwtag}
\end{equation}
This probability is defined in a control sample of multijet events by the 
fraction of jets that contain a muon within a cone of $\Delta R = 0.5$ around
a jet axis. The control sample consists of events collected with a 
multijet trigger (\progname{jet-multi}, see Table~\ref{tab:tagratettrigs}) 
that have four or more 
jets reconstructed offline ($ E_T \geq 15$ GeV, $|\eta| \leq 2$).
These events were collected under essentially the same detector and 
accelerator conditions as the signal sample. 
The multijet and untagged $W$+jets samples have 
similar jet $E_T$ and $\eta$ distributions,
and, since both samples owe their high jet multiplicity to gluon radiation, 
they should also have similar quark-flavor content.

\begin{figure}
\vbox{
\vskip -0.7cm
\centerline{\hskip 0.2in \psfig{figure=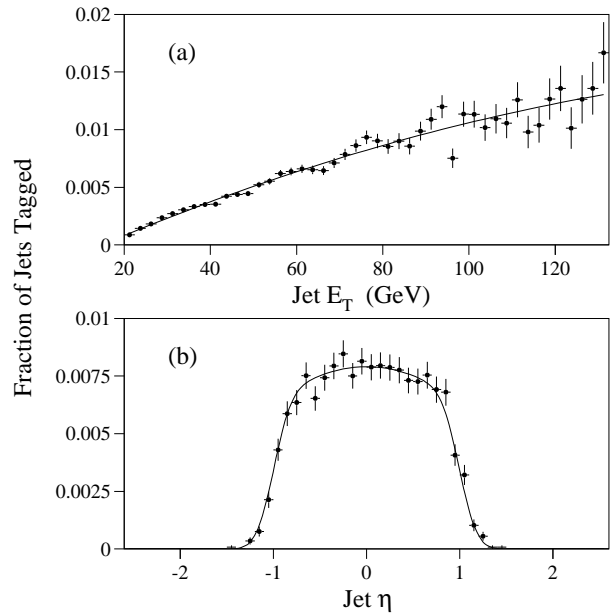,width=3.5in}}
\vskip 0.1cm
\caption{Parameterization of the muon tag rate, for muons in the CF region 
from Run 1b, as a function of (a) jet $E_T$ and (b) jet $\eta$.
\label{fig:tagrate}}}
\end{figure}

This fraction, also known as the tag rate, is parameterized explicitly
as a function of jet $E_T$ and $\eta$, and is handled separately for the
CF and EF regions of the muon system. The $\eta$ dependence is fit 
independently for the different run intervals used in the two analyses 
(see Table~\ref{tab:ljrrl}).
The tag rate as a function of jet $E_T$ and $\eta$ for muons in the CF 
region for Run 1b is shown in Fig.~\ref{fig:tagrate}. The tag rate increases
with jet $E_T$ because higher-energy jets have, on average, higher energy
muons that are more likely to penetrate the calorimeter and magnet
and be detected.
The shape of the $\eta$ distribution is primarily due to the geometrical
acceptance of the muon system, but varies somewhat over the different run
intervals. As a function of jet $E_T$, the data are fit to the functional 
form
\begin{equation}
  f(E_T) = \left\{ \begin{array}{ll}
           A_1 + A_2 \, E_T + A_3 \, E_T^2,   & \mbox{for $E_T \le \rho$} \\
           A_1 + A_2 \, \rho + A_3 \, \rho^2, & \mbox{for $E_T > \rho$}
                  \end{array}
           \right. 
  \label{eq:tagratept}
\end{equation}
where $\rho = - \frac{1}{2} A_2 / A_3$, and the parameters $A_1$, $A_2$ 
and $A_3$
are free. The resulting curves for muons in the CF and EF regions are 
denoted $f^{\rm{CF}}$ and $f^{\rm{EF}}$ respectively. As a function of $\eta$, 
the data for muons in the CF region are fit to the functional form
\begin{eqnarray}
  \label{eq:trcfeta}
  g^{\rm{CF}}(\eta,r) =   B_{1,r} \,
                      ( 1 + B_{4,r} \eta^2 ) \,
                      [ & {\rm{erf}}( \eta B_{2,r}  + B_{3,r} ) \\
                   &    - {\rm{erf}}( \eta B_{2,r}  - B_{3,r} ) ] , \nonumber
\end{eqnarray} 
where $r$ labels the three periods of the run  as specified in 
Sec.~\ref{muid}, 
${\rm{erf}}(x)=2/\sqrt{\pi} \int_0^x \exp(-t^2) \, dt$, and the parameters 
$B_{1,r}$, $B_{2,r}$, $B_{3,r}$, and $B_{4,r}$ are free to vary.
Similarly, for muons in the EF region, the data are fit to 
\begin{eqnarray}
  \label{eq:trefeta}
  g^{\rm{EF}}(\eta) = C_1
                      \left\{ \right. & {\rm{erf}}[ (|\eta|-C_4) C_2  + C_{3} ] \\
    & \left.    - {\rm{erf}}[ (|\eta|-C_4) C_2  - C_{3} ] \right\} , \nonumber 
\end{eqnarray}
with free parameters $C_{1}$, $C_{2}$, $C_{3}$, and $C_{4}$.
There is no run dependence in Eq.~\ref{eq:trefeta}, since, as noted in 
Sec.~\ref{muid}, the EF region of the muon system was only used during 
the final run period (Run 1b+c postclean). The complete tag rate function is
\begin{eqnarray}
  \label{eq:trcomplete}
  P_{\rm tag}(E_T,\eta,r) = & D^{\rm{CF}}_r \, f^{\rm{CF}}(E_T) \, g^{\rm{CF}}(\eta,r) \\
       &  + D^{\rm{EF}}_r \, f^{\rm{EF}}(E_T) \, g^{\rm{EF}}(\eta) , \nonumber
\end{eqnarray}
where $D^{\rm{CF}}_r$ and $D^{\rm{EF}}_r$ are constants that normalize the 
predicted number of tagged jets in the control sample to the actual number.
The values of the parameters in Eqs.~\ref{eq:tagratept}--\ref{eq:trcomplete} 
are given in Table~\ref{tab:trpars}.

\begin{table*}
\caption{$e$+jets+$\mu$ tag parameters from Eqs.~\ref{eq:tagratept} --
\ref{eq:trcomplete}.
\label{tab:trpars}}
 \begin{tabular}{ c | c c | c | c c c | c | c | c | c c c}
   \multicolumn{3}{c|}{ $f$ Parameters } & 
   \multicolumn{4}{c|}{$g^{\rm{CF}}$ Parameters } &
   \multicolumn{2}{c|}{$g^{\rm{EF}}$ Param. } & 
   \multicolumn{4}{c}{Normalization Param. } \\ \hline
   \multicolumn{1}{c}{ } & CF value & EF value &  
   \multicolumn{1}{c}{ } & $r=1$ & $r=2$ & $r=3$ & 
   \multicolumn{1}{c}{ } & $r=3$ &
   \multicolumn{1}{c}{ } & $r=1$ & $r=2$ & $r=3$  \\ \hline
       $A_1$           &  -0.243E-2  & -0.902E-3 & 
       $B_{1,r}$       &   0.386E-2  &  0.363E-2 &   0.395E-2 & 
       $C_1$           &   0.349E-2  & 
       $D^{\rm{CF}}_r$ &   249.6     & 248.7     & 223.4        \\ 
       $A_2$           &   0.170E-3  & 0.847E-4  & 
       $B_{2,r}$       &  11.5       & 2.26      &   4.78     & 
       $C_2$           &   3.92      &      
       $D^{\rm{EF}}_r$ &             &           & 528.8        \\
       $A_3$           &  -0.397E-6  & -0.368E-6 & 
       $B_{3,r}$       &  12.4       &  2.17     &   4.85     & 
       $C_3$           &   1.54      &  
                       &             &           &   \\
                       &             &           & 
       $B_{4,r}$       &  -0.483     & -0.477    &  -0.198    & 
       $C_4$           &   1.43      &
                       &             &           &  
\end{tabular}
\end{table*}

\begin{figure}[t]
\vbox{
\vskip -0.5cm
\centerline{\psfig{figure=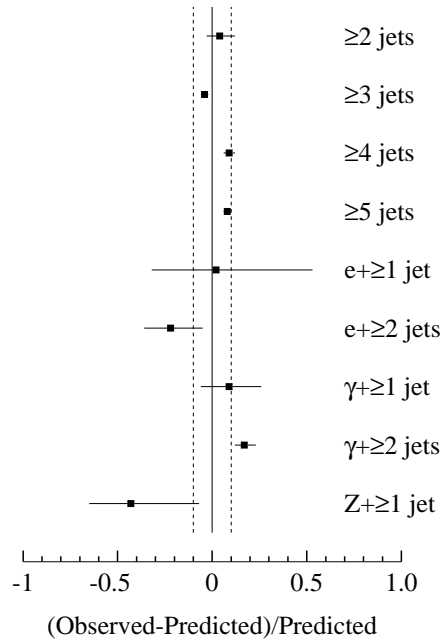,width=3.6in}}
\vskip 0.1cm
\caption{Tests of the muon tag rate. Shown are 
(Observed -- Predicted)/Predicted values for data sets that originate from
nine different triggers. Some of the scatter is due to statistics, as 
indicated by the horizontal error bars; the remainder is ascribed to 
systematic effects as described in Sec.~\ref{ss:tagrate}. The solid
vertical line is the overall mean value and the dashed vertical lines
are the uncertainty on the overall mean.
\label{fig:tagtest}}}
\end{figure}

The accuracy of this procedure has been studied by comparing the predicted 
to observed number of events having a tagged jet for a variety of data 
samples representing different trigger conditions, physics processes, and jet 
multiplicities. These studies are summarized in Fig.~\ref{fig:tagtest}, 
which shows the (Observed -- Predicted)/Predicted values for data samples
that originate from nine different triggers 
(see Table~\ref{tab:tagratettrigs} for the definitions of these triggers):
\begin{itemize}
\item The inclusive multijet samples with minimum jet multiplicity of two, 
three, four, and five were taken with the triggers \progname{jet-min}, 
\progname{jet-3-mon}, \progname{jet-4-mon}, and \progname{jet-multi},
respectively.
The last sample, with five jets selected offline, is a complete subset
of the four jet sample used in the actual tag rate calculation, 
comprising about one-third of the jets in the control sample.
\item The electron samples consist of events with a tight electron candidate, 
taken with the \progname{ele-1-mon} (\progname{gis-dijet}) trigger for the 
case of one (two) or more additional jets. Almost all of the ``electrons'' are 
false. The purpose of
examining these events is to check for an excess of tags due to $b \overline b$
or $c \overline c$ production, where one heavy quark decays to an electron and 
the other to a muon. There is no evidence of such an excess, and none is 
expected 
because of the isolation and high $E_T$ requirements imposed on the electron.
\item The photon samples consist of events with a tight photon candidate (see 
Sec.~\ref{e-id}), taken with the same triggers as the electron samples. 
About 30\% of the $\gamma + \ge 1$ jet events are from direct-photon production
and the rest are from multijet background \cite{dirphot}. 
The purity is less in the $\gamma + \ge 2$ jet data.
\item The $Z+{\rm jet}$ data were obtained with the  
\progname{em1-eistrkcc-esc} trigger, by requiring two 
loose electron candidates including at least one tight candidate. 
The invariant mass of the electron pair is required to be between 80 and
100 \gevccn. The background in this sample is low (10\%); but unfortunately 
only four events with a tagged jet survive, so the statistical uncertainty 
is quite large.
\end{itemize}
The horizontal error bars shown in Fig.~\ref{fig:tagtest} reflect
the statistical uncertainty on each comparison. As discussed in 
Sec.~\ref{syserr}, that portion of the scatter that cannot be attributed to 
the statistical uncertainty is taken as a measure of the systematic 
uncertainty of the tag rate procedure.

The functional dependence of the tag rate is 
important only to the extent that the target sample differs from the control 
sample. It should therefore be noted that the test samples with low jet 
multiplicity 
have significantly steeper jet $E_T$ spectra than either the control 
sample or the $W+\rm{jets}$ data after application of the 
\aplan\ and $H_T$ cuts.

\begin{figure}
\vbox{
\vskip -1.0cm
\centerline{\hskip 0.3 in \psfig{figure=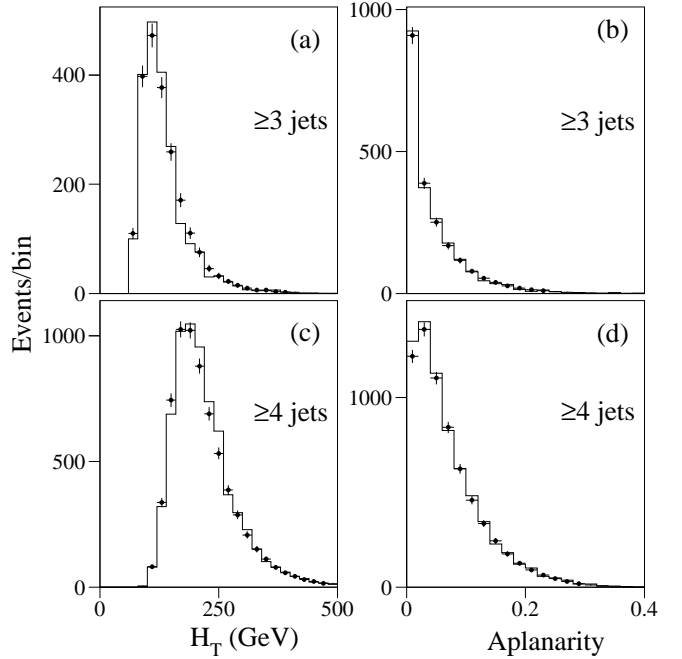,width=3.8in}}
\vskip 0.1cm
\caption{Predicted (histogram) and observed (filled circles) $H_T$ and 
$\cal{A}$ distributions in multijet data: (a) $H_T$ distributions for 
$\geq 3$ jet data, (b) $\cal{A}$ distributions for $\geq 3$ jet data, 
(c) $H_T$ distributions for $\geq 4$ jet data, 
and (d) $\cal{A}$ distributions for $\geq 4$ jet data.
\label{fig:tagshape}}}
\end{figure}

Because these analyses are concerned with the number of tagged events that 
remain in a data sample following selection cuts on $H_T$ and $\cal{A}$, 
it is important to confirm that the tag rate does not depend on
these variables in an unexpected way. Figure~\ref{fig:tagshape} 
shows a comparison of
the predicted and observed numbers of tagged events as a function of
$H_T$ and $\cal{A}$ for the $\geq 3$ jet and $\geq 4$ jet test samples.
The aplanarity distributions are in good agreement. 
Differences in the $H_T$ distributions suggest that a cut could result
in a discrepancy of a few percent between the predicted and observed number
of events. This is among the contributors to the tag-rate uncertainty that 
are discussed in Sec.~\ref{ss:tagrate}.

As noted in the outline at the beginning of this section, contamination from
QCD multijet and $t\bar t$ events requires that the background from $W+$jets
be computed via Eq.~\ref{eq:ljmwjb}. 
The QCD multijet contribution to the untagged sample is 
estimated by applying the lepton ($e$/$\mu$) misidentification rate to a 
sample of (loose-$e+\geq 3$~jet)/($\geq 4$~jet) events that have passed 
the $\met, {\cal A}$, 
and $H_T$ requirements. The $t\bar t$ contribution ($N_3^{t\bar t}$) for the 
$e+$jets/$\mu$ channel is determined from data by fitting the jet 
spectra of the QCD-multijet-subtracted $e + n$~jet data under the 
assumption of jet 
scaling and measuring the excess for $n \geq 3$. Following the hypothesis 
of jet multiplicity scaling, the number of $W+$jet events can be described
by a function of the form 
\begin{equation}
n_i = n^W_3 \alpha^{(i-3)}+n_3^{t\bar t} f_i/f_3
\label{eq:ejmwjttcon}
\end{equation}
where $n_i$ is the number of events with $i$ or more jets, $n^W_3$ is the 
number of $W$ boson events with three or more jets, $f_i$ is 
the number of events
in the $t\bar t$ MC sample with $i$ or more jets, and $\alpha$ is a free
parameter. A fit to Eq.~\ref{eq:ejmwjttcon} finds $n_3^{t\bar t}$ to be
$19.2 \pm 9.5$ events. $N_3^{t\bar t}$ is determined by applying the $t\bar t$ 
tag rate ($P_{\rm tag}^{t\bar t}$) to $n_3^{t\bar t}$.
The $t\bar t$ contribution ($N_3^{t\bar t}$) for the $\mu+$jets/$\mu$ 
channel is determined from \progname{HERWIG} MC normalized to the theoretical 
cross section~\cite{berger2}.

As given in Table~\ref{tab:tagevts}, 
the $W$+jets backgrounds for the $e$+jets/$\mu$
and $\mu$+jets/$\mu$ channels are determined via the multi-step 
procedure above to be $0.74 \pm 0.30 $ and $0.73 \pm 0.14 $ events 
respectively. Systematic uncertainties on the $W$+jets background arise
primarily from uncertainties in Berends scaling and $t\bar t$ MC tag
rate ($e+$jets/$\mu$ channel only) and the 
tag-rate parameterization. These are discussed in Sec.~\ref{syserr}.

\vskip 0.5cm

The background from $Z\rightarrow\mu\mu$ to the $\mu$+jets/$\mu$ channel 
is determined from \progname{vecbos} $Z+$jets Monte Carlo events in a
fashion similar to the Monte Carlo background calculations used for the
dilepton channels (see Eq.~\ref{eq:numllbkevts}) and is
given in Table~\ref{tab:tagevts}. 

Backgrounds from single top, $WW$, and $WZ$ production were also studied
and found to have a negligible contribution to the total combined background,
and therefore are not included in this discussion.

\begin{table}
\caption{Total observed and expected number of $\ell$+jets/$\mu$ 
events after all cuts.
\label{tab:tagevts}}
\vskip 0.5cm
\begin{tabular}{c|c|c}
                        & $e$+jets/$\mu$   & $\mu$+jets/$\mu$ \\
  Lum (${\rm pb}^{-1}$) & 112.6 &    108.0      \\
\hline 
Observed &        5        &        6       \\
\hline
\underline{$t\bar t$ MC  $m_{t}$ (\gevccn)} &  &    \\
 140     &  $6.93 \pm 1.35 $ &  $4.65 \pm 1.19 $ \\
 150     &  $6.18 \pm 1.06 $ &  $3.31 \pm 0.83 $ \\
 160     &  $4.51 \pm 0.73 $ &  $2.60 \pm 0.63 $ \\
 170     &  $3.73 \pm 0.57 $ &  $2.34 \pm 0.55 $ \\
 180     &  $3.11 \pm 0.46 $ &  $1.84 \pm 0.43 $ \\
 190     &  $2.44 \pm 0.36 $ &  $1.40 \pm 0.32 $ \\
 200     &  $1.83 \pm 0.27 $ &  $1.08 \pm 0.25 $ \\
\hline
$W$+jets            & $0.74 \pm 0.30 $ &  $0.73 \pm 0.14 $ \\
QCD multijet        & $0.32 \pm 0.26 $ &  $0.50 \pm 0.17 $ \\
\zmumu              &       --         &  $0.17 \pm 0.08 $ \\
\hline
Total background    & $1.05 \pm 0.40 $ &  $1.40 \pm 0.23 $ \\
\end{tabular}
\end{table}

\begin{figure}
\vbox{
\vskip 0.5cm
\centerline{\psfig{figure=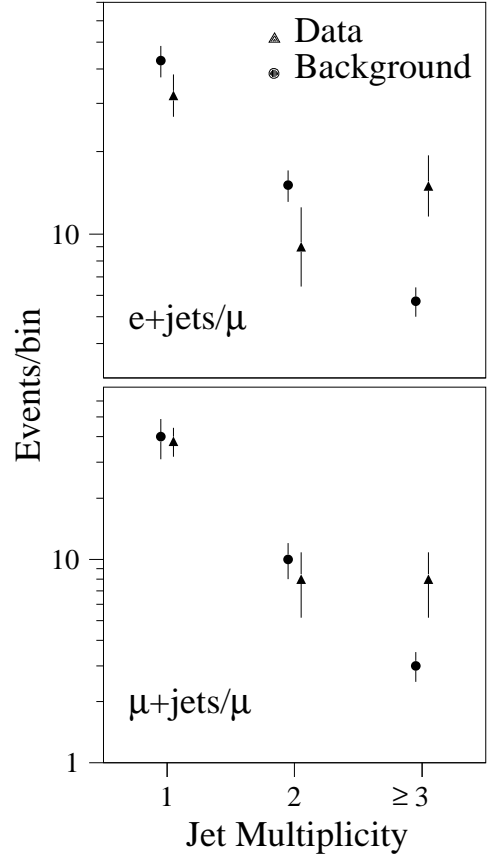,width=3.0in}}
\vskip 0.1cm
\caption{ Inclusive jet multiplicity spectra for $\ell+$jets/$\mu$ data 
(circles) and expected background (triangles) obtained prior to applying 
the $\cal{A}$ and $H_T$ requirements. Note that good agreement is seen 
for the $\geq 1$ and $\geq 2$ jet bins, but the $\geq 3$ jet bin shows
a clear excess in the data.
\label{fig:ljm_nenj}}}
\end{figure}

The inclusive jet multiplicity spectrum of the $\ell$+jets/$\mu$ 
data obtained 
prior to enforcing the ${\cal A}$ and $H_T$ requirements is compared with
that for the expected background in Fig.~\ref{fig:ljm_nenj}. Good agreement 
is seen in the background-dominated 1 and 2 jet bins, but for 3 or more jets,
the excess due to $t \bar t$ production is evident in both $\mu$-tagged 
channels.

\begin{figure}
\vbox{
\vskip -2.3cm
\centerline{\hskip 0.25 in \psfig{figure=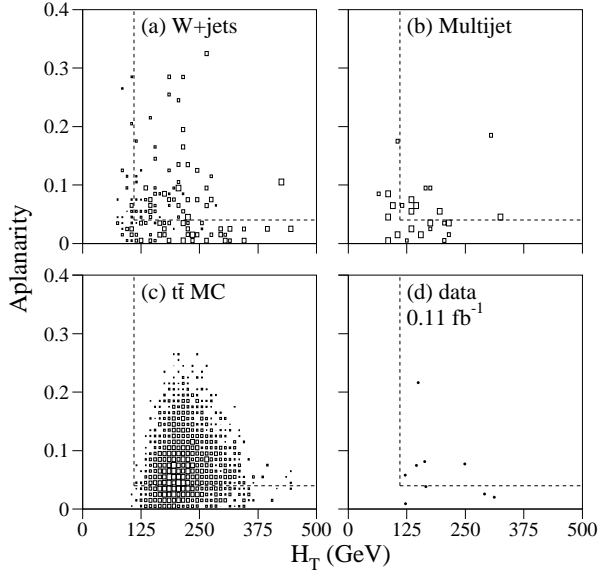,width=3.5in}}
\vskip 0.1cm
\caption{ Scatter plots of $\cal{A}$ vs $H_T$ for the $e+$jets/$\mu$ 
channel for 
(a) \progname{vecbos} $W+$jets MC background,
(b) QCD multijet background,
(c) \progname{herwig} $t\bar t$ MC events ($m_t$ = 170 \gevccn), and
(d) data. 
\label{fig:ejm_aht}}}
\end{figure}

\begin{figure}
\vbox{
\centerline{\hskip 0.25 in \psfig{figure=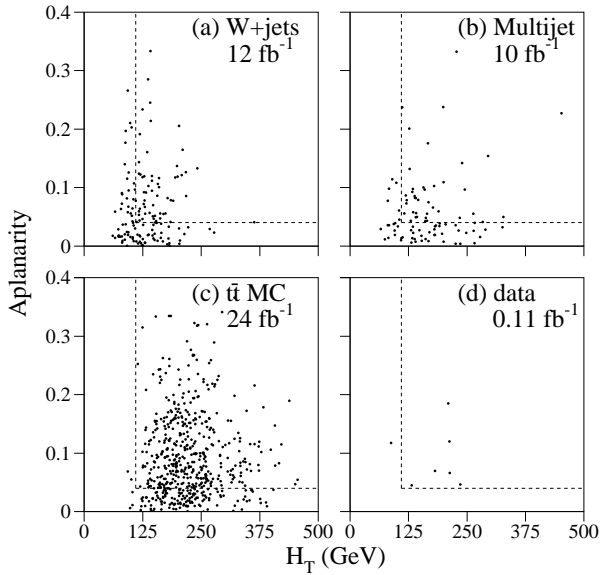,width=3.5in}}
\vskip 0.1cm
\caption{ Scatter plots of $\cal{A}$ vs $H_T$ for the $\mu$+jets/$\mu$ 
channel for 
(a) \progname{vecbos} $W+$jets MC background,
(b) QCD multijet background, 
(c) \progname{herwig} $t\bar t$ MC ($m_t$ = 170 \gevccn), 
and (d) data. The effective luminosity given for plot (b) is determined as the 
product of the luminosity of the selected multi-jet sample and the inverse 
of the muon misidentification rate.
\label{fig:mjm_aht}}}
\end{figure}

Figures~\ref{fig:ejm_aht} and \ref{fig:mjm_aht}
show the distributions of ${\cal A}$ vs $H_T$ for $e$+jets/$\mu$ and 
$\mu$+jets/$\mu$ events for 
data, \progname{HERWIG} $t \bar t$ Monte Carlo ($m_t$ = 170 \gevccn), 
QCD multijet data, and \progname{VECBOS} $W+$jets Monte Carlo events. 
From these figures it is clear that the cuts on ${\cal A}$ and $H_T$ 
provide a significant improvement in the discrimination between signal and 
background for these channels.

As described in Sec.~\ref{ljets_intro}, the $t\bar t$ 
acceptances are computed via Eq.~\ref{eq:ljaccept} 
using Monte Carlo events generated with \progname{herwig} and 
passed through the D\O\ detector simulation. 
The trigger efficiency for the $e$+jets/$\mu$ channel is obtained 
from the Trigger Simulator (see Sec.~\ref{mc}) and has been compared with
that found for $W$+jets data to estimate its systematic error, resulting
in a value of $99^{+1}_{-5} \%$.
For the $\mu$+jets/$\mu$ 
channel, the trigger efficiency is computed in the same fashion as for
the $\mu$+jets/topo channel using data-derived trigger
turn-on curves applied to $t\bar t$ Monte Carlo events and is determined 
to be $96^{+4}_{-5}$ \%.
The acceptance values after all cuts for seven different top quark masses 
(and for all channels) are given in Sec.~\ref{crsec} (Table~\ref{tab:ebrexp}). 
Following Eq.~\ref{eq:numljevts}, the expected 
number of $t\bar t$ events in the $\ell$+jets/$\mu$ channels are 
given in Table~\ref{tab:tagevts} for these same seven masses. 
Also shown in Table~\ref{tab:tagevts} are the final numbers of events 
observed in the data, 5 in the $e$+jets/$\mu$ channel and 6 in the
$\mu$+jets/$\mu$ channel. Finally, the cross section obtained from
the $e$+jets/$\mu$ and $\mu$+jets/$\mu$ channels are 
$6.0 \pm 3.6$ pb and $11.3\pm 6.6$ pb, respectively.

\section{Analysis of all-jets events}
\label{alljets}

As noted in Sec.~\ref{intro}, the all-jets channel is discussed in detail in 
Ref.~\cite{alljets} and is only summarized here.

The signature for the all-jets channel is characterized by the presence of 
six or more high transverse momentum jets. 
Given the overwhelming nature of 
the background to this channel, primarily from QCD multijet production, the 
challenge of this analysis is to develop selection criteria that 
provide maximum discrimination between signal and background, together 
with an estimate of the residual background in the signal region. 
Several kinematic and topological properties of the events were investigated, 
and neural networks employed to properly combine all possible sources of 
discrimination between signal and background. In order to improve the signal 
to background ratio, the analysis requires the presence of at least one 
muon-tagged jet in every event.
Because the data provide an almost pure sample of background events, 
the background model is determined entirely from data. The modeling uses 
untagged events that are made to represent tagged events by adding 
muon tags to one of the jets in the event. The cross section is determined
using fits to the neural network output, 
and checked using a conventional counting method.
The cross section obtained for $m_t=172.1$ \gevccn is:
\begin{equation}
\sigma_{t\bar{t}}=7.3\pm 2.8 ({\rm stat})\pm 1.5 ({\rm sys}) {\rm \ pb}.
\end{equation}
This cross section differs slightly from the value reported in 
Ref.~\cite{alljets} due to an update of the luminosity normalization.
The significance of the excess of $t\bar{t}$ signal over background is 
estimated by defining the probability $P$ of having the expected background 
fluctuate up to the observed number of events. This corresponds to a $3.2$ 
standard deviation effect, sufficient to establish the existence of a 
$t\bar{t}$ signal in multijet final states~\cite{alljets,alljprl}.

\section{Systematic uncertainties}
\label{syserr}

\def\cs{{\sigma_{t\bar t}}}

The individual uncertainties which affect the acceptance and background 
are discussed below. A discussion of the treatment of the correlations
between the uncertainties can be found in Appendix E.

\subsection{Sources}

\subsubsection{Luminosity}
\label{ss:lumerr}

As noted in Sec.~\ref{triggers}, the luminosity is determined with the
level 0 hodoscopes and is normalized to a world average total $p\bar p$ 
inelastic cross section from CDF~\cite{cdflum}, E710~\cite{e7101990}, 
and E811~\cite{e811} Collaborations. The systematic uncertainty on the 
luminosity stems 
from both the level 0 measurement and the world average total $p\bar p$ 
inelastic cross section and is found to be 4.3\%.

\subsubsection{Energy scale}
\label{ss:escaleerr}

Uncertainty in the jet energy scale affects the cross section determination
only via the uncertainty in the relative scale between data and MC. 
This uncertainty is determined by comparing $Z(\ra ee){\rm +jet}$ events in
data and MC~\cite{escalenim}. Events are selected by requiring two electrons 
with $E_T \geq 15$~GeV,  $82 \ \gevcc < m_{ee} < 102 \ \gevccn$, and at least 
one jet with 
$E_T \geq$ 15~GeV. The azimuthal bisector of the two electrons is determined
and the transverse momentum of the $Z$ boson is projected along this bisector
using the electron momentum vectors. The jet transverse momenta are also
projected along this bisector with the contribution from each jet in the 
event summed to form the jet projection. The jet energy projection versus
the $Z\ra ee$ projection is plotted for MC (\progname{herwig} 
and \progname{vecbos}) and data from Run 1b, and a linear regression fit
performed to determine the slope and offset of each sample. 
Comparison of the ratios of the slopes (MC/Data) and the differences in 
the offsets (MC -- Data) indicate an uncertainty in the jet energy scale
slope of 4\% and an uncertainty in the jet energy scale offset of 1 GeV.


\subsubsection{Electron identification}
\label{ss:eiderr}

The procedure for determining the electron identification efficiencies is
discussed in Sec.~\ref{eid-eff}. The primary source of uncertainty in this 
technique stems from the method used to subtract the background under the
$Z$ boson mass peak. Comparison of several different background subtraction 
schemes \cite{bkthesis} is used to determine the systematic uncertainties 
given in Table~\ref{tab:eid}.

\subsubsection{High $p_T$ and tag muon identification}
\label{ss:muiderr}

As described in Sec.~\ref{muid-eff}, the muon identification efficiencies 
are determined from a modified version of \progname{d{\o}geant} which has 
additional corrections to account for time dependent detector inefficiencies 
and incorrect modeling of the muon track finding efficiency. The time 
dependent correction is applied only to Run 1a and Run 1b(preclean) with an
uncertainty of 5\%, arising primarily from statistical considerations.
The track finding efficiency correction varied with detector region with an
uncertainty of 1.5\% in the CF and 2.2\% in the EF, also arising primarily 
from statistical considerations. The uncertainty arising from the detector
simulation is determined by comparing
$Z\ra\mu\mu$ MC events which are passed through the modified version of 
\progname{d{\o}geant} with $Z\ra\mu\mu$ data, the difference being a measure
of the uncertainty. This uncertainty varies with run period, detector 
region, and muon identification choice, and includes uncertainties from 
the muon trigger efficiency.
The uncertainties noted above are added in quadrature 
to determine the systematic uncertainty on the efficiencies given in 
Tables~\ref{tab:muid1a}--\ref{tab:muid1bpost}.

\subsubsection{$e$+jets trigger}
\label{ss:ejtrigerr}

This uncertainty accounts for systematic variations in the trigger 
efficiency for those signal and background MC samples that rely
primarily on electron triggers (see Table~\ref{tab:triggers_top_elex}).
The determinations of the trigger efficiencies for each channel are
discussed in the subsections of Secs.~\ref{dilep} and ~\ref{ljets_intro}.
For electron trigger efficiencies determined via the Trigger Simulator
($e\mu$: signal and all MC backgrounds; $e\nu$: signal and all MC backgrounds;
$e+$jets/$\mu$: signal), the systematic uncertainty is determined by 
comparing the trigger efficiency of $e+$jet data events (obtained from an 
unbiased trigger) with that found passing $W(\ra e\nu)+$jet MC events through 
the Trigger Simulator. For electron trigger efficiencies determined directly 
from data: for the $ee$ channel, comparison of the $Z(\ra ee)+$jets trigger 
rate obtained from unbiased data with that obtained from passing 
$Z(\ra ee)+$jet MC through the Trigger Simulator found a difference of 1\% 
which was taken as a measure of the uncertainty; for the $e+$jets channel,
studies of the efficiency variation using 
different samples and cuts led to the assignment of an uncertainty of 3\%.

\subsubsection{$\met$+jets trigger}
\label{ss:metjtrigerr}

This uncertainty accounts for systematic variations in the efficiency
of the $\MEt$ triggers (see Table~\ref{tab:triggers_top_met}).
Trigger efficiencies from the $\MEt$ triggers were obtained from measured
turn-on curves convoluted with kinematics from MC events. The systematic
uncertainty is determined from the differences in efficiency due to variations 
in top quark mass (for signal) and variations in the $\cal{A}$ and $H_T$ of 
the events (background).
Note that efficiencies for the muon triggers were determined from a
parameterization of the turn-on curves of the muon+jet triggers and the
systematics have been folded into the uncertainty on the 
muon identification efficiency.

\subsubsection{Multiple interactions}
\label{ss:mierr}

As discussed in Sec.~\ref{triggers}, there were, on average, 1.3 
$p\bar p$ interactions per bunch crossing during Run 1, giving rise to
additional minimum bias events produced along with the high-$p_T$ 
interactions of interest to the present analyses. These additional 
minimum bias events were not included in the MC models although they can
contribute to mismeasurement of the primary interaction vertex and thus
to mismeasurement of lepton and jet transverse energies/momenta. 
For $\ell+$jet 
events, such effects were found to be negligible since the presence of 
three or more hard jets from a single interaction vertex minimized any
potential confusion in determining the correct vertex. For the dilepton
channels the effect is more pronounced, and a systematic uncertainty is
estimated for all signal and MC-based backgrounds. To make this estimate,
additional signal and background MC samples were produced with one and 
two minimum bias events added. The efficiencies and background predictions
from these samples are then weighted according to the luminosity 
distribution of the Run 1 data set and compared to the samples for which
no minimum bias events had been added. The deviations, which vary 
significantly from channel to channel and between signal and background, 
are taken as an estimate of the uncertainty.

\subsubsection{$t \bar t$ Monte Carlo generator (kinematics)}
\label{ss:ttmcerr}

The uncertainty on the modeling of kinematic quantities 
(high-$p_T$ leptons, jets, and \met) due to imperfections in the MC 
generator is based on efficiency differences between the 
\progname{HERWIG} and \progname{ISAJET} generators. 
This uncertainty is calculated separately for each channel.  
The procedure, which is the same for each 
channel, is to generate a smooth curve summarizing the observed generator 
difference (\progname{ISAJET}--\progname{HERWIG}/\progname{HERWIG}) for top 
quark masses from 140~GeV/$c^2$ to 200~GeV/$c^2$, ignoring any $b$-tag or 
$b$-tag-veto cuts. As seen in Table~\ref{tab:gen-err}, the dilepton channels 
are parameterized using a constant relative uncertainty and the lepton+jets 
channels are parameterized using an exponential function of the top quark
mass. The aspect of the 
generator to which the kinematic acceptance is most sensitive is the
parton showering. \progname{HERWIG} has been shown to reproduce jet
properties well at both the Tevatron \cite{jettopo} 
and LEP \cite{delphijets}. 
Reference \cite{jettopo} describes a study of the topological properties 
(spectra of angles and 
energy distribution among jets) in inclusive three and four jet events and 
the authors find that ``[a]part from the cos($\theta^*$) distributions, the 
\progname{HERWIG} event generator provides a reasonably good description of
the data while the differences between the data and the predictions of [the]
\progname{ISAJET} and \progname{PYTHIA} event generators are large in many
distributions.''

\begin{table}[htp]
\caption{Smoothed kinematic generator uncertainties for the eight leptonic 
channels.
\label{tab:gen-err}}
\begin{tabular}{c|cc}
Channel & \multicolumn{2}{c}{Relative Uncertainty} \\
\cline{2-3}
& Fit & Applied \\
\hline
$ee$     &   5.5\% & 5\% \\
$e\mu$   &  -4.9\% & 5\% \\
$\mu\mu$ &   3.3\% & 5\% \\
$e\nu$   & -11.1\% & 12\% \\
$e+{\rm jets}$       & \multicolumn{2}{c}{$\hbox{exp}( 4.59  - 0.0407 m_t)$} \\
$\mu+{\rm jets}$     & \multicolumn{2}{c}{$\hbox{exp}( 0.546 - 0.0120 m_t)$} \\
$e+{\rm jets}/\mu$   & \multicolumn{2}{c}{$\hbox{exp}(-0.279 - 0.0150 m_t)$} \\
$\mu+{\rm jets}/\mu$ & \multicolumn{2}{c}{$\hbox{exp}(-0.293 - 0.0124 m_t)$} \\
\end{tabular}
\end{table}

\begin{table*}
\caption{Expected Run 1 dilepton backgrounds and the corresponding 
statistical and systematic uncertainties (number of events).
\label{tab:sysdilbk}}
\vskip 0.5cm
\begin{tabular}{l|ccccc|cccc|ccccc}
   & \multicolumn{5}{c|}{$ee$} & \multicolumn{4}{c|}{$e\mu$} & 
\multicolumn{5}{c}{$\mu\mu$} \\
   & $Zee$ & $Z\tau\tau$   & $WW$ & DY$ee$ & multijet
   & $Z\tau\tau$  & $WW$ &  DY$\tau\tau$ & multijet
   & $Z\mu\mu$ & $Z\tau\tau$ & $WW$  & DY$\mu\mu$ & multijet \\
\hline 
\# of evts & 0.058 & 0.081 & 0.086 & 0.056 & 0.197 & 0.103 & 0.077 & 0.006 & 0.077 & 0.579 & 0.030 & 0.007 & 0.068 & 0.068 \\
\hline
Statistical& 0.009 & 0.008 & 0.008 & 0.011 & 0.044 & 0.051 & 0.006 & 0.004 & 0.121 & 0.141 & 0.015 & 0.003 & 0.030 & 0.010 \\
Luminosity &  --   & 0.004 & 0.004 & 0.002 &  --   & 0.004 & 0.003 & 0.000 &  --   & 0.025 & 0.001 & 0.000 & 0.003 &  --   \\  
Energy Scale&  --   & 0.020 & 0.022 & 0.014 &  --   & 0.026 & 0.010 & 0.000 &  --   & 0.133 & 0.007 & 0.002 & 0.016 &  --   \\  
$e$ id     &  --   & 0.004 & 0.004 & 0.003 &  --   & 0.005 & 0.002 & 0.000 &  --   &  --   &  --   &  --   &  --   &  --   \\
High $p_{T} \ \mu$ id 
           &  --   &  --   &  --   &  --   &  --   & 0.012 & 0.007 & 0.001 &  --   & 0.040 & 0.002 & 0.001 & 0.005 &  --   \\
$e$+jets trig & -- & 0.000 & 0.000 & 0.000 &  --   & 0.005 & 0.004 & 0.000 &  --   &  --   &  --   &  --   &  --   &  --   \\
Mult. Int. &  --   & 0.008 & 0.009 & 0.006 &  --   & 0.017 & 0.012 & 0.001 &  --   & 0.018 & 0.001 & 0.000 & 0.002 &  --   \\ 
Bkg crsec &  --   & 0.010 & 0.009 & 0.028 &  --   & 0.010 & 0.008 & 0.001 &  --   & 0.059 & 0.005 & 0.001 & 0.010 &  --   \\
Other Sim  &  --   & 0.050 &  --   &  --   &  --   & 0.064 &  --   &  --   &  --   &  --   & 0.019 &  --   &  --   &  --   \\
Mis-id $e$   &  --   &  --   &  --   &  --   & 0.015 &  --   &  --   &  --   & 0.003 &  --   &  --   &  --   &  --   &  --   \\
Mis-meas $\met$ & 0.009 &  --  &  --   &  --   &  --   &  --   &  --   &  --   &  --   &  --   &  --   &  --   &  --   &  --   \\
$Z$ fitter &  --   &  --   &  --   &  --   &  --   &  --   &  --   &  --   &  --   & 0.060 & 0.001 & 0.000 & 0.002 &  --   \\
\hline
Total      & 0.013 & 0.056 & 0.027 & 0.034 & 0.046 & 0.089 & 0.021 & 0.004 & 0.121 & 0.218 & 0.026 & 0.004 & 0.036 & 0.010 \\
\end{tabular}
\end{table*}

\begin{table*}
\caption{Expected Run 1 $\ell$+jets backgrounds and the corresponding 
statistical and systematic uncertainties (number of events).
\label{tab:sysljbk}}
\vskip 0.5cm
\begin{tabular}{l|cc|cc|cc|ccc}
    & \multicolumn{2}{c|}{$e$+jets/topo} 
    & \multicolumn{2}{c|}{$\mu$+jets/topo} 
    & \multicolumn{2}{c|}{$e+{\rm jets}/\mu$} 
    & \multicolumn{3}{c}{$\mu+{\rm jets}/\mu$} \\
   & $W$+jets & multijet & $W$+jets & multijet & $W$+jets & 
multijet & $W$+jets & $Z\mu\mu$ & multijet  \\
\hline 
\# of evts            & 4.135 & 0.379 & 3.324 & 0.993 & 0.738 & 0.316 & 0.726 & 0.170 & 0.500  \\
\hline
Statistical           & 0.464 & 0.139 & 0.437 & 0.347 & 0.044 & 0.246 & 0.118 & 0.036 & 0.052  \\
Luminosity            &   --  &   --  &   --  &   --  &   --  &   --  &   --  & 0.007 & --     \\
Energy Scale          & 0.207 &   --  & 0.179 &   --  &   --  &   --  &   --  & 0.017 & --     \\  
High-$p_{T} \ \mu$ id &   --  &   --  &   --  &   --  &   --  &   --  &   --  & 0.022 & --     \\
Tag $\mu$ id          &   --  &   --  &   --  &   --  &   --  &   --  &   --  & 0.010 & --     \\
$\met$+jets trig      &   --  &   --  & 0.166 &   --  &   --  &   --  &   --  & 0.002 & --     \\
Mult. Int.            &   --  &   --  &  --   &   --  &   --  &   --  &   --  & 0.000 & --     \\ 
\progname{vecbos}     & 0.616 &   --  & 0.665 &   --  &   --  &   --  &   --  &   --  & --     \\
Bkg crsec             &   --  &   --  &   --  &   --  &   --  &   --  &   --  & 0.051 & --     \\
Berends scaling       & 0.413 &   --  & 0.369 &   --  & 0.292 &   --  &   --  &   --  & --     \\
Mis-id $e$              &   --  &   --  &   --  &   --  &   --  & 0.066 &   --  &   --  & --     \\
Tag rate              &   --  &   --  &   --  &   --  & 0.074 &   --  & 0.073 &   --  & --     \\
Mis-id $\mu$          &   --  &   --  &   --  & 0.298 &   --  &   --  &   --  &   --  & 0.100  \\
$\mu$ multijet        &   --  &   --  &   --  & 0.199 &   --  &   --  &   --  &   --  & 0.100  \\
Tag probability       &   --  &   --  &   --  &   --  &   --  &   --  &   --  &   --  & 0.075  \\
$Z$ fitter            &   --  &   --  &   --  &   --  &   --  &   --  &   --  & 0.042 & --     \\
\hline
Total                 & 0.899 & 0.139 & 0.911 & 0.498 & 0.304 & 0.255 & 0.139 & 0.081 & 0.168  
\end{tabular}
\end{table*}

\subsubsection{$t \bar t$ Monte Carlo generator ($b$-tagging)}
\label{ss:ttmcbtagerr}

In addition to kinematic quantities (high-$p_T$ leptons, jets, and \met), 
generator imperfections can contribute to the uncertainty in the 
probability that a soft muon 
will be produced and subsequently pass the identification and $p_T$ cuts
(see Sec~\ref{muid}). Potential sources of uncertainty include 
the branching fraction of $b\to\mu+X$, the branching fraction of $c\to\mu+X$ 
for cascade decays, $b$ quark fragmentation, $B$ hadron decay form factors, 
and uncertainties associated with misidentified tags. Only the effect of the 
branching fraction of $b\to\mu+X$ has been considered. In \progname{herwig},
all $b$ hadrons decay via a spectator model with a branching fraction to
muons ${\cal B}(b\ra\mu)=0.11$. The particle data book~\cite{pdg} lists 
the following inclusive measurements of $B$ hadron semileptonic branching 
fraction:

\medskip
\begin{tabular}{ll}
$\Upsilon(4S)$ inclusive $B\to\mu$ & $10.3\pm 0.5$\% \\
$\Upsilon(4S)$ inclusive $B\to\ell$ & $10.43\pm 0.24$\% \\
High energy inclusive $B\to\mu$ & $10.7\pm 0.7$\% \\
High energy inclusive $B\to\ell$ & $11.13\pm 0.29$\%
\end{tabular}
\medskip

\noindent 
The errors on the inclusive $B\to\ell$ branching fraction are
quite small, although the $\Upsilon(4S)$ and high energy measurements are 
inconsistent at two standard deviations. 
The uncertainty due to this 
variation has been increased to account for the remaining sources
of uncertainty, resulting in the assignment of a fractional uncertainty 
of 10\%.

\subsubsection{\progname{VECBOS}}
\label{ss:vberr}

As discussed in Sec.~\ref{ljets_topo}, the $\ell+$jets/topo channels 
use \progname{VECBOS} to determine the ${\cal A}(W+{\rm jets})$, $H_T$, and 
$E_T^L$ cut survival probability for $W+$jets backgrounds. The systematic
uncertainty for this procedure is estimated by comparing the 
${\cal A}(W+{\rm jets})$, $H_T$, and $E_T^L$ distributions of $\geq 2$ and
$\geq 3$ jet events in data and \progname{vecbos} (after adding contributions
from $t\bar t$ and QCD multijet production to the \progname{vecbos} 
sample in the appropriate
proportions). For $\geq 2$ jet events, a 6\% difference is seen and for
$\geq 3$ jet events, a 10\% difference is seen. Extrapolated to $\geq 4$ jet
events, a 15\% uncertainty is estimated.

\begin{table}[h]
\caption{Expected Run 1 $e\nu$ and all-jets expected backgrounds 
(number of events) and the corresponding statistical and systematic 
uncertainties. Uncertainties labelled Tag rate norm, Tag rate fn, and
$t\bar t$ corr are for the all-jets channel only and correspond 
respectively to uncertainties associated with the normalization of the 
muon tag rate, the functional form of the muon tag rate, and corrections
to the background for $t\bar t$ signal. The systematic uncertainties on 
the all-jets channel are discussed in detail in Ref.~\protect\cite{alljets}.
\label{tab:sysenuajbk}}
\begin{tabular}{l|cccc|c}
  & \multicolumn{4}{c|}{$e\nu$} & \multicolumn{1}{c}{all-jets}  \\
                  & $WW$  & $WZ$ & $W$+jets & multijet & multijet \\
\hline 
\# of evts        & 0.161 & 0.017 & 0.543 & 0.471 & 24.8 \\
\hline
Statistical       & 0.028 & 0.002 & 0.272 & 0.103 & 0.7 \\
Luminosity        & 0.007 & 0.001 & 0.023 &  --   & --  \\
Energy scale      & 0.040 & 0.004 & 0.136 &  --   & 1.0 \\
$e$ id            & 0.004 & 0.000 & 0.013 &  --   & --  \\
$e$+jets trig     & 0.005 & 0.001 & 0.016 & 0.014 & --  \\
Mult. Int.        & 0.009 & 0.001 & 0.030 &  --   & --  \\
\progname{Vecbos} &  --   &  --   & 0.086 &  --   & --  \\
Bkg crsec         & 0.016 & 0.002 &  --   &  --   & --  \\
Other Sim         &  --   &  --   &  --   & 0.104 & --  \\
Mis-id $e$        &  --   &  --   &  --   & 0.034 & --  \\
Tag rate norm     &  --   &  --   &  --   &  --   & 1.2 \\
Tag rate fn       &  --   &  --   &  --   &  --   & 1.2 \\
$t \bar t$ corr   &  --   &  --   &  --   &  --   & 1.0 \\ \hline
Total             & 0.053 & 0.005 & 0.319 & 0.151 & 2.4 \\
\end{tabular}
\end{table}

\subsubsection{Background cross section}
\label{ss:bkcserr}

As described in Secs.~\ref{dilep} and ~\ref{ljets_mutag}, backgrounds 
determined from MC have their initial cross sections normalized to either 
measured or theoretical values and the uncertainties are therefore taken 
from the cited references.

\subsubsection{Other simulation}
\label{ss:osimerr}

This uncertainty accounts for additional, channel specific, systematic 
effects due to the simulation and is only included for the $Z\ra\tau\tau$
background to the $ee$, $e\mu$, and $\mu\mu$ channels and for the QCD 
multijet background to the $e\nu$ channel. 
As described in Secs.~\ref{ee}--\ref{mumu},
the jet cut survival probabilities for the $Z\ra\tau\tau\ra \ell\ell$ 
backgrounds are obtained from $Z(\rightarrow ee)$+jet data. The primary
limitation of this technique is the limited statistics of the 
$Z(\rightarrow ee)$+jet data set, which is taken as the dominant uncertainty.
As described in Sec.~\ref{enu}, the QCD multijet background is obtained as the
mean of two independent procedures. The difference between the 
two procedures is taken as a systematic uncertainty.

\subsubsection{Berends scaling}
\label{ss:bserr}

As noted in Sec.~\ref{ljets_topo}, the assumption
of $N_{\rm jets}$ or Berends scaling (see Eq.~\ref{eq:berends}) is used by 
the $\ell+$jets/topo channels to compute the background from $W+$jets. 
In order to investigate the validity of this assumption, a number of data
sets were examined: $W$+jets, QCD multijet, $Z$+jets, photon+jets, and 
\progname{VECBOS} $W$+jets production. For each sample the number of events 
with a minimum jet 
multiplicity of $n-1$ and $n-2$ were used to predict the number of events
with a minimum jet multiplicity $\geq n$. These predictions were compared 
with observations and the maximum differences are given in 
Table~\ref{tab:berends-err}. Based on these values an uncertainty of 10\%
is assigned for the uncertainty due to Berends scaling.

\begin{table}[htp]
\caption{Maximum deviation between predictions from Berends scaling 
and observation for several data sets.
\label{tab:berends-err}}
\begin{tabular}{c|c}
 Data Set & Maximum Deviation (\%)     \\
\hline
$W$+jets                    &  3.1     \\
QCD multijet                &  $< 10$  \\
$Z$+jets                    &  $<  4$  \\
Photon + jets               &  $<  5$  \\
\progname{VECBOS} $W$+jets  &  $<  1$ 
\end{tabular}
\end{table}

As described in Sec.~\ref{ljets_mutag}, the calculation of the $W+$jets 
background for the $e+$jets/$\mu$ channel is determined via 
Eq.~\ref{eq:ljmwjb} where $N_3^{t\bar t}$ is obtained by applying the
$t\bar t$ tag rate to the measured excess for $e+$~3 or more jets as
determined from Berends scaling (Eq.~\ref{eq:ejmwjttcon}). In addition
to the uncertainty from Berends scaling of 10\%, there is a significant
uncertainty in the $t\bar t$ tag rate determined from MC, leading to a
total uncertainty of 40\% which has been included under the Berends
scaling heading for the $e+$jets/$\mu$ channel.
Note that Berends scaling is not used for the $\mu+$jets/$\mu$ channel.

\begin{table*}
\squeezetable 
\caption{Efficiency times branching fraction ($\varepsilon \times {\cal B}$) 
and statistical and systematic uncertainties (all in percent) for 
$m_t=170$ GeV$/c^2$.
\label{tab:syssig}}
\vskip 0.5cm
\begin{tabular}{l|ccccccccc}
    & $ee$ & $e\mu$ & $\mu\mu$ & $e\nu$ & $e+$jets & $\mu+$jets 
& $e+{\rm jets}/\mu$ &
$\mu+{\rm jets}/\mu$ & all-jets \\
\hline 
$\varepsilon \times {\cal B}$ & 0.165 & 0.349 & 0.106 & 0.263 & 1.288 & 0.911& 0.568 & 0.371 & 1.963 \\
\hline
Statistical   & 0.002 & 0.004 & 0.002 & 0.008 & 0.020 & 0.046& 0.017 & 0.037 & 0.151 \\
Energy scale  & 0.011 & 0.020 & 0.008 & 0.066 & 0.169 & 0.137& 0.026 & 0.008 & 0.112 \\  
Electron id  & 0.008 & 0.009 &  --   & 0.006 & 0.044 &  --  & 0.022 &  --   &  --   \\
High-$p_{T} \mu$ id &  --   & 0.033 & 0.007 &  --   &  --   & 0.098&  --   & 0.048 &  --   \\
Tag $\mu$ id  &  --   &  --   &  --   &  --   & 0.005 &  --  & 0.022 & 0.022 & 0.137 \\
$e$+jets trigger & 0.001 & 0.018 &  --   & 0.008 & 0.058 &  --  & 0.028 &  --   &  --   \\
$\met+$jets trigger &  --   &  --   &  --   &  --   &  --   & 0.046&  --   & 0.019 & 0.098 \\ 
Mult. Int. & 0.016 & 0.057 & 0.005 & 0.014 & 0.000 & 0.000& 0.000 & 0.000 &  --   \\ 
Generator (kin) & 0.008 & 0.017 & 0.005 & 0.032 & 0.126 & 0.203& 0.034 & 0.034 &  --   \\
Generator ($b$ tag) &  --   &  --   &  --   &  --   & 0.021 & 0.017& 0.057 & 0.037 &  --   \\
$Z$ fitter &  --   &  --   & 0.003 &  --   &  --   &  --  &  --   & 0.019 &  --   \\ 
\hline
Total error & 0.023 & 0.074 & 0.013 & 0.076 & 0.225 & 0.272& 0.084 & 0.086 & 0.253 \\
\end{tabular}
\end{table*}

\subsubsection{Electron misidentification rate (mis-id $e$)}
\label{ss:fakeeerr}

As described in Secs.~\ref{e-id}, ~\ref{dilep}, and ~\ref{ljets_mutag}, 
determination of the background from multijet events in which a jet is 
misidentified as an electron is based on an independent measurement of 
the electron ``misidentification rate.'' 
For the $ee$, $e\mu$, and $e\nu$ channels, 
these misidentification rates were determined by counting the number 
of loose electron
candidates found in a sample of QCD multijet events containing one 
electromagnetic cluster that passed the extra-loose electron 
identification requirements. The uncertainties on this procedure are 
dominated by the statistics of the extra-loose electron sample.
For the $e+$jets/$\mu$ sample, the misidentification rate 
described in Sec.~\ref{ljets_mutag} depends on the jet 
multiplicity from which an uncertainty of 21\% was estimated.
Note that for the $e+$jets/topo channel, the background from 
QCD multijet events is handled differently and did not make use of 
an electron ``misidentification rate.''

\subsubsection{Mismeasured $\met$}
\label{ss:fakmeterr}

As noted in Sec.~\ref{ee}, for the $ee$ channel the background from
$Z(\ra ee)+$jets is determined directly from data, but since $Z(\ra ee)+$jet 
events have no real $\met$, a $\met$ mis-measurement rate, computed from 
QCD multijet
data as a function of jet multiplicity, is applied. The uncertainty on 
this procedure is obtained by varying the triggers and selection criteria 
used to collect the initial multijet sample, and is assigned a value of 15\%.

\subsubsection{Tag rate}
\label{ss:tagrate}

The $W+$jets background to the $\ell+$jets/$\mu$ channels is 
obtained, as a function of jet $E_T$ and $\eta$, by multiplying 
the number of (QCD multijet and $t\bar t$ subtracted) 
untagged $\ell+$jets events
by a tag rate determined from multijet data.
As described in Sec.~\ref{ljets_mutag}, the accuracy of the tag rate 
was studied by applying it to a number of different 
data sets and comparing the predicted and observed values 
(see Fig.~\ref{fig:tagtest}). Variation not due to statistics is 
calculated to be 8.2\%~\cite{ptthesis} and rounded upward to 10\%.

\subsubsection{Muon misidentification rate (mis-id $\mu$)}
\label{ss:fakmuerr}

The $\mu+$jets/topo and $\mu+$jets$/\mu$ channels both employ
the use of an ``isolated muon misidentification rate'' to determine 
the background from 
QCD multijet events. As described in Sec.~\ref{ljets_topo}, 
this misidentification rate 
is dependent on the jet multiplicity and is computed from samples of QCD 
multijet events with $\MEt \leq 20$ GeV as the ratio of the number of 
{\sl isolated}-$\mu$ + $n$ jet events to the number of 
{\sl non-isolated}-$\mu$ + ($n+1$) jet events. The primary source of 
uncertainty in this measurement is the statistical precision of the control 
samples, leading to an uncertainty of 30\% for the four-jet samples used for 
the $\mu+$jets/topo channel and 20\% for the three-jet samples used 
for the $\mu+$jets$/\mu$ channel.

\subsubsection{$\mu$ multijet}
\label{ss:muqcderr}

Both the $\mu+$jets/topo and $\mu+$jets/$\mu$ channels have 
background from QCD multijet events which contain a muon 
from $b$ or $c$ quark decay that is 
misidentified as an isolated muon. Both channels rely on 
multijet control 
samples to model this background. Differences in key kinematic distributions 
between the multijet control samples and the true background are accounted for
in the uncertainty discussed here.
As discussed in Sec.~\ref{ljets_topo}, the QCD multijet background to the 
$\mu+$jets/topo channel is obtained by applying a {\sl survival
probability} to pass the $E_T^L$, ${\cal A}$, and $H_T$ cuts (determined 
from $n+1$ jet data) to an $n$ jet control sample. Comparisons of 
the ${\cal A}$ and $H_T$ distributions for the $n$ and $n+1$ jet sample
lead to an estimated uncertainty of 20\%.
Similarly, for the $\mu+$jets/$\mu$ channel, the QCD multijet background is 
determined by applying a tag probability to the jets in a multijet control 
sample of {\sl non-isolated} $\mu$ + 3 jet events on which all kinematic 
cuts (including ${\cal A}$ and $H_T$) have been applied. 
Differences in the ${\cal A}$ and $H_T$ distributions between the multijet 
control sample and the true background sample lead to the assignment of 
an uncertainty of 20\%.

\subsubsection{$\mu$ tag probability}
\label{ss:tagproberr}

As described in Sec.~\ref{ljets_mutag}, for the $\mu+$jets/$\mu$ channel
the QCD multijet background is determined by applying a tag 
probability, derived from $t\bar t$ MC events, to a multijet control sample. 
An uncertainty of 15\% is assigned 
to this tag probability to account for the fact that the probability is
averaged over the CF and EF detector regions and that the MC sample has not
been subjected to the corrections described in Sec.~\ref{muid-eff}.

\subsubsection{$Z$ boson mass fitter ($Z$ fitter)}
\label{ss:zfiterr}

As described in Secs.~\ref{mumu} and ~\ref{ljets_mutag}, the $\mu\mu$ and 
$\mu+$jets/$\mu$ channels reduce their background from $Z\ra\mu\mu$ events
by cutting on a minimized $\chi^2$ fit for the muon pair mass to give 
$M_Z$ and for $\metcal$ to equal the $p_T$ of the $Z$ boson, in effect 
``fitting for the $Z$.'' Consideration of the muon momentum resolution and 
variation of the $\met$ resolution parameterizations used for both data and
MC, lead to the estimate of a systematic uncertainty of 10\% for this
procedure.


\section{Cross Section Results}
\label{crsec}

\begin{table*}
\squeezetable 
\caption{Efficiency $\times$ branching fraction (in percent) for all eight
leptonic channels for $m_t$ = 140--200 \gevccn. Uncertainties correspond to 
statistical and systematic contributions added in quadrature.
\label{tab:ebrexp}}
\vskip 0.5cm
\begin{tabular}{c|c|c|c|c|c|c|c}
$m_t$ (GeV/$c^2$) & 140 & 150 & 160 & 170 & 180 & 190 & 200 \\
\hline 
$ee$             & $0.106 \pm 0.015$ & $0.129 \pm 0.018$ & $0.152 \pm 0.021$  
  & $0.165 \pm 0.023$ & $0.187 \pm 0.026$ & $0.199 \pm 0.028$  
  & $0.209 \pm 0.029$ \\  
$e\mu$           & $0.214 \pm 0.045$ & $0.252 \pm 0.053$ & $0.302 \pm 0.064$  
  & $0.349 \pm 0.074$ & $0.389 \pm 0.082$ & $0.429 \pm 0.092$  
  & $0.440 \pm 0.093$ \\  
$\mu\mu$         & $0.055 \pm 0.008$ & $0.069 \pm 0.010$ & $0.088 \pm 0.012$  
  & $0.106 \pm 0.013$ & $0.119 \pm 0.016$ & $0.133 \pm 0.018$  
  & $0.137 \pm 0.018$ \\  
$e\nu$           & $0.156 \pm 0.046$ & $0.201 \pm 0.058$ & $0.224 \pm 0.065$  
  & $0.263 \pm 0.076$ & $0.315 \pm 0.091$ & $0.335 \pm 0.096$  
  & $0.360 \pm 0.104$ \\  
$e+$jets/topo    & $0.597 \pm 0.256$ & $0.801 \pm 0.264$ & $1.036 \pm 0.236$  
  & $1.288 \pm 0.225$ & $1.479 \pm 0.197$ & $1.558 \pm 0.174$  
  & $1.703 \pm 0.158$ \\  
$\mu$+jets/topo  & $0.451 \pm 0.194$ & $0.621 \pm 0.235$ & $0.810 \pm 0.271$  
  & $0.911 \pm 0.272$ & $1.058 \pm 0.276$ & $1.164 \pm 0.276$  
  & $1.291 \pm 0.278$ \\  
$e+$jets/$\mu$   & $0.364 \pm 0.069$ & $0.469 \pm 0.078$ & $0.491 \pm 0.077$  
  & $0.568 \pm 0.084$ & $0.656 \pm 0.095$ & $0.709 \pm 0.099$  
  & $0.721 \pm 0.102$ \\  
$\mu+$jets/$\mu$ & $0.255 \pm 0.064$ & $0.262 \pm 0.065$ & $0.295 \pm 0.071$  
  & $0.371 \pm 0.086$ & $0.405 \pm 0.093$ & $0.423 \pm 0.096$  
  & $0.443 \pm 0.100$ \\  
\end{tabular}
\end{table*}

\begin{figure}
\vbox{
\vskip -1.2cm
\centerline{\psfig{figure=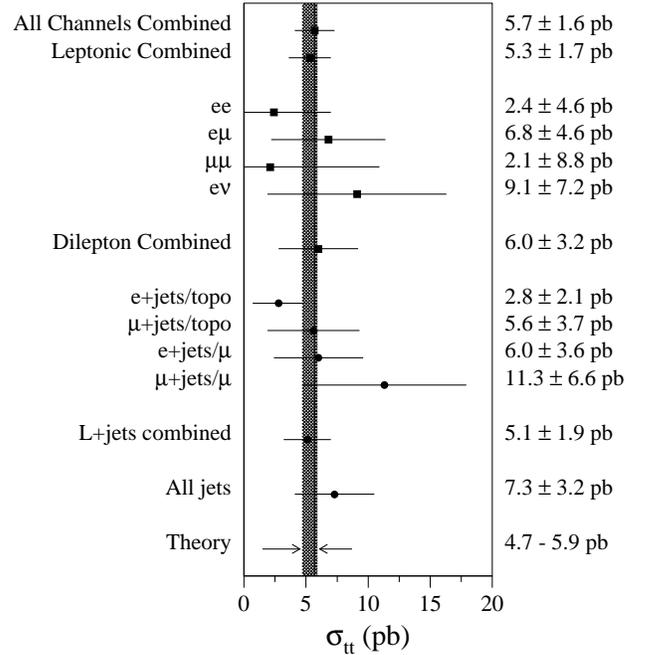,width=4.0in}}
\caption{ D\O\ measured $t \bar t$ production cross section values
for all channels, assuming a top quark mass of 172.1 \gevccn. 
The vertical line corresponds to the cross section for all channels
combined and the shaded band shows the range of theoretical 
predictions~\protect\cite{laenen,berger3,bonciani,kidonakis}.}
\label{fig:csec2item}}
\end{figure}

The preceding sections describe nine analyses that extract
data samples rich in $t \bar t$ events. 
For an individual channel $i$, the cross section is determined from the
relation 
\begin{equation}
 \sigma(m_t)_{{t\bar t},i} = \frac{N_i - 
                             \left(\displaystyle\sum_j B_j \right)_i}
                             { A(m_t)_i \cdot {\cal L}_i } 
\label{eq:cseccalc}
\end{equation}
where $A(m_t)$ is the acceptance (efficiency times branching
fraction) for a top quark mass of $m_t$, ${\cal L}_i$ is the integrated
luminosity, $N_i$ is the number of observed
events, and $B_j$ is the number of expected background events from source $j$.
The efficiency times branching fraction values for all eight leptonic 
channels for $m_t$ = 140--200 \gevcc are given in Table~\ref{tab:ebrexp}. 
The numbers of observed
events, along with those expected from signal and background, 
the integrated luminosity, and the final measured cross sections 
(for $m_t$ = 172.1 \gevccn) for each channel are
summarized in Table~\ref{tab:summary}. The value of $m_t$ = 172.1 \gevcc 
is D\O's combined dilepton and lepton+jets mass 
measurement~\cite{ljmprd,mllprd}. 
The cross section 
results for the various channels (and several combinations) are compared 
in Fig.~\ref{fig:csec2item}, and are seen to be in good agreement with one
another and with theoretical expectations~\cite{laenen,berger3,bonciani}. 
Complete details of the 39 observed leptonic events are given in 
Ref.~\cite{cand_params}.

\begin{table*}
\caption{Summary of observed number of events, expected signal and background, 
integrated luminosity, and cross section for all nine channels at 
$m_t = 172.1$ \gevccn. Uncertainties correspond to statistical and 
systematic contributions added in quadrature.
\label{tab:summary}}
\vskip 0.5cm
\begin{tabular}{c|c|c|c|c|c|c}
 &              & Total    & Signal                 & Total       
 &                                    &             \\
 & $N_{\rm obs}$& sig+bkg & ($m_t$ = 172.1)        & background  
 & $\int{\cal L}dt$ (${\rm pb}^{-1})$ & $\sigma$(pb)\\
\hline 
$ee$                & 1 & $1.68 \pm 0.23$ & $1.20 \pm 0.17$ & $0.48 \pm 0.10$ 
        & $130.2 \pm 5.6$ & $2.37 \pm 4.58$ \\  
$e\mu$              & 3 & $2.45 \pm 0.53$ & $2.19 \pm 0.47$ & $0.26 \pm 0.16$ 
        & $112.6 \pm 4.8$ & $6.81 \pm 4.59$ \\  
$\mu\mu$            & 1 & $1.39 \pm 0.30$ & $0.64 \pm 0.09$ & $0.75 \pm 0.24$ 
        & $108.5 \pm 4.7$ & $2.11 \pm 8.79$ \\  
$e\nu$              & 4 & $2.87 \pm 0.71$ & $1.68 \pm 0.49$ & $1.19 \pm 0.38$ 
        & $112.3 \pm 4.8$ & $9.12 \pm 7.23$ \\  
\hline
Dilepton combined   & 9 & $8.39 \pm 1.48$ & $5.71 \pm 1.07$ & $2.69 \pm 0.66$ 
        &        --        & $6.02 \pm 3.21$ \\  
\hline
$e+$jets/topo       & 9 & $13.16 \pm 1.67$& $8.64 \pm 1.47$ & $4.51 \pm 0.91$ 
        & $119.5 \pm 5.1$ & $2.83 \pm 2.05$ \\  
$\mu$+jets/topo     &10 & $9.84 \pm 1.62$ & $5.52 \pm 1.62$ & $4.32 \pm 1.04$ 
        & $107.7 \pm 4.6$ & $5.60 \pm 3.71$ \\  
$e+$jets/$\mu$      & 5 & $4.65 \pm 0.54$ & $3.59 \pm 0.55$ & $1.05 \pm 0.40$ 
        & $112.6 \pm 4.8$ & $5.98 \pm 3.56$ \\  
$\mu+$jets/$\mu$    & 6 & $3.62 \pm 0.52$ & $2.22 \pm 0.52$ & $1.40 \pm 0.23$ 
        & $108.0 \pm 4.6$ & $11.27\pm 6.60$ \\  
\hline
$\ell+$jets combined &30 & $31.27 \pm 3.52$& $19.98\pm 3.52$ & $11.28\pm 1.97$ 
        &        --        & $5.10 \pm 1.85$ \\  
\hline
{\bf Leptonic combined}  &{\bf 39} & {\boldmath $39.66 \pm 4.65$} & 
   {\boldmath $25.69 \pm 4.41$} & {\boldmath $13.97 \pm 2.22$} & -- & 
   {\boldmath $5.31 \pm 1.72$}  \\  
\hline
all-jets            & 41  & $37.40 \pm 2.92$ & $12.60 \pm 2.12$ & 
   $24.80 \pm 2.37$ & $117.9 \pm 5.1$    & $7.33 \pm 3.20$  \\  
\hline
{\bf All channels total}  & {\bf 80} & {\boldmath $77.06 \pm 6.19$} & 
   {\boldmath $38.29 \pm 5.34$} & {\boldmath $38.77 \pm 3.32$}  
        & -- & {\boldmath $5.69 \pm 1.60$}            \\  
\end{tabular}
\end{table*}

\begin{figure}
\centerline{\psfig{figure=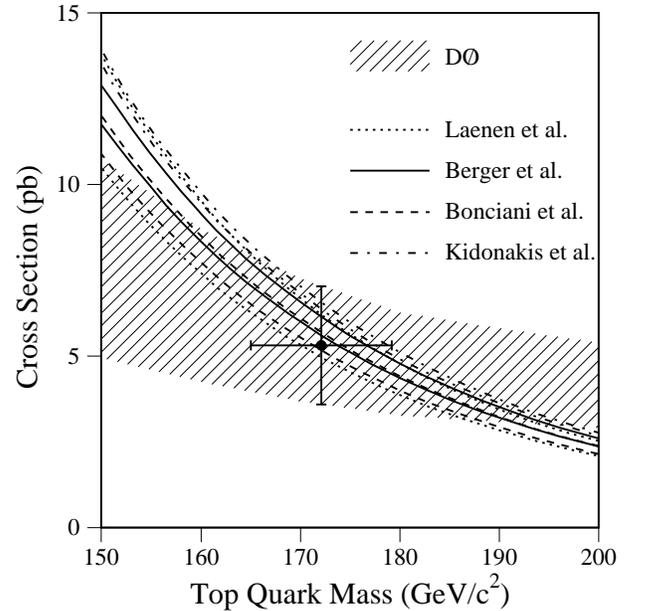,width=3.2in}}
\vskip 0.1cm
\caption{ D\O\ measured $t \bar t$ production cross section as a function 
of the top quark mass ($m_t$) for the leptonic channels (shaded band) and 
at the D\O\ measured top quark mass (point with error bars). 
Also shown are the upper ($\mu=m_t/2$) and lower ($\mu=2m_t$) bounds for 
four theoretical predictions of the $t\bar t$ cross section 
as a function of $m_t$: 
Laenen {\it et al.} [LL] (dotted lines)~\protect\cite{laenen}, 
Berger {\it et al.} [LL] (solid lines)~\protect\cite{berger3}, 
Bonciani {\it et al.} [NLL] (dashed lines)~\protect\cite{bonciani}, 
and Kidonakis [NNLL] (dot-dash lines)~\protect\cite{kidonakis},
where the labels LL, NLL, and NNLL indicate leading-log,
next-to-leading-log, and next-to-next-to-leading-log resummation 
calculations respectively.}
\label{fig:csec1}
\end{figure}

The combined $t \bar t$ production cross section is determined from the
analog of Eq.~\ref{eq:cseccalc}:
\begin{equation}
  \sigma(m_t)_{t\bar t} = {{\displaystyle\sum_i N_i - \sum_j B_j }
  \over { \displaystyle\sum_i A(m_t)_i \cdot 
  {\cal L}_i }},
\label{eq:totcseccalc}
\end{equation}
where the sum $i$ is over all nine channels and the sum $j$ is over 
all background sources in all nine channels. Recall (see Sec.~\ref{intro}) that
all channels are, by construction, orthogonal. As discussed in 
Appendix~\ref{correrr}, the determination of the cross section takes into 
account the correlated uncertainties between the inputs to 
Eq.~\ref{eq:totcseccalc}. Plotting the cross section values and their
uncertainties for a range of top quark masses gives the band shown in
Fig.~\ref{fig:csec1}. Also shown are the theoretical expectations
for the $t \bar t$ cross section as a function of 
$m_t$~\cite{laenen,berger3,bonciani,kidonakis}. Combining this cross section 
result with the combined D\O\ dilepton and lepton+jets mass 
measurement~\cite{ljmprd} gives the point with
error bars shown in Fig.~\ref{fig:csec1}.

\begin{figure*}
\vbox{
\vskip -0.6cm
\centerline{\hskip 0.25 in \psfig{figure=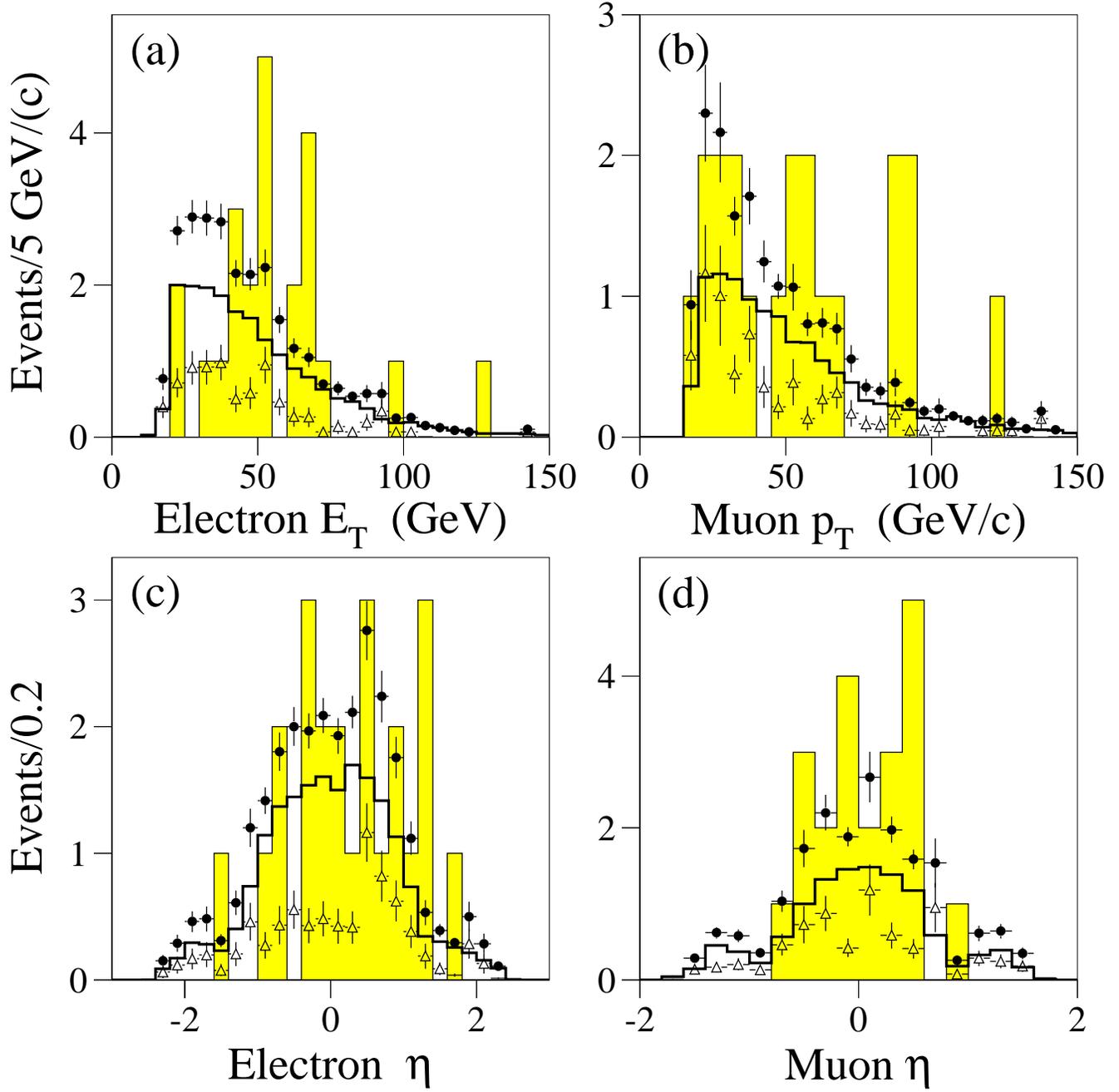,width=7.6in}}
\caption{ $E_T$($p_T$) and $\eta$ distributions for electrons (muons) for
leptonic $\ttb$ candidates (shaded histogram), $\ttb$ MC [\progname{HERWIG}, 
$m_t$ = 170 \gevccn] (unshaded histogram), expected background 
(open triangles), and expected signal plus background (solid circles). 
The measured muon $p_T$ for $e\mu$ candidate 58796-7338(417) is 
280.0 GeV/c and is therefore off scale in plot (b). As given in 
Ref.~\protect\cite{cand_params}, the event label corresponds to run 
number and event 
number (in an early event numbering scheme, this event became well know as
``event 417'' and thus retains this label parenthetically).
\label{fig:dvmcptlep}}}
\end{figure*}

\begin{figure*}
\vbox{
\vskip -1.0cm
\centerline{\hskip 0.25 in \psfig{figure=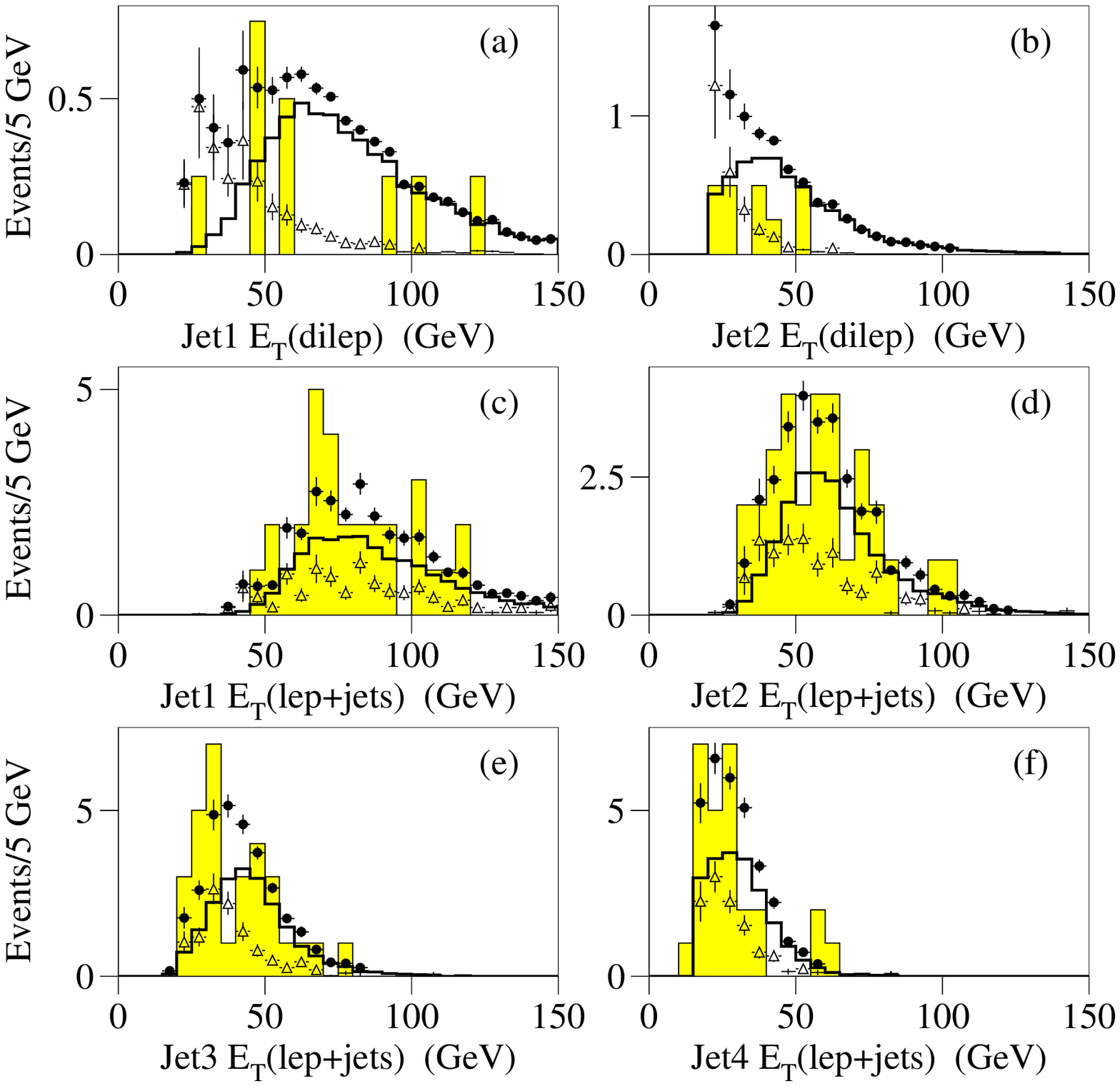,width=7.6in}}
\caption{ Jet $E_T$ distributions for dilepton (a)--(b) and $\ell$+jets 
(c)--(f) $\ttb$ candidates (shaded histogram), $\ttb$ MC [\progname{HERWIG}, 
$m_t$ = 170 \gevccn] (unshaded histogram),
expected background (open triangles), and expected signal plus background
(solid circles). The dilepton candidate histograms [(a)--(b) shaded] have
been multiplied by a factor of 0.25 for presentational clarity.
\label{fig:dvmcjetet}}}
\end{figure*}

\begin{figure*}
\vbox{
\vskip -0.4cm
\centerline{\hskip 0.25 in \psfig{figure=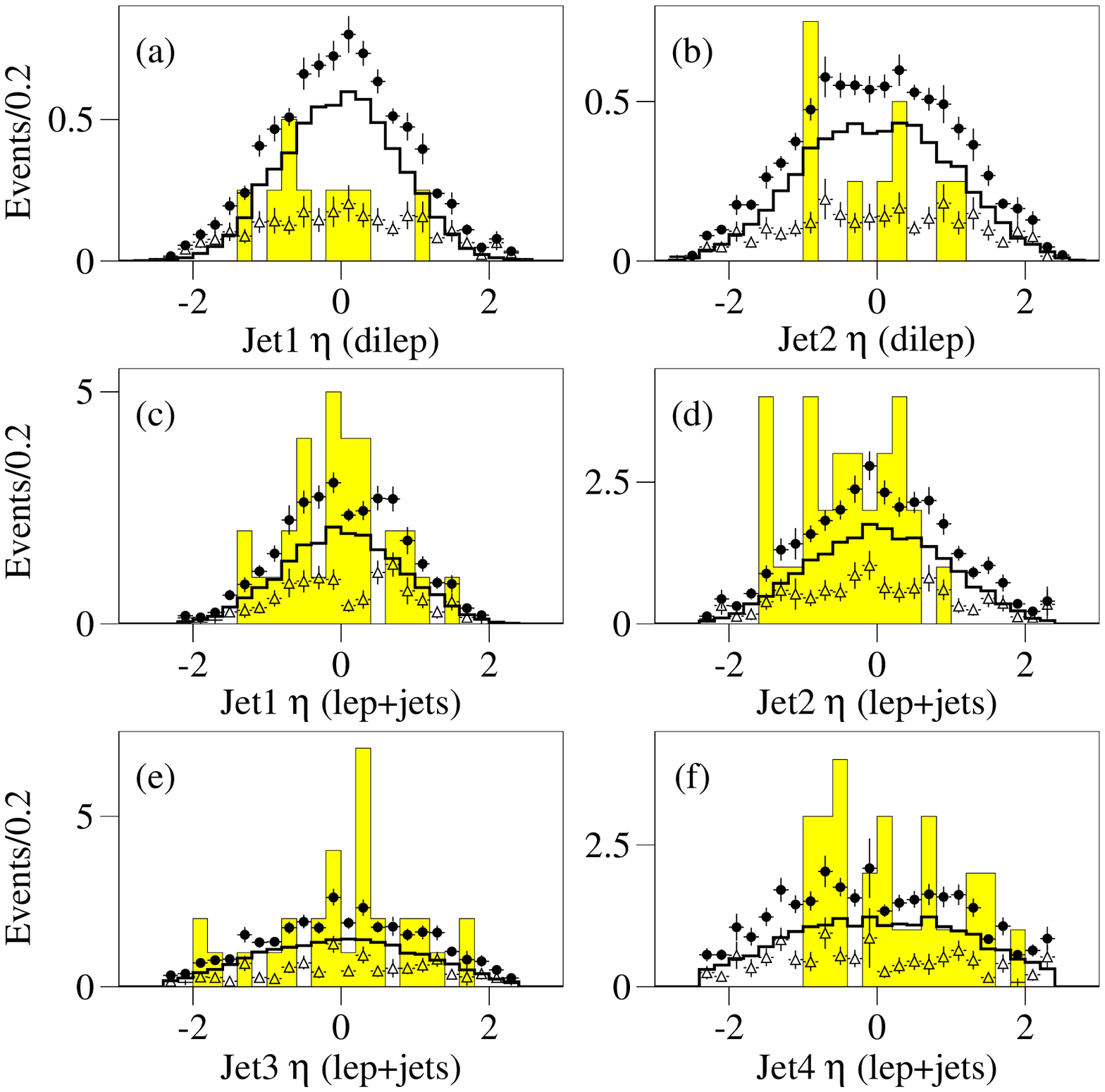,width=7.6in}}
\caption{ Jet $\eta$ distributions for dilepton (a)--(b) and $\ell$+jets 
(c)--(f) $\ttb$ candidates (shaded histogram), $\ttb$ MC [\progname{HERWIG}, 
$m_t$ = 170 \gevccn] (unshaded histogram),
expected background (open triangles), and expected signal plus background
(solid circles). The dilepton candidate histograms [(a)--(b) shaded] have
been multiplied by a factor of 0.25 for presentational clarity.
\label{fig:dvmcjeteta}}}
\end{figure*}

\begin{figure*}
\vbox{
\vskip -0.7cm
\centerline{\hskip 0.25 in \psfig{figure=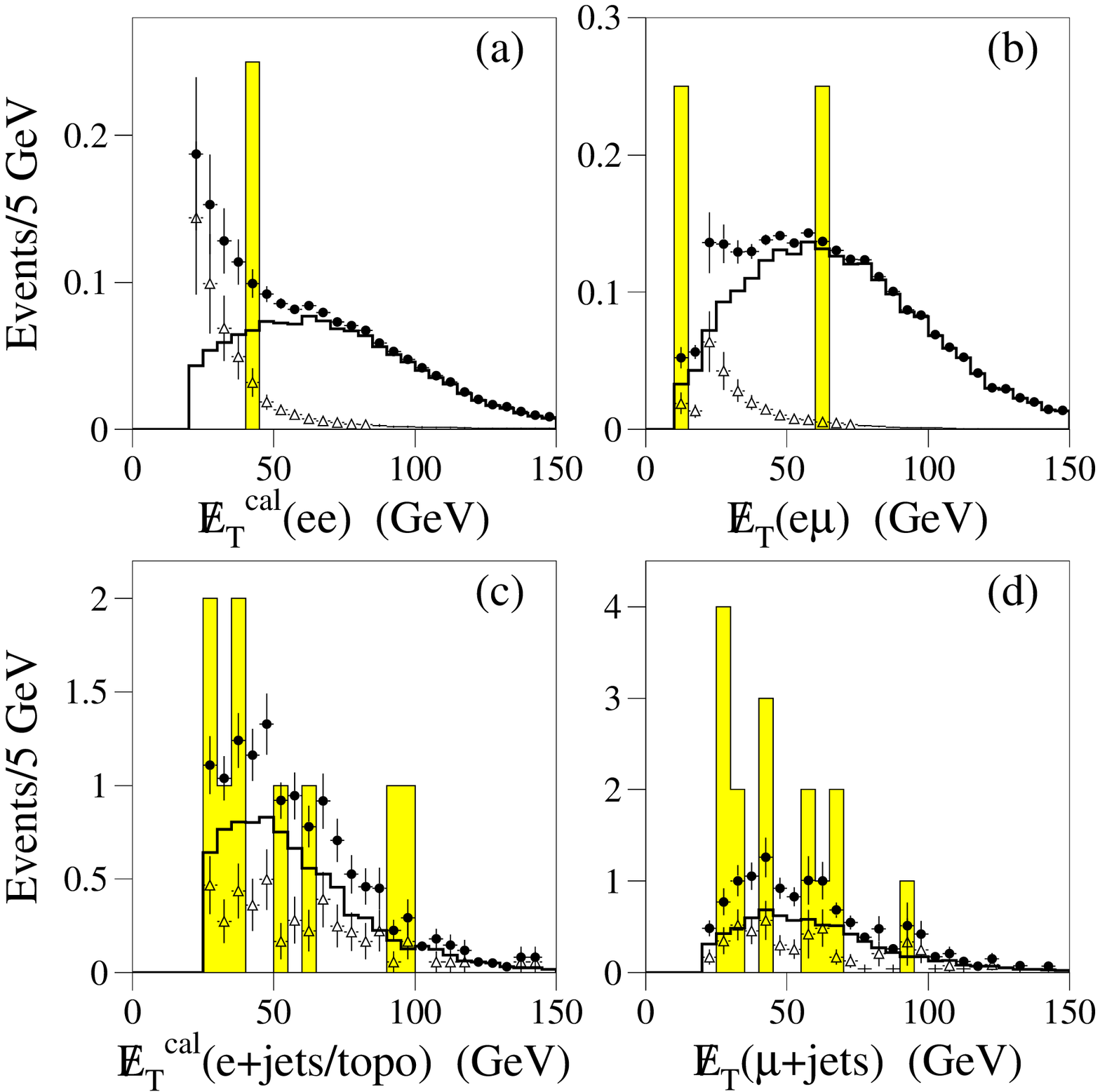,width=7.6in}}
\caption{  $\metcal$ distributions for the $ee$ (a) and $e+$jets/topo
(c) channels, and $\met$ distributions for the $e\mu$ 
(b) and $\mu+$jets/topo and $\mu+$jets/$\mu$ (d) channels: $\ttb$ candidates 
(shaded histogram), $\ttb$ MC [\progname{HERWIG}, 
$m_t$ = 170 \gevccn] (unshaded histogram), expected background 
(open triangles), and expected signal plus background
(solid circles). The $ee$ and $e\mu$ candidate histograms [(a)--(b) shaded] 
have been multiplied by a factor of 0.25 for presentational clarity.
The measured \MEt for $e\mu$ candidate 58796-7338(417) is 
182.9 GeV and is therefore off scale in plot (b).
\label{fig:dvmcmet}}}
\end{figure*}

In addition to the final cross section and mass result, it is also 
instructive to compare the properties of the $t\bar t$ candidate events 
with expectations. These is examined in Figs.~\ref{fig:dvmcptlep} -- 
\ref{fig:dvmcmet} which show the distributions
of the $t\bar t$ candidates (shaded histograms), $\ttb$ Monte Carlo 
(unshaded histogram), expected background (open triangles), and expected 
signal plus background (solid circles) for various quantities. Overall, 
these plots show better
agreement between the candidate and $t\bar t$ + background distributions
than between the candidate and the background-only distributions.

\section{Conclusions}
\label{concl}

Nine analyses have been described which select event samples dominated by
$t \bar t$ production. A total of 39 events are found in the leptonic
channels with an expected background of $14.0 \pm 2.2$.
Combining these results with the integrated luminosity and signal efficiency 
(at $m_t$ = 172.1 \gevccn), the $t \bar t$ production cross section for
the leptonic channels is determined to be 
\begin{equation}
5.31 \pm 1.34 ({\rm stat}) \pm 1.08 ({\rm sys}) \ {\rm pb}.
\label{eq:leptocsec}
\end{equation}
This cross section differs slightly from the value reported in 
Ref.~\cite{d0csec1997} due primarily to an updated luminosity normalization,
and to a lesser extent to minor changes in the background estimation for some 
channels and to the use of a slightly different top mass.

For the all-jets channel, summarized in Sec.~\ref{alljets} and described in 
detail in Ref.~\cite{alljets}, a total of 41 events are found with 
an expected background of $24.8 \pm 2.4$ events. 
Combining the leptonic and all-jets channels
gives a total of 80 candidates with an expected background of 
$38.8 \pm 3.3$ events. This combination results in a $t \bar t$ production 
cross section of
\begin{equation}
5.69 \pm 1.21 ({\rm stat}) \pm 1.04 ({\rm sys}) \ {\rm pb}. 
\label{eq:allchancsec}
\end{equation}

As can be seen in Fig.~\ref{fig:csec2item}, the $t\bar t$ production cross 
sections obtained for the individual channels are in good agreement with the 
combined cross section and with that from 
theory~\cite{laenen,berger3,bonciani}. And as shown in 
Fig.~\ref{fig:top_his}(b), the combined cross section is in excellent 
agreement with D\O's previously reported values. The current level of 
uncertainty on QCD predictions for the $t\bar t$ production cross 
section~\cite{berger3,bonciani} is seen in Fig.~\ref{fig:csec1} to be 
about $\pm 0.3$ pb, less than 20\% of the current experimental uncertainty.
Run II of the Fermilab Tevatron is expected to provide an experimental 
uncertainty on the $t\bar t$ cross section of around 
$\pm 9\%$ ($\approx 0.6$ pb) in 2 ${\rm fb}^{-1}$, limited by 
systematic uncertainties~\cite{tev2000}. This will begin to place 
restrictions on the various QCD predictions 
and provide stringent tests for non-standard production and decay mechanisms.
In the longer term, the systematic limitations on the measurement of the 
$t\bar t$ production cross section at the CERN Large Hadron Collider is 
expected to be less than 10\%~\cite{lhc}.

\section*{Acknowledgements}
%
We thank the staffs at Fermilab and collaborating institutions, 
and acknowledge support from the 
Department of Energy and National Science Foundation (USA),  
Commissariat  \` a L'Energie Atomique and 
CNRS/Institut National de Physique Nucl\'eaire et 
de Physique des Particules (France), 
Ministry for Science and Technology and Ministry for Atomic 
   Energy (Russia),
CAPES and CNPq (Brazil),
Departments of Atomic Energy and Science and Education (India),
Colciencias (Colombia),
CONACyT (Mexico),
Ministry of Education and KOSEF (Korea),
CONICET and UBACyT (Argentina),
The Foundation for Fundamental Research on Matter (The Netherlands),
PPARC (United Kingdom),
Ministry of Education (Czech Republic),
A.P.~Sloan Foundation,
NATO, and the Research Corporation.
\appendix
\section{Energy Scale Corrections}
\label{escale}

Gluon radiation and fragmentation can alter a parton's original energy 
and direction before its remnants interact and are measured in the  
calorimeter (\emeas). Also, accompanying spectator interactions, not 
associated with the hard scattering, can deposit energy within a jet.
In addition, fluctuations in interactions in the detector can provide 
changes to \emeas.  For example, emitted particles, especially hadrons, 
can produce very wide showers in the calorimeter that can affect the 
fraction of energy ($1-S$) contained within any fixed size cone.
Also, most of the absorber is composed of uranium, the radioactive decay 
of which can deposit significant energy in the calorimeter. Finally,
the signal response ($R$) of the calorimeter to a jet is dominated by any 
difference of response to electrons (or photons) relative to charged hadrons
\cite{lowepi,calor96}, and by any energy deposited in uninstrumented
or non-uniform parts of the detector. The energy from spectator interactions
and uranium noise provides a total offset ($O$) that must be corrected.

Other than correcting for spectator interactions, only detector effects 
are considered in the energy calibration of jets. A jet's particle level 
energy (\eptcl) is defined as the energy of a jet found from final state 
particles using a similar cone algorithm to that used at the calorimeter 
level.
The calibration 
procedure~\cite{calor96} provides \eptcl\ from \emeas through the 
relationship:
\begin{equation}
E_{\rm ptcl}^{\rm jet} = \frac{(E_{\rm meas}^{\rm jet}-O)}{R(1-S)}.
\label{eq:totscale}
\end{equation}

The calibration is performed separately but identically in data and in
the Monte Carlo, with the $O$ and $S$ corrections applied to jet energies 
to extract the particle-level values \eptcl.

The offset $O$ is estimated as follows. The difference in \et\ density 
in ($\eta,\phi$) space between single and double-interaction events, which was
obtained with a minimum bias trigger, is defined to be the contribution
of the underlying event to single interactions.  The contribution from 
noise is obtained from this same sample by subtracting the \et\ for the 
underlying event from the \et\ density in single interactions. The total 
systematic uncertainty for the offset in \et\ density varies from 100 
$\mev$\ to 300 $\mev$, depending on the value of $\eta$.

The showering of a jet's fragments in the calorimeter causes energy 
to leak out of, or into, any jet cone. To quantify this effect, jets 
are generated using the \progname{herwig} program \cite{herwig}, and
reconstructed from their original final-state particles. These are 
subsequently replaced with electron or hadron showers from test beam data, 
and reconstructed using our cone algorithm, thereby defining a jet shower. 
The total shower energy is normalized to that of the original final-state 
particles. The ratio of the contained shower energy to that of the original
energy ($\equiv 1-S$) is calculated as a function of $\Delta R$. For central
jets with $\Delta R=0.5$, $S$ lies between 0.01 and 0.03, 
depending on jet energy, with a systematic uncertainty of 1\%  on $1-S$.


The \met\ in direct-photon candidate events (composed of true direct photon
events and background dijet events where a $\pi^0$ is back to back with a 
hadronic jet) is used to determine the response 
of the calorimeter to jets. Differences in response between the photon and the 
recoiling hadronic system produce an overall imbalance in transverse energy 
in the calorimeter, giving rise to \met. In these events, the absolute 
response $R$ of the leading jet can be determined from other well-measured 
quantities in the event:
\begin{equation}
R = 1 + \frac{\vec{\met} \cdot {\hat{n}}_{T}^{\gamma} }{\et^{\gamma}}, 
\label{eq:relres}
\end{equation}
where $E_T^{\gamma}$ ($> 15$ GeV) is the transverse energy of the photon and 
$\hat{n}_{T}$ is the unit vector along the photon's transverse momentum.
Since both the \et\ of the photon and the direction of the probe
jet are well-measured, the energy estimator $E'$ can be defined
\begin{equation}
    E' = E_T^{\gamma}{\rm cosh}(\eta_{\rm jet}).
\label{eq:eprime_calc}
\end{equation}
Measuring the correlation of $R$ with $E'$, and \emeas\ with $E'$, 
determines the dependence of $R$ on \emeas. 


Backgrounds to direct photons are a source of uncertainty for this analysis, 
particularly in collider data. Instrumental background from highly 
electromagnetic jets is limited by tight isolation criteria.
The residual bias to the measured response is 1.4\%.  The remaining 
background consists mostly of $W(\to e \nu)+$jets production, and 
corresponds to about 0.5\%.

In the calibration, because of the rapidly falling photon cross section,
energies of central jets are limited to $< 150$ GeV. Exploiting the 
uniformity of the detector, events with EC jets are used to measure the
response to higher energy jets. Sensitivity to the number of multiple 
interactions in an event results in a 2\% systematic uncertainty.  Because 
uncertainties in the measurement of the energy scale of low \et\ jets 
are quite large, a Monte Carlo direct-photon sample is used for this 
region, and provides a systematic uncertainty of about 3.5\%.

\begin{figure}[t]
\vskip 0.1cm
\centerline{\hskip -0.6cm \psfig{figure=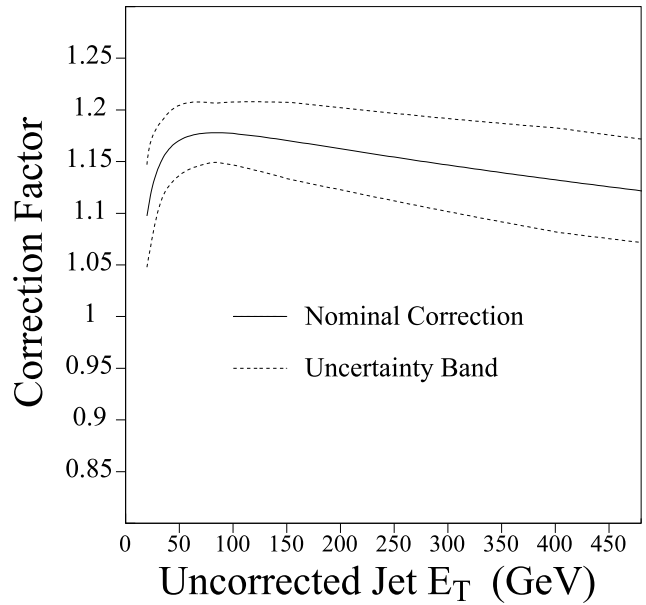,height=3.7in}}
\vskip -1.0cm
\caption{Total correction to the energy scale for central ($|\eta|<0.5$) jets.}
\label{fig:escale}
\end{figure}

The total correction is shown in Fig.~\ref{fig:escale}.  It rises to a 
maximum of 1.18 at $E_T \simeq 70$ GeV, followed by a slow fall to 1.12 
at $E_T \simeq 500$ GeV.  The upper and lower dashed lines correspond to 
one standard deviation upper and lower excursions on the total uncertainty,
taken as the addition in quadrature of the independent
effects discussed above.

\section{Main-Ring Veto}
\label{mrveto}

As noted in Sec.~\ref{triggers}, particles lost from the Main Ring can 
affect the measurements of the outer hadronic calorimeter and muon system.
The primary losses occur every 2.4 seconds when the protons are injected 
into the Main Ring and 0.3 seconds later as the beam, which is being
accelerated, passes through transition~\cite{jttm}.  The injection
from the Booster into the Main Ring causes the bunch to widen, and,
consequently, a greater 
amount of beam leaks out of the beampipe. After a few full circuits of 
the beam in the Main Ring, the bunch coalesces and is mainly confined 
to the beampipe. Additional losses need to be accounted for in the case 
when the passage of the proton beam coincides with the $p\bar p$ crossing 
in the Tevatron (which occurs every 3.5 $\mu$s).  With each pass, errant
particles from the bunch scatter outside the beampipe causing 
energy deposition in the outer layers of the calorimeter and 
multiple tracks in the muon system. Because the 
electromagnetic calorimeter and tracking systems are shielded from these 
losses, the electron triggers are not significantly impacted.
However, jet and especially muon triggers are affected, and it is 
necessary to veto events from jet, $\metcal$, and muon triggers that occur
during periods of Main-Ring activity. During the course of the run, several
schemes were used to eliminate such events without introducing unnecessary
deadtime:
\begin{itemize}
\item \progname{mrbs-loss (mrbs)}: The trigger is disabled for 
0.4 s after a proton bunch is injected into the Main Ring. This vetoes events
during injection and transition and provides a brief recovery time for the
muon and calorimeter systems. The typical deadtime for \progname{mrbs-loss} 
veto is $\approx 17$\%.
\item \progname{micro-blank (mb)}: The trigger is disabled for 
events where Main-Ring bunches are present during the livetime of the muon 
system which is $\approx \pm 800$ ns centered on the $p\bar p$ crossing time.
The calorimeter livetime is somewhat longer ($\approx 2 \mu$s), so this is 
therefore not completely efficient for vetoing events with Main-Ring energy 
in the
calorimeter. The typical deadtime for \progname{micro-blank} is $\approx 7$ \%.
\item \progname{max-live (ml)}: The trigger is disabled during periods
of overlap between \progname{mrbs} and \progname{mb}. This corresponds to
the first few passes of newly injected beam through the detector.
\item \progname{good-cal (gc)}: The trigger is disabled during periods
of overlap between \progname{mrbs} and \progname{mb} and during \progname{mb} 
periods of highest intensity beam leakage.  This leakage is measured by a 
set of scintillator arrays surrounding the Main-Ring beampipe upstream of 
the D\O\ detector.   
\item \progname{good-beam (gb)}: The trigger is disabled during 
periods of either \progname{mrbs} or \progname{mb}. \progname{good-beam} 
is the cleanest possible running condition.
\end{itemize}
The Main-Ring veto used for each trigger is given in 
Tables~\ref{tab:triggers_top_elex}--\ref{tab:tagratettrigs}.
However, by default, all channels required \progname{good-beam} for the 
offline analyses. As will be noted in Secs.~\ref{mrrecov}, \ref{ee}, and 
\ref{ljets_topo}, for the $ee$ and $e$+jets/topo channels it is
possible to remove this offline requirement on \progname{good-beam} and 
recover a significant fraction of the data lost to it.

\section{Main-Ring Recovery}
\label{mrrecov}

As noted in Sec.~\ref{triggers}, all triggers used in the
present analyses, being combinations of electron, muon, jet, and $\metcal$ 
triggers, suffer some loss from the vetoing of
events that coincide with activity in the Main Ring. 
Due to the location of the Main-Ring beam pipe within the detector, the
fine hadronic (FH) and electromagnetic sections of the calorimeter and the 
tracking systems are well shielded from 
this background, so electron and photon measurements are not 
significantly affected. However, hadronic jet (and thus $\metcal$) and muon 
measurements are affected. The effect on the hadronic calorimeter gives 
rise to fake jet backgrounds and mismeasured $\metcal$ arising from either
large positive signals, if the Main-Ring losses coincide with the Tevatron 
beam crossing (\progname{micro-blank}), or from large negative signals for 
Tevatron beam crossings that were preceded by Main-Ring losses
(\progname{mrbs}). In the latter case, the output 
voltage of the calorimeter preamps slowly decreases toward zero, causing 
the difference between a peak and the baseline to become negative.
As discussed in the following paragraphs, these effects on the hadronic 
calorimetry can be
minimized with the proper corrections. The effect on the muon system is to
decrease the overall muon-finding efficiency by less than 10\% during periods 
of Main-Ring activity, with most of the inefficiency localized to the 
regions near the Main Ring.

\begin{figure} 
\vskip 0.1cm
\centerline{\hskip -0.5cm \psfig{figure=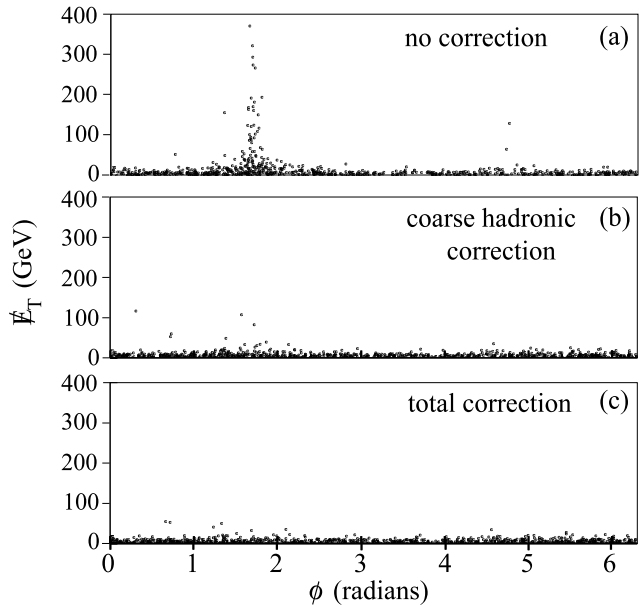,height=4.8in}}
\vskip -3.9cm
\caption{Effect of Main-Ring $\metcal$ corrections: $\metcal$ vs $\phi$ for 
\progname{mrbs} events for (a) no correction, (b) CH correction, and (c)
total correction.}
\label{fig:mrrecov}
\end{figure}

The $ee$, $e$+jets/topo, and $\mu$+jets/topo channels 
all retrieve Main-Ring events and correct jets and $\metcal$ in the same way.

Normal jets (those from periods when the Main Ring is not active 
[\progname{good-beam}]) typically have at 
most 10\% of their energy in the outer, coarse hadronic (CH) region of the
calorimeter. During periods of Main-Ring activity (\progname{micro-blank}), 
a significant enhancement is seen in the number of jets with $\phi$ values 
close to that of the Main Ring ($\phi \approx 1.7$), and the vast majority 
of these jets have CH energy fractions between 60 and 90\%. Therefore, for 
jets in the vicinity of the Main Ring ($1.5 < \phi < 2.0$) that have 
CH energy fractions greater than 20\%, the CH energy is simply removed 
\cite{bkthesis}. 
This correction causes the jet $E_T$ to be biased low due to the fact that 
some ``real'' CH energy is also removed, but as this only affects a small 
fraction ($< 2\%$) of jets in Main-Ring events, it is not a significant 
concern. Since jets in top quark events are very energetic, the removal of the 
CH energy typically leaves the jet $E_T$ well above threshold. 
Therefore, the loss in efficiency is minimal, affecting only a small fraction 
of the 2\% of jets in Main-Ring events that are corrected. For events with
large negative signals (\progname{mrbs}) there is also only a small reduction 
in efficiency, so jets in these events are not corrected.

For $\metcal$ the situation is more complicated and requires corrections for
both the large positive signals in \progname{micro-blank} events and the large 
negative signals in \progname{mrbs-loss} events.
The vast majority of these events are corrected simply by removing the
CH energy from the $\metcal$ calculation.
This can be seen in Fig.~\ref{fig:mrrecov} which shows $\metcal$ vs $\phi$ 
for \progname{mrbs} events. Fig.~\ref{fig:mrrecov}(a) is without any 
correction and shows a large number of
events with large $\metcal$ pointing towards the Main Ring. 
As shown in Fig.~\ref{fig:mrrecov}(b), where the CH energy has been 
removed, most of the events with large $\metcal$ pointing towards the Main 
Ring have been corrected.
Although this procedure does remove some positive energy that 
would normally be included, it does not degrade the $\metcal$ resolution 
appreciably due to the fact that normal (non-Main-Ring) events 
characteristically have a low ($< 10\%$) CH energy fraction. Unfortunately, 
some events with large $\metcal$ in the vicinity of the Main Ring persist
after the removal of the CH energy. These events appear primarily in the
region of the intercryostat detector (ICD) and massless gap (MG). To correct 
such events, a vector sum is calculated for all cells in the ICD and MG
that have negative energy below a given threshold, and this vector is then
subtracted from the $\metcal$ vector. These thresholds were determined from
comparisons of the negative energy spectra of the ICD and MG cells for 
\progname{good-beam} (non-Main-Ring) and \progname{mrbs-loss} events
\cite{bkthesis}. In addition to removing all unwanted negative energy, 
as seen in Fig.~\ref{fig:mrrecov}(c), this procedure brings the $\metcal$
resolution to an approximately normal level.

\section{$\lowercase{e}\mu$ Neural Network Analysis}
\label{emuNN}

\begin{table*}
\caption{Number of observed and expected events passing at each cut level of 
the $e\mu$ neural network analysis. Expected number of $\ttb$ events 
are for $m_t = 170 \gevcc$. Uncertainties correspond to statistical 
and systematic contributions added in quadrature.
\label{tab:emunncuts}}
\vskip 0.5cm
\begin{tabular}{l|c|c|c|c|c}
\multicolumn{6}{c}{Number of events passing $e\mu$ NN selection} \\
\multicolumn{1}{l|}{}&       & Total       & Mis-id & Physics &  \\
\multicolumn{1}{l|}{}& Data  & sig + bkg  & bkg & bkg    & $\ttb$ \\
\hline 
$E_{T}^{e} > 15$ GeV, $p_{T}^{\mu} > 15$ GeV       &   &   &   &   &   \\

+ $e$ id + $\mu$ id + trig & 130 & $98  \pm 12  $ & $54   \pm 2  $ & 
$40  \pm 9  $ & $4.3 \pm 0.9$ \\

+ $\Delta R({\mu,{\rm jet}}) > 0.5 $, $\Delta R({e,\mu}) > 0.25 $ 
     &  58 & $54  \pm 9  $ & $12 \pm 1$ & $39  \pm 8  $ & $3.4 \pm 0.7$ \\

+ $\metcal > 15$ GeV            & 44 & $42 \pm 8$ & $5.8 \pm 0.5$ & $32 \pm 7$ 
     & $3.3 \pm 0.7$ \\

+ 2 jets, $E_{T}^{\rm jet} > 15$ GeV  & 6 & $4.4 \pm 0.9$ & $0.68 \pm 0.17$ 
     & $0.85\pm 0.21$ & $2.8 \pm 0.7$ \\

+ $O^{\rm comb}_{NN} \geq 0.88$ & 4 & $2.5 \pm 0.7$ & $0.04 \pm 0.12$ 
     & $0.19 \pm 0.07$ & $2.3 \pm 0.5$

\end{tabular}
\end{table*}

To further explore this channel, a more sophisticated, 
independent analysis is also performed.
The basic scheme begins with a loose selection and then uses a
neural network (NN) to maximize the significance.

This newer analysis is based on the same data set and trigger requirements
described in Sec.~\ref{emu}, and the initial selection is similar.
After passing the trigger requirement, events are 
required to have at least 1 loose electron with $E_T > 15$ GeV, 
$|\eta| \leq 2.5$ and at least 1 loose muon with $p_T > 15$ GeV/$c$. 
A cut of $\Delta R(\mu,{\rm jet}) > 0.5$ is then applied to reduce 
background from QCD multijet events containing a misidentified electron and a 
non-isolated muon. To remove QCD multijet events in which a misidentified 
electron and an isolated muon arise from the same jet, a cut of 
$\Delta R(e,\mu) > 0.25$ is applied. 
As can be seen in Table~\ref{tab:emunncuts}, at this stage the backgrounds
from QCD multijet events containing a misidentified electron and an 
isolated muon 
from the semi-leptonic decay of a $b$ or $c$ quark and $W(\ra\mu\nu)+$jets
events in which one of the jets is misidentified as an electron is still 
non-negligible.
A cut $\metcal\geq15$ GeV eliminates the multijet events with low \metcal
and rejects the majority of the $W(\ra\mu\nu)+$jet events (as noted above,
for $W(\ra\mu\nu)+$jet events, $\metcal$ is a measure of the transverse 
momentum of the $W$ boson).
The background at this stage consists primarily of dijet events with a
misidentified electron and an isolated muon from semileptonic 
$b$ or $c$ quark decay
(note that the muon momentum contributes to the measured \metcal).
This background is effectively eliminated by requiring two jets with 
$E_T \geq 15$ GeV. At this stage the background is a mixture of QCD 
multijet (including $W(\ra\mu\nu)+$jet events), 
$Z/\gamma^* \ra\tau\tau\ra e\mu$, and $WW\ra e\mu$ events.
For the remaining stages of event selection, neural network techniques 
are used. 

The optimal discrimination between signal and background can be achieved 
using three separate networks~\cite{hsthesis}. Each of these discriminates 
between the signal and one of the dominant backgrounds:
\begin{itemize}
\item Network 1 (NN1): $t\bar t$ vs QCD multijet events
\item Network 2 (NN2): $t\bar t$ vs $WW\ra e\mu$ events
\item Network 3 (NN3): $t\bar t$ vs $Z\ra\tau\tau\ra e\mu$ events
\end{itemize}
Training is performed on large samples of data (QCD multijet) and MC 
($t\bar t$, $WW$, $Z\ra\tau\tau$) events. To reduce bias, these samples are 
prepared with a minimal selection criteria of $E_T^e \geq 10$ GeV, 
$p_T^\mu \geq 10$ GeV/$c$, and $N_{\rm jets} \geq 1$ with 
$E_T^{\rm jet} \geq 10$ GeV. From these, a small sub-sample of 1000--2000
events is selected at random to provide the training sample.
The number of nodes and the input parameters for each network are selected
to maximize discrimination between signal and background. 
The best results are obtained using three identical networks, each with 
six input nodes, seven hidden nodes, and one output node.
The input parameters, which consist of five energy and one topological 
variable for each of the three networks, are listed below:
\begin{itemize}
\item Variables used in NN1 and NN2
  \begin{itemize}
  \item $E_T^e$, transverse energy of leading electron
  \item $E_T^{\rm jet2}$, transverse energy of next to leading jet
  \item $\metcal$, missing transverse energy as measured by the calorimeter
  \item $H_T^{\rm jets}$, scalar sum of jet transverse energies
     \begin{eqnarray*}
       H_T^{\rm jets}  =  & 
                    \displaystyle{\sum_{\rm all \ jets} E_T^{\rm jet}}, \ 
                     & {\rm with \ } | \eta^{\rm jet} | \leq 2.5   \\  
                  &  & {\rm and \ }  E_T^{\rm jet} \geq 15 {\rm \ GeV}
     \end{eqnarray*}
  \item $M({e\mu})$, electron-muon invariant mass
  \item $\Delta\phi({e\mu})$, azimuthal separation of the leading electron
        and muon
  \end{itemize}
\item Variables used in NN3
  \begin{itemize}
  \item same as NN1 and NN2 except that $E_T^{\rm jet1}$ replaces 
        $E_T^{\rm jet2}$ (transverse energy of leading jet)
  \end{itemize}
\end{itemize}
Each of the three networks is trained for 2000 training cycles. Training
is started with a set of random weights and thresholds which are adjusted 
using back propagation as the training proceeded. During training the target 
outputs are set to unity for signal and zero for background. 
For simplicity, the outputs of the three networks, $O_{NN1}, O_{NN2}, 
O_{NN3}$, are combined into an overall discriminant, 
\begin{equation}
O_{NN}^{\rm comb} = 
      {3\over{ {1\over{O_{NN1}}} + {1\over{O_{NN2}}} + {1\over{O_{NN3}}} } },
\end{equation}
which gives the probability that a given event is signal.
The output from such a combination is equivalent to that from a single network
that was trained on each of the three different backgrounds and the 
signal~\cite{hsthesis}. Testing on independent samples found that a 
requirement of
$O_{NN}^{\rm comb} \geq 0.88$ maximized the relative significance (which is
defined to be the ratio of expected number of signal events to the measured 
uncertainty on the number of background events).

\begin{table}
\caption{Expected number of signal and background events 
after all cuts in 112.6 ${\rm pb}^{-1}$ for the $e\mu$ neural network 
analysis. Uncertainties are statistical and systematic contributions 
added in quadrature. The systematic uncertainty on the total background 
includes the correlations among the different background sources.
\label{tab:emunnexp}}
\vskip 0.5cm
\begin{tabular}{c|c}
\multicolumn{2}{c}{Expected number of $e\mu$ NN evts in 
112.6 ${\rm pb}^{-1}$ } \\
\hline 
\underline{$t\bar t$ MC  $m_{t}$ (GeV/$c^2$)} &     \\
                   150       &  $3.51 \pm 0.86 $ \\
                   160       &  $2.84 \pm 0.68 $ \\
                   170       &  $2.30 \pm 0.53 $ \\
                   180       &  $1.81 \pm 0.41 $ \\
\hline
$Z\ra \tau\tau \ra e\mu$          & $0.10 \pm 0.07 $ \\
multijet (mis-id $e$)             & $0.04 \pm 0.12 $ \\ 
$WW\ra e\mu$                      & $0.08 \pm 0.02 $ \\
${\rm DY}\ra \tau\tau \ra e\mu$   & $0.01 \pm 0.01 $ \\
\hline
Total background   & $0.23 \pm 0.14 $ \\
\end{tabular}
\end{table}

After this selection four candidate events remain, three of which are also
selected by the conventional analysis. 

\begin{figure}[t]
\vbox{
\centerline{\hskip -0.8cm \psfig{figure=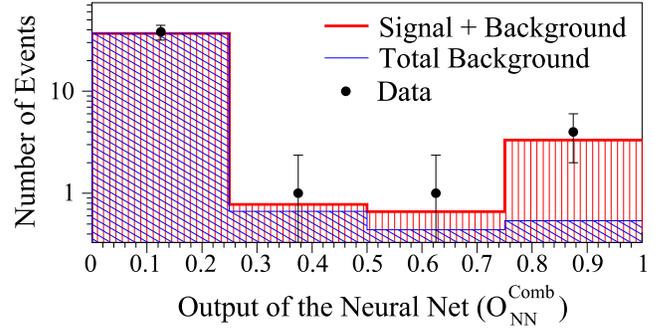,width=3.6in}}
\vskip -3.5cm
\caption{Distribution of signal+background (vertical hatching), 
background (diagonal hatching), and data (circles) as a function of 
neural network output. All initial cuts have been made except the 
requirement of 2 jets with $E_T > 15$ GeV.
\label{fig:emunn}}}
\end{figure}

Backgrounds and acceptances are estimated in much the same way as is done 
for the conventional analysis. The only real difference is that an 
additional correction is made for the effect of multiple 
interactions. This correction is obtained by comparing special MC samples
with one and two minimum bias events added with the standard MC samples which 
have no minimum bias events added. The acceptance variation is parameterized 
as a linear function of the number of interactions and a correction factor is 
obtained by applying this function to the distribution of the number of
interactions throughout Run 1. A correction factor of 9\% was found for 
$t\bar t$ events; since the $Z/\gamma^*$ and $WW$ backgrounds are 
kinematically and topologically similar, they receive the same correction.
The QCD multijet background, being derived from data, does not require such a 
correction.
The expected numbers of signal and background events passing the full selection
are given in Table~\ref{tab:emunnexp}. Figure~\ref{fig:emunn} shows a 
comparison of data
and the expected signal and background as a function of neural network output
after all initial cuts except the requirement of 2 jets with $E_T > 15$ GeV.
A cross section of $ 9.75 \pm 5.18 ({\rm stat}) \pm 1.95 ({\rm sys})$ pb
is obtained for the NN-based analysis which is in agreement with the
value of $ 7.1 \pm 4.8 $ pb obtained for the conventional analysis.
Comparison of acceptances and background expectations between the two analyses
finds the NN analysis with an increase in acceptance of 10\% 
(for $m_t$ = 172 \gevccn) for the same background. 
Table~\ref{tab:emunncuts} shows the number of data events, expected signal 
($m_t = 170$ GeV/$c^2$), and expected background surviving at each stage of 
the selection. As with the conventional analysis, good agreement is seen 
between what is observed and what is expected.

\begin{table}
\caption{Efficiency times branching fraction 
($\varepsilon \times {\cal B}$) and
statistical and systematic uncertainties (in percent), and expected background
and corresponding statistical and systematic uncertainties (in number of
events), for the $e\mu$ neural network analysis.
\label{tab:emunnerr}}
\vskip 0.5cm
\begin{tabular}{l|c|cccc}
           & 
           & \multicolumn{4}{c}{Expected \# of Bkg events}   \\
   & $\varepsilon \times {\cal B}$(\%)  
                       &   $Z\tau\tau$  & $WW$ &  DY$\tau\tau$ & Mis-id $e$ \\
\hline 
                       & 0.351 & 0.095 & 0.077 & 0.006 & 0.044 \\
\hline
Statistical            & 0.004 & 0.055 & 0.006 & 0.004 & 0.117 \\
Luminosity             &   --   & 0.005 & 0.004 & 0.000 &  --  \\
Energy Scale           & 0.035 & 0.026 & 0.011 & 0.004 &   --  \\
$e$ id                 & 0.010 & 0.004 & 0.002 & 0.000 &   --  \\
High-$p_{T} \ \mu$ id  & 0.037 & 0.011 & 0.008 & 0.001 &   --  \\
$e$+jets trig          & 0.020 & 0.005 & 0.004 & 0.000 &   --  \\
MC generator           & 0.025 & 0.005 & 0.004 & 0.000 &   --  \\
Top quark mass         & 0.020 &   --  &   --  &   --  &   --  \\
Bkg crsec              &   --  & 0.010 & 0.008 & 0.001 &   --  \\
Other sim              &   --  & 0.022 &   --  &   --  &   --  \\
Mis-id $e$             &   --  &   --  &   --  &   --  & 0.003 \\
\hline
Total                  & 0.065 & 0.067 & 0.018 & 0.005 & 0.117 \\
\end{tabular}
\end{table}

Systematic uncertainties are handled the same way as in the conventional
analysis and are summarized in Table~\ref{tab:emunnerr} and, with the
exception of the uncertainty on the efficiency times branching fraction 
due to the top quark mass, are discussed in 
Sec.~\ref{syserr}. The value of $m_t$ measured 
by D\O\ is $172.1 \pm 7.1 \gevcc$ \cite{ljmprd,mllprd} and the central 
value is used in the 
calculation of the efficiency times branching fraction.
This uncertainty of $\pm 7~\gevcc$ is composed of an 
uncertainty of 4.0~\gevcc due to jet energy scale, 1.9~\gevcc due to the
$t\bar t$ MC generator, and 6.1~\gevcc due to statistics and other sources.
The effect of these uncertainties is determined by parameterizing the
efficiency times branching fraction as a linear function of top quark mass 
in the region between 165 and 180 \gevccn. This parameterization is used
to convert the above uncertainties on $m_t$ into uncertainties on efficiency
times branching fraction. The uncertainties on $m_t$ due to the jet energy 
scale and $t\bar t$ generator translate into uncertainties on 
$\varepsilon \times {\cal B}$ of 3.3\% and 1.6\%, respectively. These 
uncertainties are combined with the other jet energy scale and $t\bar t$ 
generator uncertainty contributions (described in Sec.~\ref{syserr}).
The uncertainties on $m_t$ due to statistics and from other sources translate
into an uncertainty of 5.8\% on $\varepsilon \times {\cal B}$, and are 
included as a separate source of uncertainty in Table~\ref{tab:emunnerr}.
As discussed for the conventional dilepton analyses, there
is a significant discrepancy between data and MC for the jet $E_T$ spectra 
in $Z+$jet events. The conventional analyses correct for this 
by taking the jet cut survival probabilities from data and applying them to
the MC. Such a procedure is not possible with a NN analysis. 
Fortunately the primary disagreement between data and MC is in 
$E_T^{\rm jet 2}$. 
It is for this reason that the variables used for NN3 differ from those
for NN1 and NN2 in that NN3 employs $E_T^{\rm jet 1}$ instead of 
$E_T^{\rm jet 2}$. To account for the uncertainty due to the initial
jet cuts of $N_{\rm jets} \geq 2$ with $E_T^{\rm jet} \geq 15$~GeV,
a data versus MC comparison was made and a difference of 21\% was found.
This uncertainty 
is listed under the category ``Other simulation'' in Table~\ref{tab:emunnerr}
and is applied only to the $Z\ra\tau\tau$ background.

%
\section{Treatment of Uncertainties}
\label{correrr}

\def\cs{{\sigma_{t\bar t}}}


As shown in Eq.~\ref{eq:totcseccalc},
calculation of the $t \bar t$ production cross section requires as input 
the number of observed events found in all channels, the total expected 
background, the individual channel acceptance for $t \bar t$  
events, and the integrated luminosity for each channel.
To simplify the discussion, Eq.~\ref{eq:totcseccalc} can be written in the
form
\begin{equation}
\label{eq:totcseccalc2}
  \sigma(m_t)_{t\bar t} = {{\displaystyle N - B }
  \over { \displaystyle A {\cal L} }} 
\end{equation}
where 
\begin{eqnarray}
   N &=& \sum_i N_i \\
   B &=& \sum_j B_j 
\label{eq:syserrbtot} \\
   A &=& {1 \over {\cal L} } \sum_i A_i {\cal L}_i 
\label{eq:syserratot} \\
   {\cal L} &=& \sum_i {\cal L}_i.
\label{eq:syserrlum}
\end{eqnarray}
with the sum $i$ being over all nine channels and the sum $j$ being over
all backgrounds.


It is assumed that the backgrounds 
and acceptances are subject to the same kinds of uncertainties
(see Secs.~\ref{ss:lumerr} -- ~\ref{ss:zfiterr}) and that no 
correlation exists among the different uncertainties. 
With these assumptions, the background error matrix is given by 
\begin{equation}
\label{eq:syserrbmat}
\delta^2_{Bij} = \sum_\mu \delta^2_{B\mu ij} ,
\end{equation}
where $i$ and $j$ represent the various backgrounds in the different channels
({\it e.g.} $W$+jets background in $e$+jets/topo channel), $\mu$
represents the source of uncertainty ({\it e.g.} electron identification), 
and the
error matrix for a given uncertainty, $\delta^2_{B\mu ij}$, is computed from
first principles according to the equation
\begin{equation}
\label{eq:syserrbmatij}
\delta_{B\mu ij}^2 =  \langle B_i B_j \rangle_\mu - 
                      \langle B_i \rangle_\mu \langle B_j \rangle_\mu , 
\end{equation}
where the symbol $\langle ... \rangle_\mu$ represents the average of the 
enclosed quantity when it is varied according to the uncertainty $\mu$.
Accordingly, the correlation matrix for a given uncertainty is given by
\begin{equation}
\label{eq:syserrbcormat}
C_{B\mu ij} = {{\delta^2_{B\mu ij}} \over { \delta_{B\mu i} \delta_{B\mu j} }},
\end{equation}
where $\delta_{B\mu i}$ is the uncertainty on background $i$ due to source
$\mu$.

With these definitions, the uncertainty on the total background is obtained
from Eq.~\ref{eq:syserrbtot} by propagation of errors
\begin{eqnarray}
\delta_{B}^2 &=& 
  \sum_{ij} \left( \partial B \over \partial B_i \right)
            \left( \partial B \over \partial B_j \right)
            \delta_{Bij}^2  \\
&=& \sum_{ij} \delta_{Bij}^2  \\
&=& \sum_{\mu} \sum_{ij} \delta_{B\mu ij}^2  \\
&=& \sum_{\mu} \left(\delta_B\right)_\mu^2,
\label{eq:syserrbcormat1}
\end{eqnarray}
where
\begin{eqnarray}
\left(\delta_B\right)_\mu^2 &=& 
  \sum_{ij} \delta_{B\mu ij}^2  \\
&=& \sum_{ij} C_{B\mu ij} \delta_{B\mu i} \delta_{B\mu j}.
\label{eq:syserrbcormat2}
\end{eqnarray}

In the case of uncorrelated errors ($C_{B\mu ij} = $ the unit matrix), 
Eq.~\ref{eq:syserrbcormat2} reduces to the usual quadratic sum formula,
\begin{eqnarray}
\left(\delta_B\right)_\mu^2 &=& 
  \sum_{i}  \delta_{B\mu i}^2.
\label{eq:syserrbcormat3}
\end{eqnarray}
In the case of maximal positive correlation ($C_{B\mu ij}$ populated
entirely by 1's), Eq.~\ref{eq:syserrbcormat2} reduces to a linear sum of
errors,
\begin{eqnarray}
\left(\delta_B\right)_\mu^2 &=& 
  \left( \sum_{i}  \delta_{B\mu i} \right)^2. 
\label{eq:syserrbcormat4}
\end{eqnarray}

For the analyses in this paper, all uncertainties are handled according to
one of these two limiting cases. Statistical uncertainties are handled by
the quadratic sum formula (Eq.~\ref{eq:syserrbcormat3}) and systematic
uncertainties are handled according to the linear sum formula 
(Eq.~\ref{eq:syserrbcormat4}). The total uncertainty on the background is
therefore 
\begin{equation}
\label{eq:syserrbcormat5}
\delta_{B}^2 =   \sum_{\mu={\rm stat}} \sum_{i} \delta_{B\mu i}^2 + 
                 \sum_{\mu={\rm sys}} 
                 \left( \sum_{i} \delta_{B\mu i} \right)^2.
\end{equation}
The importance of correlations for the background calculation as a whole 
depend on the extent to which different backgrounds are affected by the 
same systematic uncertainties. The systematic uncertainties for all 
backgrounds to all
channels are given in Tables~\ref{tab:sysdilbk} -- \ref{tab:sysenuajbk}.


Applying the steps above to Eq.~\ref{eq:syserratot},
the uncertainty on $A$ is found to be
\begin{eqnarray}
\delta_A^2 &=& 
  {1\over {{\cal L}^2}} \sum_{\mu} \sum_{ij} 
  C_{A\mu ij}\delta_{A\mu i}\delta_{A\mu j} {\cal L}_i {\cal L}_j \\
 &=& {1\over {{\cal L}^2}} \sum_{\mu={\rm stat}} \sum_{i} 
     \delta_{A\mu i}^2 {\cal L}_i^2 \nonumber \\
 & & + {1\over {{\cal L}^2}}
     \sum_{\mu={\rm sys}} \left( \sum_{i} \delta_{A\mu i} {\cal L}_i \right)^2.
\label{eq:syserrda2}
\end{eqnarray}
Systematic uncertainties on the acceptance ($\varepsilon \times {\cal B}$) 
are given for all channels in Table~\ref{tab:syssig}.
Note that the acceptance uncertainties are highly correlated 
due to the fact that the calculation for each channel
is affected by essentially the same set of systematic uncertainties.
The same relative uncertainty on the luminosity has been assumed for all
channels (see Sec.~\ref{ss:lumerr}).
 
The total uncertainty on the top quark cross section is obtained by 
propagation of errors using Eq.~\ref{eq:totcseccalc2}.  
The four inputs to the cross section
can, in principle, give rise to six different correlation terms.
However, the signal ($N$) has only a statistical uncertainty and
the uncertainties on the acceptance ($A$) and 
the integrated luminosity ($\cal L$) are uncorrelated.
Therefore, the only correlation terms are those between the background
($B$) and the acceptance ($A$) and between the background ($B$) and the 
integrated luminosity ($\cal L$). The corresponding uncertainties are
given by the equations
\begin{eqnarray}
\delta_{BA}^2 &=&  \sum_{\mu} \left( \delta_{BA} \right)_\mu^2 
{\rm \ \ \ \ and} \\
\delta_{B{\cal L}}^2 &=&  \sum_{\mu} \left( \delta_{B{\cal L}} \right)_\mu^2 .
\end{eqnarray}
The error corresponding to a given uncertainty ($\mu$) is 
calculated from first principles according to the equations
\begin{eqnarray}
\left( \delta_{BA} \right)_\mu^2 &=&  \langle BA \rangle_\mu - \langle B \rangle_\mu \langle A \rangle_\mu {\rm \ \ \ \ and} \\ 
\left( \delta_{B{\cal L}} \right)_\mu^2 &=&  \langle B{\cal L} \rangle_\mu - \langle B \rangle_\mu \langle {\cal L} \rangle_\mu ,
\label{eq:syserrdbldef}
\end{eqnarray}
where the symbol $\langle ... \rangle$ represents the average of the 
enclosed quantities when they are varied according to the uncertainties $\mu$. 
The correlation coefficients are given by 
\begin{equation}
\label{eq:syserrbablcormat}
C_{BA} = {{\delta^2_{BA}} \over { \delta_{B} \delta_{A} }}
{\rm \ \ \ and \ \ \ }
C_{B{\cal L}} = {{\delta^2_{B{\cal L}}} \over { \delta_{B} \delta_{\cal L} }}.
\end{equation}
In the linear/quadratic approximation, these correlation coefficients 
simplify to
\begin{eqnarray}
C_{BA} &=& {1\over {\delta_B \delta_A}} \sum_\mu C_{BA\mu} 
           \left(\delta_B\right)_\mu \left(\delta_A\right)_\mu , \\
       &=& {1\over {\delta_B \delta_A}} \sum_{\mu = {\rm sys}} 
           \left(\delta_B\right)_\mu \left(\delta_A\right)_\mu, 
\end{eqnarray}
and
\begin{eqnarray}
C_{B{\cal L}} &=& {1\over {\delta_B \delta_{\cal L}}} \sum_\mu C_{B{\cal L}\mu} 
           \left(\delta_B\right)_\mu \left(\delta_{\cal L}\right)_\mu , \\
       &=& {1\over {\delta_B \delta_{\cal L}}} \sum_{\mu = {\rm sys}} 
           \left(\delta_B\right)_\mu \left(\delta_{\cal L}\right)_\mu, \\
       &=& {1\over {\delta_B \delta_{\cal L}}} 
           \left(\delta_B\right)_{\cal L} \left(\delta_{\cal L}\right), \\
       &=& {1\over {\delta_B}} 
           \left(\delta_B\right)_{\cal L}.
\label{eq:cc2}
\end{eqnarray}


The total uncertainty on the top quark cross 
section is therefore given by
\begin{eqnarray}
\delta_{\cs}^{2} &=& 
  \nonumber
  \left( \partial\cs \over \partial N \right)^2 \delta_N^2 + 
  \left( \partial\cs \over \partial B \right)^2 \delta_B^2 + 
  \left( \partial\cs \over \partial A \right)^2 \delta_A^2 \\ \nonumber
&&+\left( \partial\cs \over \partial {\cal L} \right)^2 \delta_{\cal L}^2 +
  \left( \partial\cs \over \partial B \right)
  \left( \partial\cs \over \partial A \right)
  \delta_{BA}^2 \\ 
\label{eq:toterr1}
&&+\left( \partial\cs \over \partial B \right)
  \left( \partial\cs \over \partial {\cal L} \right)
  \delta_{B\cal L}^2, \\ \nonumber
  &=&\cs^2  \left[
  {\delta_N^2 + \delta_B^2 \over \left( N-B \right)^2 } + 
  {\delta_A^2 \over A^2} + {\delta_{\cal L}^2 \over {\cal L}^2} \right. \\
&& \ \ \ \ \ \ \ +{ \delta_B \over N-B } \left. \left( C_{BA} 
  {\delta_A \over A} + C_{B\cal L} {\delta_{\cal L} \over {\cal L}} \right) 
  \right] .
\label{eq:toterr2}
\end{eqnarray}

\bibliographystyle{csecsty}
\bibliography{top_prd26}

\begin{thebibliography}{100}
\input list_of_visitor_addresses_r1.tex
\bibitem{cdfdisc}
{CDF} Collaboration, F. Abe {\it et~al.}, Phys. Rev. Lett. {\bf 74},  2626
  (1995).

\bibitem{d0disc}
{\dzero} Collaboration, S. Abachi {\it et~al.}, Phys. Rev. Lett. {\bf 74},
  2632  (1995).

\bibitem{taudisc}
M.~L. Perl {\it et~al.}, Phys. Lett. {\bf 70B},  487  (1977).

\bibitem{bdisc}
S.~W. Herb {\it et~al.}, Phys. Rev. Lett. {\bf 39},  252  (1977).

\bibitem{KM}
M. Kobayashi and T. Maskawa, Prog. Theor. Phys. {\bf 49},  652  (1973).

\bibitem{qb1}
C. Berger {\it et~al.}, Phys. Lett. {\bf 76B},  243  (1978).

\bibitem{qb2}
C.~W. Darden {\it et~al.}, Phys. Lett. {\bf 76B},  246  (1978).

\bibitem{qb3}
J.~K. Bienlein {\it et~al.}, Phys. Lett. {\bf 78B},  360  (1978).

\bibitem{i31}
W. Bartel {\it et~al.}, Phys. Lett. {\bf 146B},  437  (1984).

\bibitem{i32}
G.~L. Kane and M.~E. Peskin, Nucl. Phys. {\bf B195},  29  (1982).

\bibitem{i33}
C. Matteuzzi {\it et~al.}, Phys. Lett. {\bf 129B},  141  (1983).

\bibitem{petra1}
{PLUTO} Collaboration, C. Berger {\it et~al.}, Phys. Lett. {\bf 86B},  413
  (1979).

\bibitem{petra2}
{CELLO} Collaboration, H. Behrend {\it et~al.}, Phys. Lett. {\bf 144B},  297
  (1984).

\bibitem{tristan}
{AMY} Collaboration, H. Sagawa {\it et~al.}, Phys. Rev. Lett. {\bf 60},  93
  (1988).

\bibitem{markII}
{MARKII} Collaboration, G.~S. Abrams {\it et~al.}, Phys. Rev. Lett. {\bf 63},
  2447  (1989).

\bibitem{opal}
{OPAL} Collaboration, M. Akrawy {\it et~al.}, Phys. Lett. B {\bf 236},  364
  (1990).

\bibitem{ua11988}
UA1 Collaboration, C. Albajar {\it et~al.}, Z. Phys. C {\bf 37},  505  (1988).

\bibitem{ua11990}
UA1 Collaboration, C. Albajar {\it et~al.}, Z. Phys. C {\bf 48},  1  (1990).

\bibitem{ua21990}
UA2 Collaboration, T. {\AA}kesson {\it et~al.}, Z. Phys. C {\bf 46},  179
  (1990).

\bibitem{cdf1990}
{CDF} Collaboration, F. Abe {\it et~al.}, Phys. Rev. Lett. {\bf 64},  142
  (1990).

\bibitem{cdf1992}
{CDF} Collaboration, F. Abe {\it et~al.}, Phys. Rev. Lett. {\bf 68},  447
  (1992).

\bibitem{d01994}
{\dzero} Collaboration, S. Abachi {\it et~al.}, Phys. Rev. Lett. {\bf 72},
  2138  (1994).

\bibitem{cdf1994}
{CDF} Collaboration, F. Abe {\it et~al.}, Phys. Rev. D {\bf 50},  2966  (1994).

\bibitem{glasgowd0}
{\dzero} Collaboration, P. Grannis,  in {\em Proceedings of the XXVII
  International Conference on High Energy Physics, Glasgow, Scotland} (IOP,
  Bristol, UK, 1994).

\bibitem{d01997}
{\dzero} Collaboration, S. Abachi {\it et~al.}, Phys. Rev. Lett. {\bf 79},
  1197  (1997).

\bibitem{cdf1998}
{CDF} Collaboration, F. Abe {\it et~al.}, Phys. Rev. Lett. {\bf 80},  2773
  (1998).

\bibitem{cdf2001}
{CDF} Collaboration, T. Affolder {\it et~al.}, Phys. Rev. D {\bf 64},  032002
  (2001).

\bibitem{singtop}
S. Willenbrock and D. Dicus, Phys. Rev. D {\bf 34},  155  (1986).

\bibitem{d0singt1}
{\dzero} Collaboration, B. Abbott {\it et~al.}, Phys. Rev. D {\bf 63},  031101
  (2001).

\bibitem{d0singt2}
{\dzero} Collaboration, V. Abazov {\it et~al.}, Phys. Lett. B {\bf 517},  282
  (2001).

\bibitem{css}
J.~C.~D. Soper and G. Sterman, Nucl. Phys. {\bf B263},  37  (1986).

\bibitem{georgi78}
H. Georgi, S. Glashow, M. Machacek, and D. Nanopoulos, Ann. Phys. {\bf 114},
  273  (1978).

\bibitem{jones78}
L. Jones and H. Wyld, Phys. Rev. D {\bf 17},  1782  (1978).

\bibitem{gluck78}
M. Gl{\"{u}}ck, J. Owens, and E. Reya, Phys. Rev. D {\bf 17},  2324  (1978).

\bibitem{babcock78}
J. Babcock, D. Sivers, and S. Wolfram, Phys. Rev. D {\bf 18},  162  (1978).

\bibitem{hagiwara79}
K. Hagiwara and T. Yoshino, Phys. Lett. {\bf 80B},  282  (1979).

\bibitem{combridge79}
B. Combridge, Nucl. Phys. {\bf B151},  429  (1979).

\bibitem{dawson88}
P. Nason, S. Dawson, and R.~K. Ellis, Nucl. Phys. {\bf B303},  607  (1988).

\bibitem{dawson89}
P. Nason, S. Dawson, and R.~K. Ellis, Nucl. Phys. {\bf B327},  49  (1989).

\bibitem{beenakker89}
W. Beenakker, H. Kuijf, W. van Neerven, and J. Smith, Phys. Rev. D {\bf 40},
  54  (1989).

\bibitem{beenakker91}
W. Beenakker {\it et~al.}, Nucl. Phys. {\bf B351},  507  (1991).

\bibitem{mangano92}
M. Mangano, P. Nason, and G. Ridolfi, Nucl. Phys. {\bf B373},  295  (1992).

\bibitem{altarelli88}
G. Altarelli, M. Diemoz, G. Martinelli, and P. Nason, Nucl. Phys. {\bf B308},
  724  (1988).

\bibitem{ellis91}
R. Ellis, Phys. Lett. B {\bf 259},  492  (1991).

\bibitem{laenen}
E. Laenen, J. Smith, and W.~L. van Neerven, Phys. Lett. B {\bf 321},  254
  (1994).

\bibitem{berger3}
E. Berger and H. Contopanagos, Phys. Rev. D {\bf 57},  253  (1998).

\bibitem{bonciani}
R. Bonciani, S. Catani, M. Mangano, and P. Nason, Nucl. Phys. {\bf B529},  424
  (1998).

\bibitem{kidonakis99}
N. Kidonakis and R. Vogt, Phys. Rev. D {\bf 59},  074014  (1999).

\bibitem{kidonakis0105}
N. Kidonakis, Phys. Rev. D {\bf 64},  014009  (2001).

\bibitem{kidonakis}
N. Kidonakis, E. Laenen, S. Moch, and R. Vogt, Phys. Rev. D {\bf 64},  114001
  (2001).

\bibitem{d0chiggs}
{\dzero} Collaboration, B. Abbott {\it et~al.}, Phys. Rev. Lett. {\bf 82},
  4975  (1999).

\bibitem{d0chiggs2}
{\dzero} Collaboration, V.~M. Abazov {\it et~al.}, Phys. Rev. Lett. {\bf 88},
  151803  (2002).

\bibitem{cdfchiggs}
{CDF} Collaboration, F. Abe {\it et~al.}, Phys. Rev. Lett. {\bf 79},  357
  (1999).

\bibitem{herwig}
G. Marchesini {\it et~al.}, Comp. Phys. Comm. {\bf 67},  465  (1992).

\bibitem{vecbos}
F.~A. Berends, H. Kuijf, B. Tausk, and W.~T. Giele, Nucl. Phys. {\bf B357},  32
   (1991).

\bibitem{alljets}
{\dzero} Collaboration, B. Abbott {\it et~al.}, Phys. Rev. D {\bf 60},  12001
  (1999). The all-jets $t\bar t$ cross section given in this paper is $7.1
  \pm 2.8 {\rm (stat)} \pm 1.5 {\rm (sys)}$ pb.

\bibitem{topstrat}
H. Baer, V. Barger, and R.~J.~N. Phillips, Phys. Rev. D {\bf 39},  3310
  (1989).

\bibitem{prd1}
{\dzero} Collaboration, S. Abachi {\it et~al.}, Phys. Rev. D {\bf 52},  4877
  (1995).

\bibitem{d0nim}
{\dzero} Collaboration, S. Abachi {\it et~al.}, Nucl. Instrum. Methods Phys.
  Res. A {\bf 338},  185  (1994).

\bibitem{wwidth}
{\dzero} Collaboration, B. Abbott {\it et~al.}, Phys. Rev. D {\bf 61},  072001
  (2000).

\bibitem{cdflum}
{CDF} Collaboration, F. Abe {\it et~al.}, Phys. Rev. D {\bf 50},  5550  (1994).

\bibitem{e7101990}
{E710} Collaboration, N. Amos {\it et~al.}, Phys. Rev. Lett. {\bf 68},  2433
  (1992).

\bibitem{e811}
{E811} Collaboration, C. Avila {\it et~al.}, Phys. Lett. B {\bf 445},  419
  (1999).

\bibitem{elecres1}
{\dzero} Collaboration, B. Abbott {\it et~al.}, Phys. Rev. Lett. {\bf 80},
  3008  (1998).

\bibitem{elecres2}
{\dzero} Collaboration, B. Abbott {\it et~al.}, Phys. Rev. D {\bf 58},  092003
  (1998).

\bibitem{d0geant}
{\dzero} Collaboration, J. Womersley,  in {\em Proceedings of the XXVI
  International Conference on High Energy Physics, Dallas, Texas}, edited by
  J.~R. Sanford (AIP, New York, 1993).

\bibitem{ewv}
E. Varnes, Ph.D. thesis, University of California, Berkeley, 1997
  (unpublished), available from {\tt http://www-\\
  d0.fnal.gov/results/publications\_talks/thesis-\\/varnes/thesis.ps}. Top
  quark theses for the \dzero \\ experiment can be found from {\tt
  http://www-d0.fnal.-\\gov/results/publications\_talks/thesis/Top.html}.

\bibitem{bkthesis}
R. Kehoe, Ph.D. thesis, University of Notre Dame, 1997 (unpublished), available
  from {\tt http://www-d0.fnal.-\\
  gov/results/publications\_talks/thesis/kehoe}.

\bibitem{ptthesis}
P. Tamburello, Ph.D. thesis, University of Maryland, 1997 (unpublished),
  available from {\tt http://www-d0.-\\
  fnal.gov/results/publications\_talks/thesis/tam-\\
  burello/thesis\_2sided.ps}.

\bibitem{geant}
R. Brun {\it et~al.}, {\em {\progname{GEANT}} Detector Description and
  Simulation Tool}, 1993, {CERN} Program Library Number Q123.

\bibitem{ajm}
A.~J. Milder, Ph.D. thesis, University of Arizona, 1993 (unpublished),
  available from {\tt http://www-d0.fnal-\\
  .gov/results/publications\_talks/thesis/milder/-\\ diss.ps}.

\bibitem{rva}
R.~V. Astur, Ph.D. thesis, Michigan State University, 1992 (unpublished).

\bibitem{cteq3m}
H.~L. Lai {\it et~al.}, Phys. Rev. D {\bf 51},  4763  (1995).

\bibitem{mrsap}
A. Martin, W. Stirling, and R. Roberts, Phys. Lett. B {\bf 354},  155  (1995).

\bibitem{dglap}
G. Altarelli and G. Parisi, Nucl. Phys. {\bf B126},  298  (1977).

\bibitem{isa}
H. Baer, F.~E. Paige, S.~D. Protopopescu, and X. Tata, {BNL} Report
  {HET-99/43}, {Brookhaven} (unpublished), {\tt
  http://xxx.lanl.gov/abs/hep-ph/0001086}.

\bibitem{feynfield}
R. Field and R. Feynman, Nucl. Phys. {\bf B136},  131  (1978).

\bibitem{colorcoh}
{\dzero} Collaboration, B. Abbott {\it et~al.}, Phys. Lett. B {\bf 414},  419
  (1997).

\bibitem{d0jetshape}
{\dzero} Collaboration, S. Abachi {\it et~al.}, Phys. Lett. B {\bf 357},  500
  (1995).

\bibitem{orr1}
L. Orr, T. Stelzer, and W. Stirling, Phys. Rev. D {\bf 52},  124  (1995).

\bibitem{orr2}
L. Orr, T. Stelzer, and W. Stirling, Phys. Lett. B {\bf 354},  442  (1995).

\bibitem{pythia}
T. Sj{\"{o}}strand, Comp. Phys. Comm. {\bf 82},  74  (1994).

\bibitem{d0zcsec}
{\dzero} Collaboration, B. Abbott {\it et~al.}, Phys. Rev. D {\bf 60},  052003
  (1999).

\bibitem{d0ptz}
{\dzero} Collaboration, B. Abbott {\it et~al.}, Phys. Rev. D {\bf 61},  032004
  (2000).

\bibitem{pdg}
R.~M. Barnett {\it et~al.}, Phys. Rev. D {\bf 54},  1  (1996).

\bibitem{jmthesis}
J. McKinley, Ph.D. thesis, Michigan State University, 1996 (unpublished),
  available from {\tt http://www-d0.-\\
  fnal.gov/results/publications\_talks/thesis/mckin-\\ ley/jtm\_thesis.ps}.

\bibitem{wwwz}
J. Ohnemus, Phys. Rev. D {\bf 50},  1931  (1994).

\bibitem{pbthesis}
P. Bloom, Ph.D. thesis, University of California, Davis, 1998 (unpublished),
  available from {\tt http://www-d0.-\\ fnal.gov/~pbloom/thesis.html}.

\bibitem{randgrid}
{\dzero} Collaboration, N. Amos,  in {\em Proc. Intl. Conf. on Computing in
  High Energy Physics, Rio de Janeiro, Brazil}, edited by R. Shellard and T.~D.
  Nguyen (World Scientific, Singapore, 1995).

\bibitem{momentens}
V. Barger, J. Ohnemus, and R.~J.~N. Phillips, Phys. Rev. D {\bf 48},  3953
  (1993).

\bibitem{optimize}
{\dzero} Collaboration, M. Narain,  in {\em Proc. Les Recontres de Physique de
  la Vall{\'{e}}e d'Aoste, La Thuile, Italy}, edited by M. Greco (World
  Scientific, Singapore, 1993).

\bibitem{ljmprd}
{\dzero} Collaboration, B. Abbott {\it et~al.}, Phys. Rev. D {\bf 58},  052001
  (1998).

\bibitem{berends1}
S.~D. Ellis {\it et~al.}, Phys. Lett. {\bf 154B},  435  (1985).

\bibitem{berends2}
F.~A. Berends {\it et~al.}, Nucl. Phys. {\bf B357},  32  (1991).

\bibitem{berger2}
E. Berger and H. Contopanagos, Phys. Rev. {\bf D54},  3085  (1996).

\bibitem{dirphot}
{\dzero} Collaboration, S. Abachi {\it et~al.}, Phys. Rev. Lett. {\bf 77},
  5011  (1996).

\bibitem{alljprl}
{\dzero} Collaboration, B. Abbott {\it et~al.}, Phys. Rev. Lett. {\bf 83},
  1908  (1999).

\bibitem{escalenim}
{\dzero} Collaboration, B. Abbott {\it et~al.}, Nucl. Instrum. Methods Phys.
  Res. A {\bf 424},  352  (1999).

\bibitem{jettopo}
{\dzero} Collaboration, S. Abachi {\it et~al.}, Phys. Rev. D {\bf 53},  6000
  (1996).

\bibitem{delphijets}
DELPHI Collaboration, P. Abreu {\it et~al.}, Z. Phys. C {\bf 73},  11  (1996).

\bibitem{mllprd}
{\dzero} Collaboration, B. Abbott {\it et~al.}, Phys. Rev. D {\bf 60},  052001
  (1999).

\bibitem{cand_params}
Detailed information on the candidates used in this analysis is available at
  {\tt http://www-d0.fnal.gov/-\\public/top/csec-candidates.ps}.

\bibitem{d0csec1997}
{\dzero} Collaboration, S. Abachi {\it et~al.}, Phys. Rev. Lett. {\bf 79},
  1203  (1997). The leptonic $t\bar t$ cross section given in this paper is
  $5.5 \pm 1.4 {\rm (stat)} \pm 0.9 {\rm (sys)}$ pb.

\bibitem{tev2000}
D. Amidei and R. Brock, {\em Future ElectroWeak Physics at the Fermilab
  Tevatron: Report of the TeV 2000 Study Group}, {FNAL} Report: Pub-96/082.

\bibitem{lhc}
F. Gianotti and M. Pepe-Altarelli, Nucl. Phys. B (Proc. Suppl.) {\bf 89},  177
  (2000), hep-ex/0006016.

\bibitem{lowepi}
H. Aihara {\it et~al.}, Nucl. Instrum. Methods Phys. Res. A {\bf 325},  393
  (1993).

\bibitem{calor96}
{\dzero} Collaboration, R. Kehoe,  in {\em Proceedings of the Sixth
  International Conference on Calorimetry in High Energy Physics, Frascati,
  Italy}, edited by A. Antonelli, S. Bianco, A. Calcaterra, and F. Fabbri
  (World Scientific, River Edge, NJ, 1996).

\bibitem{jttm}
J. Thompson, {FNAL} Report {TM-1909}, {Fermilab}.

\bibitem{hsthesis}
H. Singh, Ph.D. thesis, University of California, Riverside, 1999
  (unpublished), available from {\tt http://-\\
  www-d0.fnal.gov/results/publications\_talks-\\ /thesis/singh/thesis.ps}.

\end{thebibliography}

%
%
\end{document}